%% file: main.tex
\documentclass[a4paper12pt]{article}
\pdfoutput=1

\usepackage[english]{babel}
\usepackage{authblk}
\usepackage{longtable}
\usepackage{soul}
\usepackage{cite}
\usepackage{url}
\usepackage{amsmath}
\usepackage{amssymb}
\usepackage{hyperref}
\usepackage[T1]{fontenc}
\usepackage[utf8]{inputenc}
\usepackage{graphicx}
\usepackage{subfig}
\usepackage{multirow}
\usepackage{booktabs}
\usepackage{siunitx}
\usepackage{lineno}
\usepackage{multicol}
\usepackage{ulem}
\usepackage{float}
\hypersetup{
    colorlinks=true,
    linktoc=all,
    linkcolor=black,
    filecolor=magenta,      
    urlcolor=blue,
}
\expandafter\def\expandafter\UrlBreaks\expandafter{\UrlBreaks
  \do\a\do\b\do\c\do\d\do\e\do\f\do\g\do\h\do\i\do\j%
  \do\k\do\l\do\m\do\n\do\o\do\p\do\q\do\r\do\s\do\t%
  \do\u\do\v\do\w\do\x\do\y\do\z\do\A\do\B\do\C\do\D%
  \do\E\do\F\do\G\do\H\do\I\do\J\do\K\do\L\do\M\do\N%
  \do\O\do\P\do\Q\do\R\do\S\do\T\do\U\do\V\do\W\do\X%
  \do\Y\do\Z\do\*\do\-\do\~\do\'\do\"\do\-}%

\renewcommand\Affilfont{\fontsize{10}{10.8}\itshape}
\renewcommand\Authfont{\fontsize{13}{14.4}\selectfont}

\input{preamble.tex}

\begin{document}

\maketitle

\begin{abstract}
\noindent
The demand for underground labs for neutrino and rare event search experiments has been increasing over the last few decades. Yemilab, constructed in October 2022, is the first deep ($\sim$1~km) underground lab dedicated to science in Korea, where a large cylindrical cavern (D: 20~m, H: 20~m) was excavated in addition to the main caverns and halls. The large cavern could be utilized for a low background neutrino experiment by a liquid scintillator-based detector (LSC) where a 2.26 kiloton LS target would be filled. 
It's timely to have such a large but ultra-pure LS detector after the shutdown of the Borexino experiment so that solar neutrinos can be measured much more precisely. Interesting BSM physics searches can be also pursued with this detector when it's combined with an electron linac, a proton cyclotron (IsoDAR source), or a radioactive source. This article discusses the concept of a candidate detector and the physics potential of a large liquid scintillator detector. 
\end{abstract}

\clearpage

\tableofcontents

\newpage

\setcounter{section}{0}
\input{intro.tex}

\input{Site/site.tex}

\input{Detector/det.tex}
\input{Detector/LS.tex}

\input{Detector/WbLS.tex}
\input{Detector/slowLS.tex}
\input{Detector/PMT.tex}

\input{Facility/facility}
\input{Facility/LS_purification.tex}
\input{Facility/isodar_facility.tex}

\input{Facility/linac.tex}

\clearpage

\section{Physics Potential}
\input{Physics/physics.tex}
\clearpage
\input{Physics/Solar.tex}
\clearpage
\input{Physics/Geo_Reactor.tex}
\clearpage

\input{Physics/SN.tex}
\input{Physics/DSNB.tex}
\input{Physics/IsoDAR.tex}

\input{Physics/Source.tex}

\input{Physics/DP.tex}
\input{Physics/CosmogenicBSM.tex}

\input{summary.tex}

\vspace{0.5cm}

\noindent
{\bf \Large{Acknowledgments:}} \\

\noindent
S. H. Seo is supported by the National Research Foundation of Korea (NRF) grant funded by the Korea Ministry of Science and ICT (MSIT) (No. 2017R1A2B4012757, IBS-R016-D1-2019-b01) and IBS-R016-D1.  
P. Bakhti, M. Rajaee, and S. Shin are supported by NRF [NRF-2020R1I1A3072747 and NRF-2022R1A4A5030362]. 
S.S. is also supported by IBS-R018-D1.
D.~Kim is supported by the DOE Grant No. DE-SC0010813.
J. C. Park is supported by NRF [NRF-2019R1C1C1005073 and NRF-2021R1A4A2001897] and IBS under the project code, IBS-R018-D1.

\bibliography{references}

\end{document}

%% file: preamble.tex
\makeatletter  
\def\@maketitle{
\begin{center}
{\Huge \bfseries \sffamily \@title }\\[4ex] 
{\Large  \@author}\\[4ex] 
\end{center}}
\makeatother

\title{{\bf \Large Physics Potential of a Few Kiloton Scale Neutrino Detector at a Deep Underground Lab in Korea}}

\renewcommand{\Authands}{, }

\author[5]{\bf Seon-Hee Seo,\footnote{Corresponding author: 1.sunny.seo@gmail.com}}
\author[11]{\bf Jose Alonso}
\author[8]{\bf Pouya Bakhti}
\author[11]{\bf Janet Conrad}
\author[4]{\bf Steve Dye}
\author[14]{\bf Doojin Kim}
\author[9]{\bf Jost Migenda}
\author[7]{\bf Marco Pallavicini}
\author[3, 6]{\bf Jong-Chul Park}
\author[8]{\bf Meshkat Rajaee}
\author[2]{\bf Mike Shaevitz}
\author[6, 8]{\bf Seodong Shin}
\author[13]{\bf Joshua Spitz}
\author[11]{\bf Daniel Winklehner}
\author[12]{\bf Slawomir Wronka}
\author[10]{\bf Michael Wurm}
\author[1]{\bf Minfang Yeh}

\affil[1]{\it Chemistry Division, Brookhaven National Laboratory, USA}
\affil[2]{\it Dept.~of Physics, Columbia University, New York, NY, USA}
\affil[3]{\it Dept.~of Physics and Institute of Quantum Systems (IQS), Chungnam National University, Daejeon 34134, Republic of Korea}
\affil[4]{\it Dept.~of Physics and Astronomy, University of Hawaii, Honolulu, HI 96822, USA}
\affil[5]{\it Center for Underground Physics, Institute for Basic Science (IBS), Daejeon, 34126, Republic of Korea}
\affil[6]{\it Particle Theory and Cosmology Group, Center for Theoretical Physics of the Universe, Institute for Basic Science (IBS), Daejeon, 34126, Republic of Korea}
\affil[7]{\it Dept.~of Physics, University of Genoa, Italy}
\affil[8]{\it Laboratory for Symmetry and Structure of the Universe, Department of Physics, Jeonbuk National University, Jeonju, Jeonbuk 54896, Korea}
\affil[9]{\it Dept.~of Physics, King's College, London, UK}
\affil[10]{\it Dept.~of Physics, Mainz University, Germany}
\affil[11]{\it Dept.~of Physics, Massachusetts Institute of Technology, Cambridge, MA 02139, USA}
\affil[12]{\it Dept.~of Accelerator Physics and Technology, National Center for Nuclear Physics, Poland}
\affil[13]{\it Dept.~of Physics, University of Michigan, Ann Arbor, MI 48109, USA}
\affil[14]{\it Mitchell Institute for Fundamental Physics and Astronomy, Dept.~of Physics and Astronomy, Texas A\&M University, College Station, TX 77843, USA}
\affil[]{\it }

\newpage

\date{\today}

\newpage

%% file: intro.tex
\section{Introduction}
\label{sec:intro}

Neutrino and rare event search experiments need a low background environment, and an underground lab can provide such an environment by reducing the muon flux produced in the atmosphere. Several discovery measurements in the neutrino field were achieved when the experiments were performed under some overburden or in underground labs. The discovery of the neutrino at the Savanna River plant in 1956~\cite{Cowan:1956rrn} was possible due to some overburden reducing cosmic muon background, thanks to the lesson from its pre-experiment at Hanford in 1953~\cite{Reines:1953pu}. The first observation of atmospheric neutrino was performed in 1965 at the Kolar Gold Fields (3~km depth) in southern India~\cite{Achar:1965ova} and East Rand Priority Mines (3.2~km depth) in S. Africa~\cite{Reines:1965qk}, independently. Ray Davis’s radio-chemical experiment at Homestake mine ($\sim$1.5~km depth) observed solar neutrinos for the first time in the late 1960s and continued until 1992. The KamiokaNDE experiment~\cite{Kamiokande-II:1987idp} observed neutrinos from a supernova burst, SN1987a, for the first time in the Kamioka mine (1~km depth). Neutrino mixing angles, $\theta_{23}$ and $\theta_{12}$, were first measured by Super-Kamiokande~\cite{Super-Kamiokande:1998kpq} and SNO~\cite{SNO:2002tuh} in SNOLAB (2~km depth), respectively. The smallest mixing angle was first measured by Daya Bay~\cite{DayaBay:2012fng} and RENO~\cite{RENO:2012mkc}, independently, and their detectors were located under small mountains/hills.

Demand for more underground spaces has been increasing more and more due to dark matter and 0$\nu\beta\beta$ experiments as well as other modern neutrino experiments.  Upcoming flagship neutrino experiments, DUNE~\cite{DUNE:2015lol} and Hyper-K~\cite{Hyper-Kamiokande:2018ofw}, are currently excavating underground caverns at SURF ($\sim$1.5~km depth) and Tochibora mine (700~m overburden), respectively. Civil construction work of the JUNO underground site (650~m overburden) was completed in 2021, and its detector is being installed, aiming to take data in 2024.   

Yemilab is the 1$^{\mathrm{st}}$ deep underground lab in Korea dedicated to science, and its civil-engineering construction work was completed in late 2022, where COSINE-200~\cite{COSINE:2020egt} dark matter and AMoRE-II~\cite{Lee:2020rjh} 0$\nu\beta\beta$ experiments will be installed in late 2023. Additionally, a large cylindrical cavern (D: 20~m, H: 20~m) was also excavated in Yemilab. In this cavern, a kiloton scale liquid scintillator detector, called LSC (Liquid Scintillator Counter), for a neutrino experiment could be installed. 

This white paper is to demonstrate the best use cases of the biggest cavern in Yemilab by building a neutrino detector based on liquid scintillator technology, where an exciting physics program could be launched, and some of the research topics would be competitive. In principle, the kilo-ton scale LS detector itself could be installed in any underground labs, not limited to Yemilab. In the following sections, deep underground labs, the kton scale neutrino detector, its facilities, and the physics potential for various research topics are presented.

%% file: Site/site.tex
\section{Underground Labs}
\label{sec:site}

Currently, about 17 underground labs are operating or under construction in the US (SURF and Soudan), Canada (SNOLAB), South America (ANDES), Italy (Gran Sasso), UK (Boulby), France (Modane), Spain (Canfranc), Finland (CallioLab), Russia (Baksan), Japan (Kamioka and Tochibora), China (Jinping lab and JUNO), Korea (Y2L\footnote{Y2L will be emptied by the end of 2023.} and Yemilab), and Australia (SUPL). SUPL and Yemilab are constructed recently, SURF is being upgraded and expanded to host DUNE, and ANDES is planned. INO in India is funded but couldn’t start its construction because of a veto from environmentalists. 

Each underground lab has a different physical environment, i.e., different muon fluxes and natural radioactivities, which could affect experimental results significantly. In the following subsections, muon fluxes and natural radioactivities for major underground sites are discussed.

\subsection{Muon fluxes}

Muons are produced by cosmic ray interactions in the atmosphere. Most of the muons decay before reaching sea level. Survived muons reaching underground can produce neutrons and unstable light isotopes by interacting with materials surrounding a detector or within the detector itself. These neutrons and isotopes are direct sources of fatal background for neutrino and rare event search experiments. Therefore, reducing the muon flux is crucial for these experiments.  

\begin{figure}[h]
\begin{center}
  \includegraphics[width=0.9\textwidth]{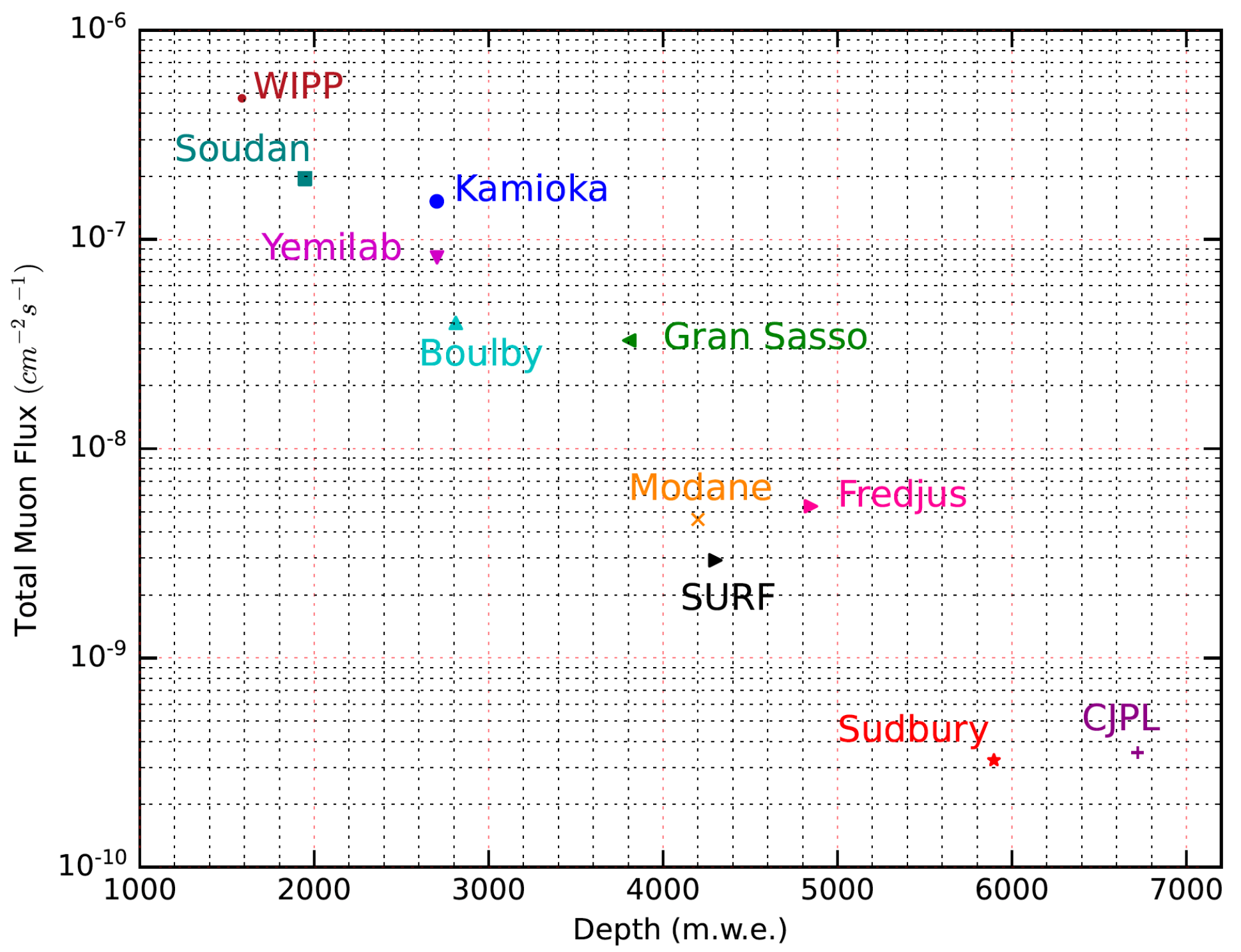}
\end{center}
\caption{Muon fluxes vs. depth of the major underground labs in the world.}
\label{f:muonFlux}
\end{figure}

Usually, the deeper (more overburdened) the underground, the less muon flux, and Table~\ref{t:reactor_mu_flux} (Figure~\ref{f:muonFlux}) lists (shows) major underground labs with their depths and muon fluxes. However, depending on the geographical shape of the overburden, muon flux can be different for the same depth. For example, Kamioka, Yemilab, and Boulby have almost the same depth, but their muon fluxes differ. Jinping lab (CJPL) is the deepest underground lab in the world, but its muon flux is not much different from that in SNOLAB, the 2$^{\mathrm{nd}}$ deepest. 

\begin{table}[!tbp]
\centering
\begin{tabular}{|c|c|c|c|}
\hline
Sites & Depth (m.w.e.) & Muon flux (cm$^{-2}$sec$^{-1}$) & Reactor $\nu$ IBD rate (NIU) \\
\hline
\hline
WIPP & 1585.7 & 4.80e-7 & 33 \\
Soudan & 1946.65 & 1.98e-7 & 66 \\
Kamioka & 2702.61 & 1.31e-7 & 121 \\
Yemilab & 2704 & 8.20e-8 & 696 \\
Boulby  & 2811.58 & 4.80e-8 & 921 \\
Gran Sasso & 3805.9 & 2.18e-8 & 75 \\
Modane & 4200 & 4.59e-9 & 921 \\
SURF & 4300 & 2.93e-9 & 29 \\
Frejus & 4847.9 & 5.31e-09 & -- \\
Sudbury & 5896.71 & 3.77e-10 & 174 \\
CJPL & 6720.77 & 2.12e-10 & 19 \\
\hline
\end{tabular}
\captionof{table}{\label{t:reactor_mu_flux}
Some of the underground lab sites and their depths, muon flux, and IBD rates (in NIU) for reactor neutrinos. 1 NIU (Neutrino Interaction Unit) = 1 interaction/10$^{32}$ targets/year.
}
\end{table}

\begin{figure}[h]
\begin{center}
\includegraphics[width=1.\textwidth]{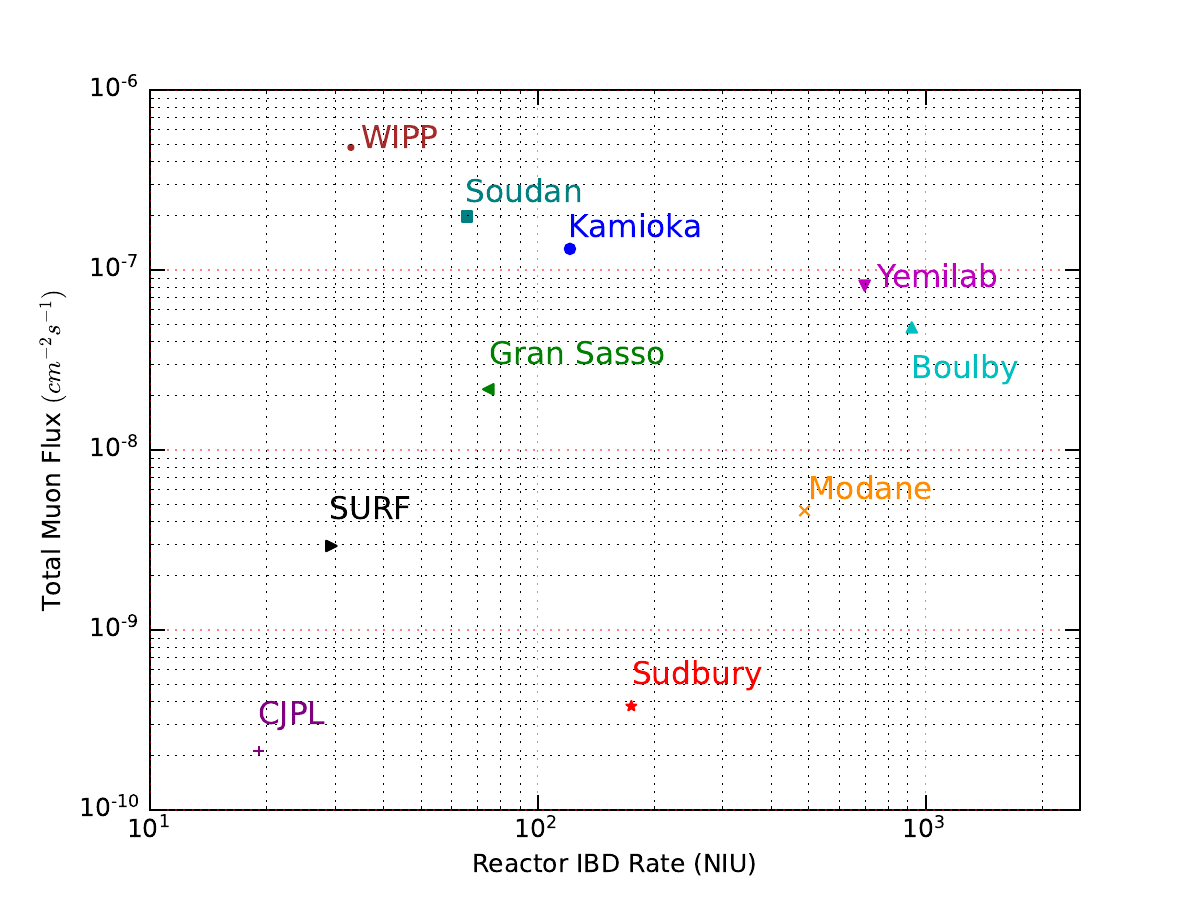}
\end{center}
\caption{Muon flux vs. reactor $\nu$ flux of the major underground labs in the world, where the reactor $\nu$ flux is represented as IBD rate in NIU (Neutrino Interaction Unit) = 1 interaction/10$^{32}$ targets/year.}
\label{f:reactor_mu_flux}
\end{figure}

Depending on the physics that is being pursued, for example, solar and geoneutrino measurements, reactor neutrinos can produce a severe background. Table~\ref{t:reactor_mu_flux} (Figure~\ref{f:reactor_mu_flux}) lists (shows) major underground labs with their depths and the expected reactor neutrino rate.

\subsection{Rock radioactivity}

Natural radioactivity on the Earth originates mainly from $^{238}$U, $^{232}$Th, and $^{40}$K isotopes that produce fission gammas below 3~MeV. Additionally, fast neutrons are also produced by spontaneous fissions and ($\alpha$, n) interactions, and these neutrons contribute to the background through their scatterings and/or capture on hydrogen (or any neutron affinitive nuclei that is a component or part of the detector material). 

Natural radioactivities are unavoidable but can be reduced by choosing a cleaner environment and materials as well as purification of the detector material. Environmental radioactivity depends on geographical locations and rock types, and therefore the concentration of $^{238}$U and $^{232}$Th can vary significantly depending on where the underground labs are located. Table~\ref{t:radioactivity} lists the rock radioactivity of the major underground labs.

\begin{table}[!tbp]
\centering
\begin{tabular}{|c|c|c|c|}
\hline
Sites & $^{238}\mathrm{U}$ & $^{232}\mathrm{Th}$ & $^{40}\mathrm{K}$\\
\hline
\hline
Jinping~\cite{Serenelli:2012zw} & $1.8\pm0.2$ ($^{226}\mathrm{Ra}$) & < 0.27 & < 1.1 \\
Sudbury~\cite{Adelberger:1998qm} & $13.7\pm1.6$ & $22.6\pm2.1$ & $310\pm40$ \\
Gran Sasso hall A~\cite{Adelberger:2010qa}& $116\pm12$ & $12\pm0.4$ & $307\pm8$ \\
Gran Sasso hall B~\cite{Adelberger:2010qa} & $7.1\pm1.6$ & $0.34\pm0.11$ & $7\pm1.7$ \\
Gran Sasso hall C~\cite{Adelberger:2010qa} & $11\pm2.3$ & $0.37\pm0.13$ & $4\pm1.9$ \\
Kamioka~\cite{Zhang:2010yua} & $\sim12$ & $\sim10$ & $\sim520$ \\
Boulby-halite~\cite{Smith:2005se} & $0.413$ & $0.650$ & $9.77$ \\
Homestake~\cite{Mei:2009py} & $18.7$ & $30.0$ & $303$ \\
Yemilab  &  18.6 & 37 & 835 \\
\hline
\end{tabular}
\captionof{table}{\label{t:radioactivity}
Radioactivity (Bq/kg) of the rocks in the major underground labs in the world. 
}
\end{table}

%% file: Detector/det.tex
\section{Detector}
\label{sec:detector}

\begin{figure}[h]
\begin{center}
\includegraphics[width=.5\textwidth]{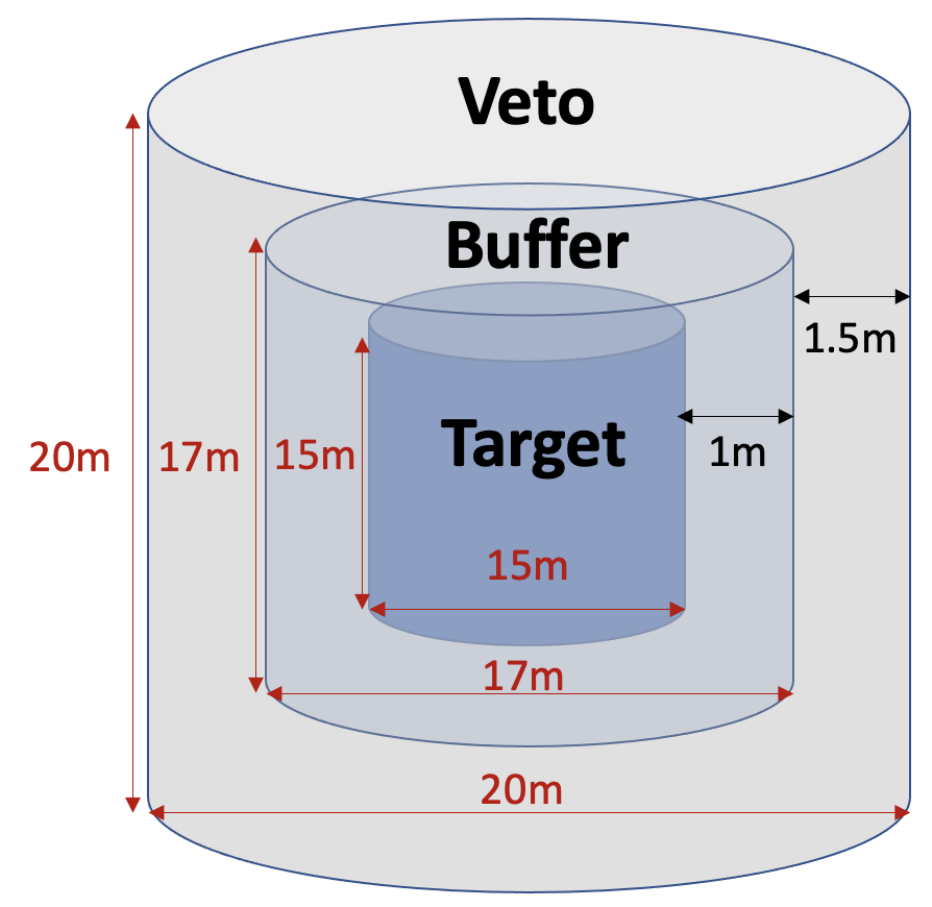}
\end{center}
\caption{A candidate design of the LSC in Yemilab.}
\label{f:det_LSC}
\end{figure}

In the current design, the LSC detector consists of three layers of cylindrical volumes, and they are target, buffer, and veto from the inner to outer volumes. The target material would be either LS or water-based LS (WbLS) of $\sim$2.26 kton to be filled in an acrylic vessel of 15~m diameter and 15~m height. The buffer region will be filled with mineral oil of $\sim$1.14 kton in a stainless steel vessel of 17~m diameter and 17~m height, and a few thousand 20-inch photo-multiplier tubes (PMT) will be mounted at the stainless steel wall. The Veto region will be filled with purified water of $\sim$2.41 kton in the cavern rock with lining (or in a stainless steel vessel) of 20~m diameter and 20~m height, and a few hundred 20-inch PMT will be installed at the wall.
Figure~\ref{f:det_LSC} illustrates a candidate design of the LSC detector just described. 
Details on target materials, LS and WbLS, are discussed in the following subsections.  

%% file: Detector/LS.tex
\subsection{LS}

LS has been used in neutrino experiments since the 1950s for the discovery of neutrinos, and it is still widely used in low energy (solar, supernova, geo, reactor, etc.) neutrino physics thanks to its high light yield. With an LS detector, it's possible to access the sub-MeV energy region, which is critical for precise measurements of solar neutrino fluxes, including $pp$, $^7Be$, $pep$, and CNO neutrinos which were not measured by any other experiments except Borexino that used ultra-low background LS as a target material. 

LS is a cocktail of dominant ($\sim$60 to $\sim$99\%) solvent and sub-dominant fluors. Solvent absorbs energy deposited by particles and transfers it to fluors which absorb and re-emit photons. PC (Pseudo Cumin), PXE (P-Xylene), and LAB (Linear Alkylbenzene) are commonly used solvents. Among them, more and more modern experiments use LAB due to the high flash point (140 $^{\mathrm{o}}$C) and its environment-friendliness. Fluors are classified into primary and secondary ones, and the secondary one (aka wavelength shifter) is employed to collect more photons by matching the wavelength that is sensitive to the photomultiplier tubes (PMTs). Examples of primary (secondary) fluors are PPO, butyl-PBD, p-Terpenyl, and Naphthalene (bis-MSB, POPOP, and TBP). 
Table~\ref{t:LS-exp} lists the composition of the LS for some modern neutrino experiments. 
Modern neutrino experiments using LS detectors often use LAB, PPO, and bis-MSB as solvent, primary, and secondary fluors, respectively. Gd loading is made for some IBD-seeking experiments to reduce accidental backgrounds efficiently~\cite{Yeh:2016cby}. 

The advantages of LS detectors are high light yield, fast signals, and high purity, while the disadvantages are no directional information, luminescence background, and quenching. 


\begin{table}[!tbp]
\centering
\begin{tabular}{|c|c|c|c|c|c|}
\hline
Experiment & Solvent & Primary fluor & Secondary fluor & Metal loading & LS Target \\
  & & (PPO) & (bis-MSB) & & \\
\hline
\hline
KamLAND & PC + dodecane & 1.36 g/l & -- & -- & 1 kton \\
Borexino & PC & 1.5 g/l & -- & -- & 300 ton \\
Double Chooz & dodecane + PXE & 7 g/l & 20 mg/l  & 0.2\% Gd & 8 ton x 2 \\
Daya Bay & LAB & 3 g/l  & 15 mg/l  & 0.1\% Gd & 20 ton x 4 \\
RENO & LAB & 3 g/l  & 30 mg/l & 0.11\% Gd & 16 ton x 2 \\
NEOS & LAB  & 3 g/l & 30 mg/l & 0.5\% Gd & 1 ton \\
SNO+ & LAB & 2 g/l & 15 mg/l & 0.3\% Te & 780 ton \\
JUNO & LAB & 2.5 g/l  & 4 mg/l & -- & 20 kton \\
\hline
\end{tabular}
\captionof{table}{\label{t:LS-exp}
Composition of the LS for some modern LS-based neutrino detectors. 
Note that in NEOS 10\% of UG-F was added to LS for more light yield and better pulse shape discrimination. 
}
\end{table}

%% file: Detector/WbLS.tex
\subsection{WbLS}
A pure target of multiple kilotons of liquid scintillator has great sensitivity for sub-MeV neutrinos and other rare-event physics. However, on some occasions, large, pure LS detectors might not be a viable choice due to cost, ESH, and chemical safety. Water-based Liquid Scintillator (WbLS), first developed at Brookhaven National Laboratory (BNL) in 2010~\cite{Yeh:2010mye}, is a novel detection medium that bridges organic scintillating materials and water to form scintillator liquids ranging from almost pure water to almost pure organic scintillator. By introducing varying amounts (typically 1-10\%) of liquid scintillator into the water, the light yield can be adjusted to allow the detection of particles below the Cherenkov threshold while maintaining directional capability. The WbLS delivers a new generation detector medium linking features of the Cherenkov radiation and scintillation process that is cost-effective and environmentally friendly, which largely improve affordability and essentially reduce chemical usage and waste for kilotons of particle detectors.

A metal-loaded liquid scintillator is an advanced detector liquid developed for several nuclear and particle physics experiments. The metal of choice depends on the physics implications, such as gadolinium to provide a delayed neutron-capture signal for reactor-based antineutrino experiments ~\cite{DayaBay:2012fng,RENO:2012mkc,DCZ:2012dcc} and for serving as a veto detector for dark matter searches~\cite{Has:2019myh}, tellurium as a target for neutrinoless double-beta-decay ~\cite{SNO+:2017tel}, or indium to measure low-energy neutrinos from the sun, including the pp, pep, $^7$Be and CNO, by the LENS experiment~\cite{LENS:2004rlh}. Procedures have been established to transfer metals into organic liquids by solvent extraction or direct dissolution of organometallic complexes. The WbLS also offers a new opportunity allowing organic scintillators to receive inorganic metallic ions (particularly hydrophilic) at nearly 100\% efficiency. The metal-doped WbLS, eliminating the chemical complexation and/or other chelation steps, is a transformative technology that deviates from the preparation of conventional metal-doped liquid scintillators. Recent work has been performed to load gadolinium in WbLS with long scattering length (WATCHMAN) and lithium in a superior PSD WbLS-based scintillator ~\cite{PRO:2019ja}.

WbLS is a leading candidate as the target medium for next-generation particle detectors. There is an active development effort to realize the novel WbLS liquid target being considered as (1) an antineutrino detection medium to permit investigation of advanced stand-off methods to improve sensitivity to the existence and operation of nuclear reactors for nonproliferation science and (2) a “hybrid” event detection of particle interactions to combine the unique topology of Cherenkov light with the increased low-threshold scintillation light yield to improve energy and vertex resolution and obtain particle identification for optical neutrino detectors, such as THEIA ~\cite{Theia:2019non}. The ability to distinguish the two signals is facilitated by new developments in the scintillator, photon detection, and readout technology as well as sophisticated analysis methods. 

\begin{figure}[h]
\begin{center}
\includegraphics[width=1.\textwidth]{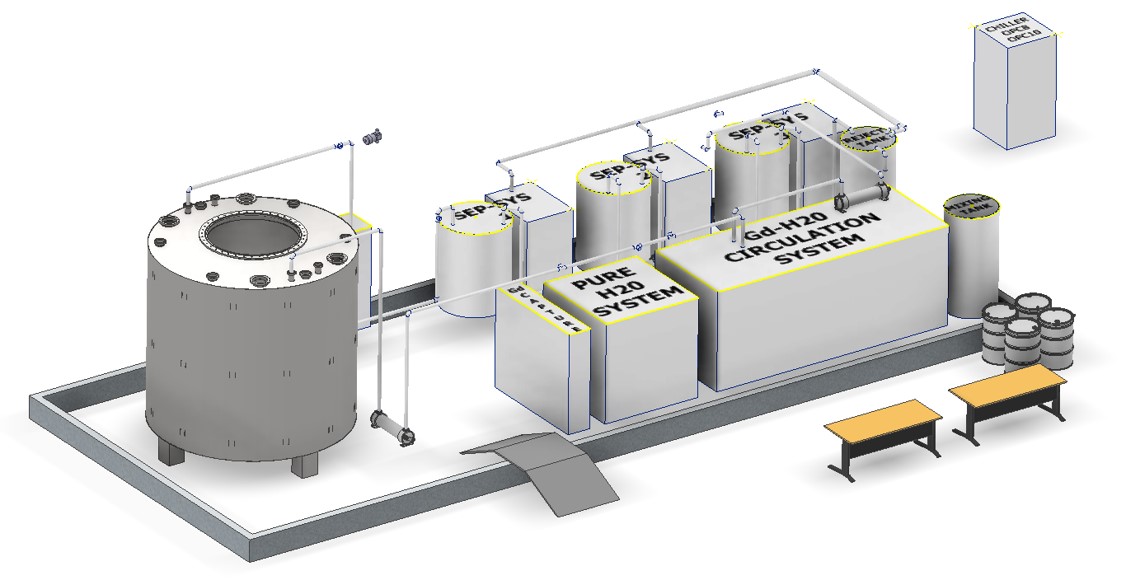}
\end{center}
\caption{30-ton BNL Demonstrator equipped with Gd-water and nanofiltration system, projected data-taking in 2024.}
\label{f:30TBNL}
\end{figure}

Several WbLS testbeds and demonstrators are now funded and being built at several US (LBL and BNL) and European labs, in addition to research programs at several universities. The BNL 30-ton WbLS demonstrator (Figure~\ref{f:30TBNL}) is a 3-year development effort aiming to construct a fully operational testbed directed towards a precision measurement of optical properties at long distances, demonstration of material compatibility with detector components, and characterization of simulations as a next step towards the deployment of a kiloton-scale detector. This proposed demonstrator also examines the feasibility, in terms of formulation, fabrication, and deployment, of WbLS with or without loading of gadolinium. A newly developed in-situ deployment scheme to transform any present water Cherenkov detectors to WbLS detectors will be installed and exercised at ton-scale deployment. This sequential mixing technology allows the detector to be first filled with high-purity (18 Megohm) water, followed by introducing organic scintillators sequentially to form WbLS by in-situ circulation. This in-situ sequential mixing largely simplifies the manufacture of large volume WbLS by direct formation of target liquids at the detector vessel avoiding extensive laboratory space and labor requirements as often demanded by the deployment of nominal kiloton-scale liquid scintillator detectors. 

Two vital subsystems, the Gd-water purification system (a.k.a. that for SK) and organics separation system, are included in the BNL 30T demonstrator. Both subsystems use an industrial-engineering nanofiltration technology, a pressure-driven membrane process, which has a high rejection of multivalent salts and larger molecules. In general, the Cherenkov detector requires constant purification to moderate microorganic activity that is known to nurture in water, leading to degradation of optical transparency, while the liquid scintillator detector doesn’t require constant circulation during operation. This BNL demonstrator will determine the circulation requirement and, if crucial, demonstrate the circulation efficacy to maintain the optical stability of WbLS. 

\begin{figure}[h]
\begin{center}
\includegraphics[width=1.\textwidth]{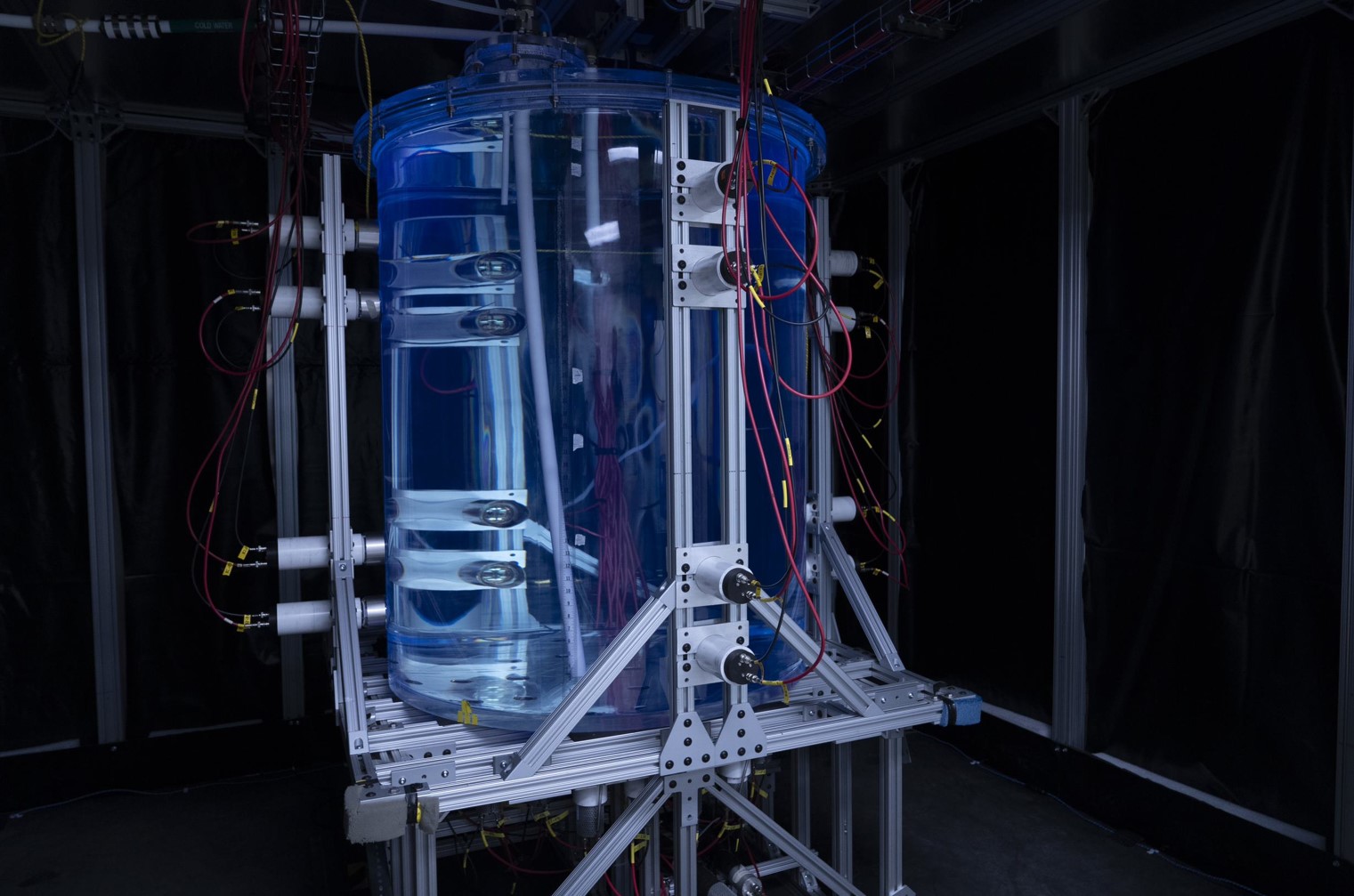}
\end{center}
\caption{1-ton BNL Testbed operation since 2022.}
\label{f:1TBNL}
\end{figure}

The performance and stability in addition to nanofiltration separation and recombination efficacy of WbLS at different scintillator loadings (1-5\%) will be first investigated at the existing BNL 1-ton testbed (Figure~\ref{f:1TBNL}), commissioned in early 2022. This 1T testbed, instrumented with batches of different photodetectors, contains comprehensive water-purification, LS-injection and circulation, and drained systems and is capable of measuring the precise performance of different WbLS liquids with a fast turn-around time leveraging existing the BNL ton-scale scintillator production facility. The lessons learned and experience gained from the operation of this 1T testbed are to be extended to the 30T demonstrator.

Upon maturity, this 30T demonstrator will retire risks derived from the deployment and operation of a kiloton-scale WbLS detector. The cleanliness and material compatibility (with PMTs, cable, calibration, etc.) of WbLS detectors under different operation configurations will also be concluded. The 1T and 30T testbeds provide liquid-handling and ESH training and are open for research activity to collaborators from the international scientific community. The full operation of the 30T BNL demonstrator is scheduled to start in early 2024. 

\begin{figure}[h]
\begin{center}
\includegraphics[width=0.5\textwidth]{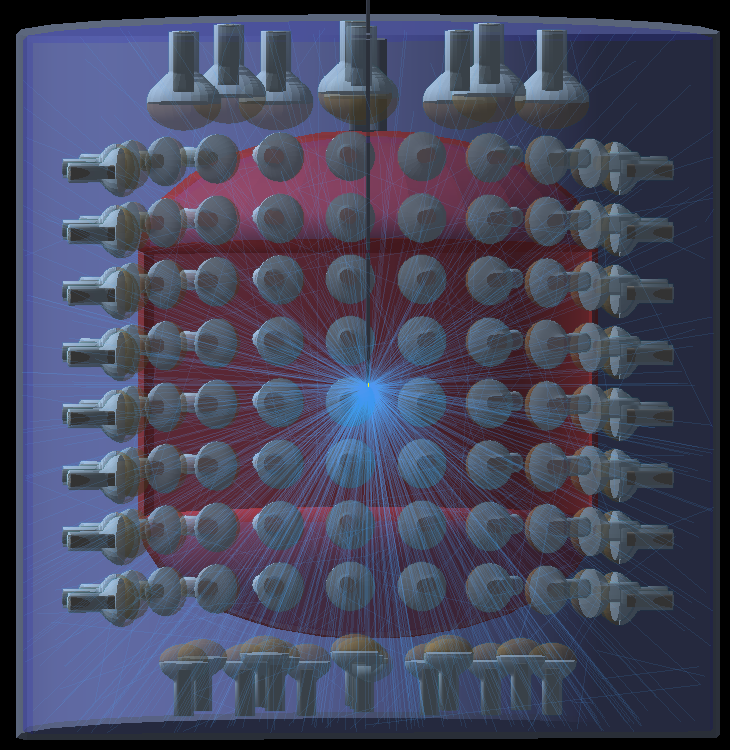}
\end{center}
\caption{4-ton Eos detector at LBL, projected data-taking in 2024.}
\label{f:Eos}
\end{figure}

The LBL-proposed Eos (Figure~\ref{f:Eos}) is another prototype designed to hold a range of novel scintillators at a few-ton scale and coupled with an array of photon detection options with the ability to deploy a range of low-energy calibration sources. Eos will be sufficiently large to use time-of-fight-based reconstruction and to fully contain a range of low-energy events for detailed event-level characterization. Eos 
represents a balance of sufficient size for full event characterization, complemented with economy of scale, and flexibility to adapt for multiple target materials and photon detection options, such as the fraction of LS in a WbLS target cocktail, or by using PMTs with differing TTS, or deployment of dichroicons. 

The primary goal of Eos is to validate performance predictions for large-scale hybrid neutrino detectors by performing a data-driven demonstration of low-energy event reconstruction leveraging both Cherenkov and scintillation light simultaneously. By comparing data to model predictions, and allowing certain detector configuration parameters to vary, the predictive power of the model can be validated. This validated microphysical model of hybrid neutrino detectors can then be used by the neutrino community for the design optimization of next-generation hybrid detectors. Eos represents significant risk reduction for a large-scale deployment of (Wb)LS and novel detection technology.

Assuming a successful surface deployment, Eos could later be re-deployed underground, for example, at SURF or SNOLAB, or an alternative location such as a reactor or test beam for further characterization of detector response to a range of particle interactions. 

%% file: Detector/slowLS.tex
\subsection{Slow LS} 

Discrimination between Cherenkov and scintillation can also be achieved by slowing down the scintillation emission to further separate from the prompt Cherenkov radiation. Unlike the WbLS combining angular and timing information allows separation between Cherenkov and scintillation light for high-energy events in a standard scintillation mechanism, slow scintillator utilizes slow fluors or wavelength shifter to provide excellent separation in MeV-scale energy region. 

\begin{figure}[h]
\begin{center}
\includegraphics[width=0.9\textwidth]{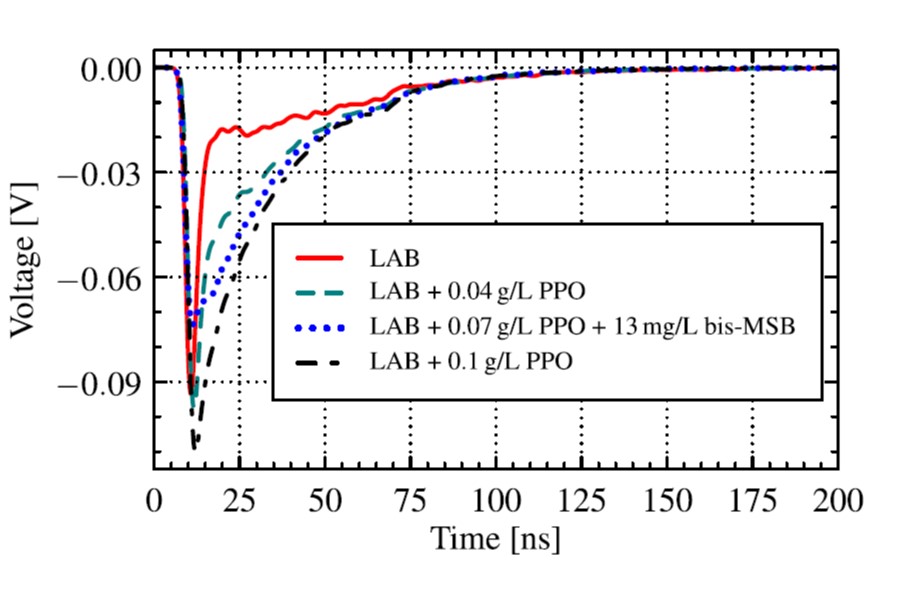}
\end{center}
\caption{The measured waveforms for LAB with different combinations of PPO and bis-MSB ~\cite{Guo:2019slo}.}
\label{f:slowLAB}
\end{figure}

\begin{figure}[h]
\begin{center}
\includegraphics[width=0.7\textwidth]{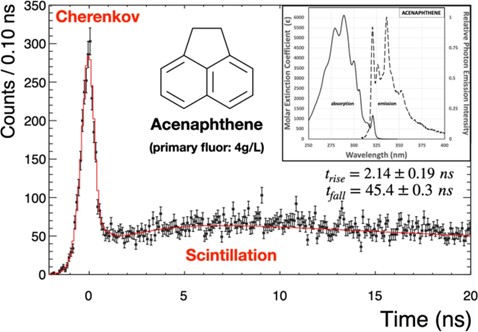}
\end{center}
\caption{Zoomed in time spectrum for 4 g/L acenaphthene in LAB with clear Cherenkov peak~\cite{Bil:2020slo}.}
\label{f:acenaph}
\end{figure}

The concept of slow scintillator allowing a directional cut for the enhancement of particle ID has been proposed on several occasions ~\cite{Yeh:2015tim}. A liquid scintillator mixture of LAB, PPO and bis-MSB solution with different compounding ratios was investigated by ~\cite{Guo:2019slo}, and is presented in Figure~\ref{f:slowLAB}. An inverse relationship between the light yields and decay time constants for these samples was observed. The relationship is understood by the mechanism of the energy transfer between scintillator molecules. The properties of slow fluors and wavelength shifters have also been studied in the context of LAB-based liquid scintillator mixtures to provide a means to effectively separate Cherenkov light in time from the scintillation signal by ~\cite{Bil:2020slo}. An example of showing the measured time spectrum for acenaphthene (4 g/L in LAB) displaying the separation of Cherenkov and scintillation is given in Figure~\ref{f:acenaph}.  

Different slow scintillator combinations could be suitable for different applications and have potentially important consequences for a variety of future physics experiments using large scale liquid scintillation detectors. An accurate measurement of the energy of charged particles may provide extra discriminating power to the background suppression, while the solar angle cut on the direction of charged particles can be a powerful selection criterion for solar neutrino events. Atmospheric neutrino through the neutral and charged current interactions, which is one of the major backgrounds in the search of supernova relic neutrino events, can be effectively suppressed if electrons and muons are distinguished from non-Cherenkov induced neutrons and protons by particle identification. In addition, it is possible to further perform particle identification based on the ratio of Cherenkov light yield to scintillation light yield, which could open the possibility of obtaining good directional information for elastic scattering events from supernovae neutrinos and reactor anti-neutrinos. For a multiphysics program, a balance between vertex resolution, Cherenkov/scintillation ratio and energy resolution need to be achieved for particular physics objectives.

%% file: Detector/PMT.tex
\subsection{Photo Sensors}
The stainless steel vessel of the buffer is expected to be installed with a few thousand 20-inch photomultipliers (PMTs) on its wall. This vessel will be filled with mineral oil which will immerse the PMTs with a thickness of 1 meter. This buffer region will reduce the background events originating from the PMTs and function as additional passive shielding for external gammas and neutrons. The photo coverage of the PMTs is aimed to be about 65 (or 49)\% with 4\,000 (or 3\,000) 20-inch PMTs.  

%% file: Facility/facility.tex
\section{Facilities}
\label{sec:facility}

The LSC will be a multi-purpose detector. Depending on the physics pursued by the LSC, either IsoDAR or linac facilities are additionally required. In this section, the LS purification facility required for low energy physics and the two additional facilities are discussed. 

%% file: Facility/LS_purification.tex
\subsection{LS purification facility}

Several purification techniques are available to improve the performance of an LS detector. In addition to using gas purging to remove spurious gases dissolved in the liquid, such as oxygen, which is a known quencher of the scintillator light yield and could introduce solvent instability due to oxidation reactions, other efficient methods that are used to remove the optical or radioactive impurities are distillation, water extraction or column purification, and most experiments use combinations of them. The applied purification methods aim to be as efficient as possible in the removal of potential trace contaminants, even for impurities below the sensitivity (i.e. ~10$^{-16}$~g for most lanthanides and actinides using ICP-MS) of chemical analysis techniques in the laboratory. Experiments like Borexino and KamLAND have demonstrated the capabilities of LS detectors with very low count rates in the MeV region. The purification techniques ~\cite{Yeh:2016cby} for the LSC in Yemilab are described below.

\subsubsection{Gas Purging}

Volatile components dissolved in the liquid such as oxygen or radioactive noble gases can be depleted by nitrogen or argon stripping. The method is based on differences in the equilibrium composition between liquid and vapor. Approximations on the purification
factor can be made by use of Henry's Law. The technique is very effective, in particular
by applying the counter-current flow of the liquid and the gas. The efficiency can be further
increased by using columns with structured packing or higher temperatures, implying
reduced solubility of gas contaminants. To avoid that impurities such as radon or krypton are
added by gas purging, the purification and selection of clean gas is essential.

\subsubsection{Distillation}

Distillation is an effective process to improve the scintillator's transparency and to reduce metallic impurities in the solvent liquid. All impurities which are less volatile (higher boil point or viscosity) than the solvent can be separated. When distillation is applied to the full scintillator mixture, one can carefully design an operational temperature range either to separate the solvent from the fluor for reprocessing or to distill both solvent and fluor together for one-step purification based on boiling point of each component in the mixture. 

\begin{figure}[h]
\begin{center}
\includegraphics[width=1.\textwidth]{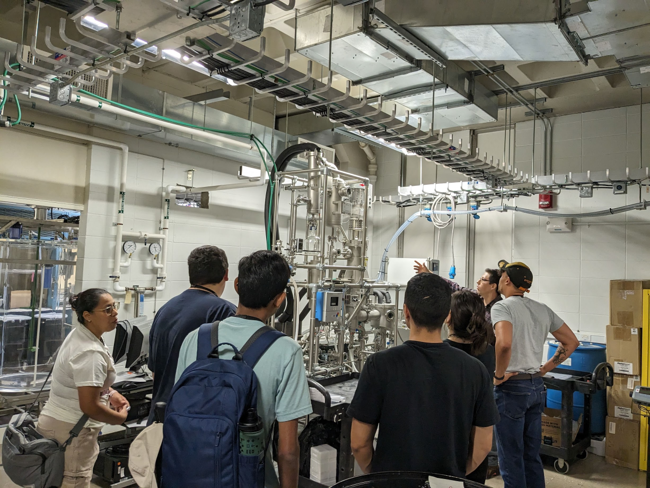}
\end{center}
\caption{BNL ton-scale thin-film vacuum distillation system.}
\label{f:thin-film}
\end{figure}

There are a few distillation configurations available in the market. A multi-staged vacuum distillation is applied as a purification step operating at a fractional mode. In particular, the vacuum distillation column is composed of many theoretical plates that can effectively separate the organic solvents fractionally with different boiling points, e.g., at 10C intervals. On the other hand, a single-staged vacuum distillation unit runs at a portion of liquid that forms a thin layer around the heated surface using a jet-like feed vessel, short-path evaporator, and internal condenser with computer-controlled heated oil bath, condenser cooling system, cold trap, and vacuum system. This short-path evaporator only heats a small fraction of solvent per injection, thus avoiding large-bucket solvent heating to shorten the distillation time. A pilot-scale (50 liters per hour) thin-film short-path distillation system installed at BNL is shown in Figure~\ref{f:thin-film}.

The vacuum distillation system is operated in closed mode with very little exposure of the organic fumes to the environmental air. In a specific vacuum distillation procedure, distillation is usually set to finish when 90\% of the solvent is collected. The purity of the organic solvents before and after purification often shows an improvement in optical clarity. The removal of metallic isotopes in most cases is better than 95\%.

\subsubsection{Water Extraction}

The water extraction technique is a known purification method utilized in several neutrino experiments. The impurities in the organic LS are transferred into an immiscible aqueous phase based on their polarities. An advantage of water extraction over distillation is the ability to directly process the full scintillator mixture allowing for in-situ repurification of the detector liquid. This method is highly effective at removing polar or charged substances and significantly lowers the quantity of radioactive impurities such as uranium, thorium, or potassium, which typically enter the scintillator mixture via the primary fluor. Water extraction in concentrated PPO solutions was found to be a promising way of removing potassium. On the other side, water extraction is less suited to reducing optical
impurities, which are mainly of organic type and require phase separation from the
scintillator solvent.

\subsubsection{Exchange Column}

Traces of chemical impurities can be removed from the scintillator by passing the liquid through a packed column of an adsorber material such as silica gel or aluminum oxide. The important parameters of absorber materials are surface-to-volume ratio, pore size, and surface conditions (acidic, basic, or neutral). Many adsorber materials are hygroscopic and should be activated at elevated temperatures before usage to remove water molecules blocking adsorption sites. The ratio of impurities retained on the adsorber surface and in the liquid is determined by a specific equilibrium constant. 

Since the chemical composition of most of the different impurities inherited from manufacturers is not well defined, several layers of activated column packing differing in absorbers can be used. In this multi-stage column approach, acids, bases as well as neutralized particles can be effectively removed. An example of an array of six sequential columns filled with different absorber materials has been built and deployed at BNL for LZ and other neutrino experiments.

\subsubsection{Purification of Water-based Liquid Scintillators}

In general, the purification stage for LS detectors is to cleanse all scintillator components before deployment, and there is no in-situ circulation, except water-extraction, when the detector is in operation. For a water Cherenkov detector, it is known that the pure water has to be constantly circulated using fairly standard off-the-shelf techniques to maintain optical stability. The WbLS is a binary liquid medium and typical water systems cannot handle organic concentrations at these levels; thus, the organics must be separated from the water before optical contaminants can be removed using standard techniques. 

Nanofiltration (NF) is a pressure-driven membrane process, which is often used for water softening (i.e., separation of divalent and monovalent cations) by pushing the liquid through a series of nano filters that takes out the micelles but leaves the other contaminants intact. The mechanism lies between ultrafiltration and reverse osmosis in terms of its ability to reject molecular or ionic species. The size of nanofiltration membranes allows a transition between microporous and dense phases and can be in the range of 0.5–1 nm leading to a nominal cutoff between 200 and 1000 Da. The nanofiltration system has specific features of nanometer membranes that allows a very high rejection for multivalent ions with low to moderate rejections for monovalent ions, and high rejection of organic compounds with a molecular weight greater than the pore size of the membrane. 

The mechanism of mass transport depends strongly on the membrane structure to form the interactions between the membrane and transported molecules. The separation efficiency can be governed by the sieving effect (size of the nanopores to that of the solute molecules) or by the solution and diffusion properties of the solute molecules. Three parameters are crucial for the operation of a nanofiltration unit: solvent permeability or flux through the membrane, rejection of solutes, and yield of recovery. 

A laboratory (0.5 GPM) nanofiltration system (MaxiMem) from UCD, to be installed as a phase-I separator for a 1-ton BNL testbed, is currently tested at BNL. To date, this system is able to separate 99\% of the organics from (as measured via fluorescence spectrometry) the injection of a 1\% WbLS. A new pilot-scale NF system (7 GPM) is under construction to serve as a phase-II separator for 1T BNL testbed or to use as a development unit for the 30-ton BNL Demonstrator. While these efforts are ongoing at BNL and UCD, the attention now is given to further reducing organics by another order of magnitude, to the level of a few ppm. This development might take 1-2 years but would be the last step in finalizing this system for use in the design of a kiloton scale device, such as THEIA, at SURF or Yemilab.

%% file: Facility/isodar_facility.tex
\subsection{The IsoDAR facility}
\label{sec:isodar}

\begin{table}[tb]
\begin{center}
      \begin{tabular}{|c|c|} \hline 
Runtime  &  5 calendar years  \\ \hline
IsoDAR duty factor  &  80\%  \\ \hline
Livetime  &  4 years  \\ \hline
Protons on target/year  &  $1.97\cdot 10^{24}$  \\ \hline
$^8$Li/proton ($\bar{\nu}_e$/proton) &  0.0146  \\ \hline
$\bar{\nu}_e$ in 4 years livetime  &  $1.15\cdot 10^{23}$  \\ \hline
IsoDAR@Yemilab mid-baseline  &   17~m  \\ \hline
IsoDAR@Yemilab depth  & 985~m (2700 m.w.e.) \\ \hline
\end{tabular}
\end{center}
\caption{Plan for the IsoDAR running.}
\label{assumptions_table_1}
\end{table}

\begin{figure}[t!]       
\begin{center}
{\includegraphics[width=1.0\textwidth]{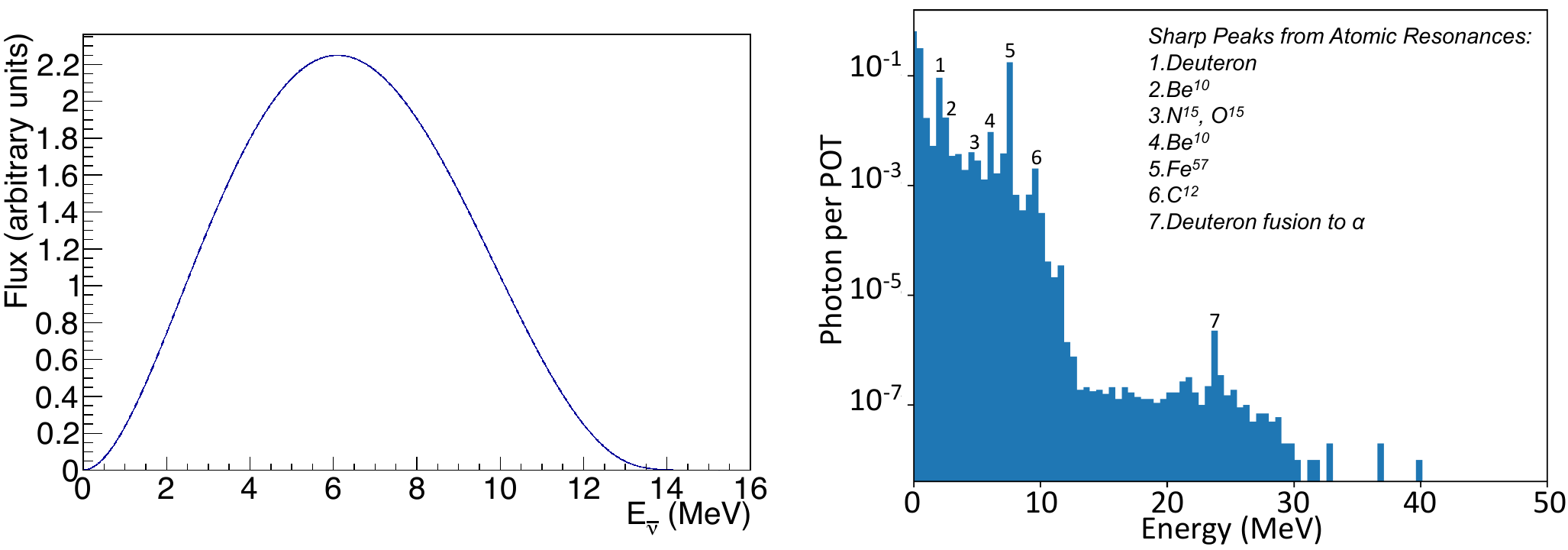}}
\end{center}  
\vspace{-.5cm}
\caption{\label{IsoDARfluxes}  Left: $\bar \nu_e$ flux from the IsoDAR source, unit normalized.   Right: photons produced in the IsoDAR target/sleeve, normalized per proton on target.     
        }
\end{figure}

The IsoDAR (Isotope Decay-At-Rest) facility produces an intense source of electron antineutrinos as well as a unique flux of monoenergetic photons, shown in Fig.~\ref{IsoDARfluxes}.   This source is a first-of-its kind underground accelerator-driven neutrino source.   When paired with the LSC, this facility opens a range of new physics searches that are unavailable at any other facility.   The physics potential of IsoDAR, based on the assumptions for running the facility described in Table~\ref{assumptions_table_1}, is
described in Sec.~\ref{IsoPhysText}.

\begin{figure}[tb]       
\begin{center}
\includegraphics[width=3.in]{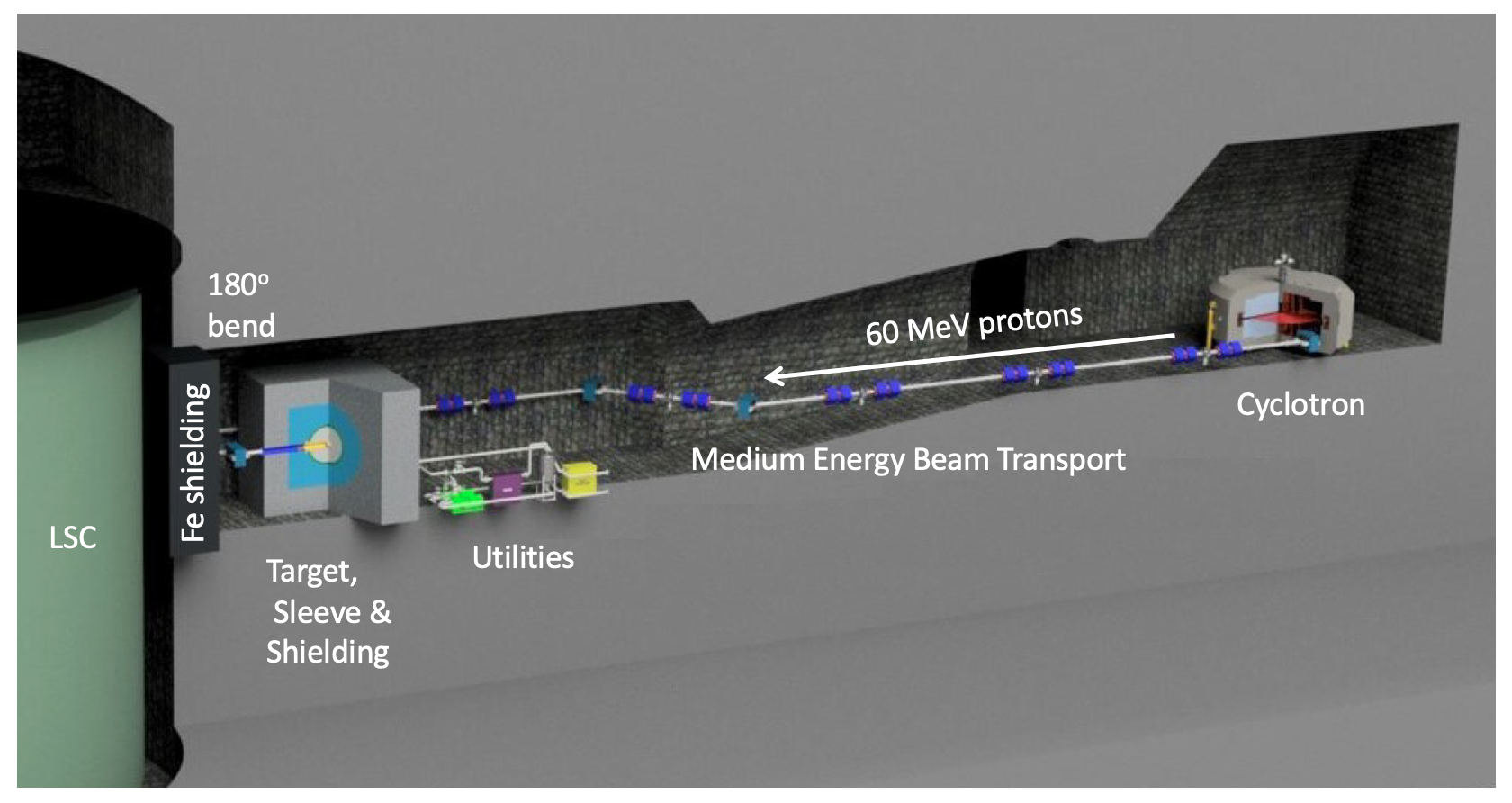}~\includegraphics[width=3.in]{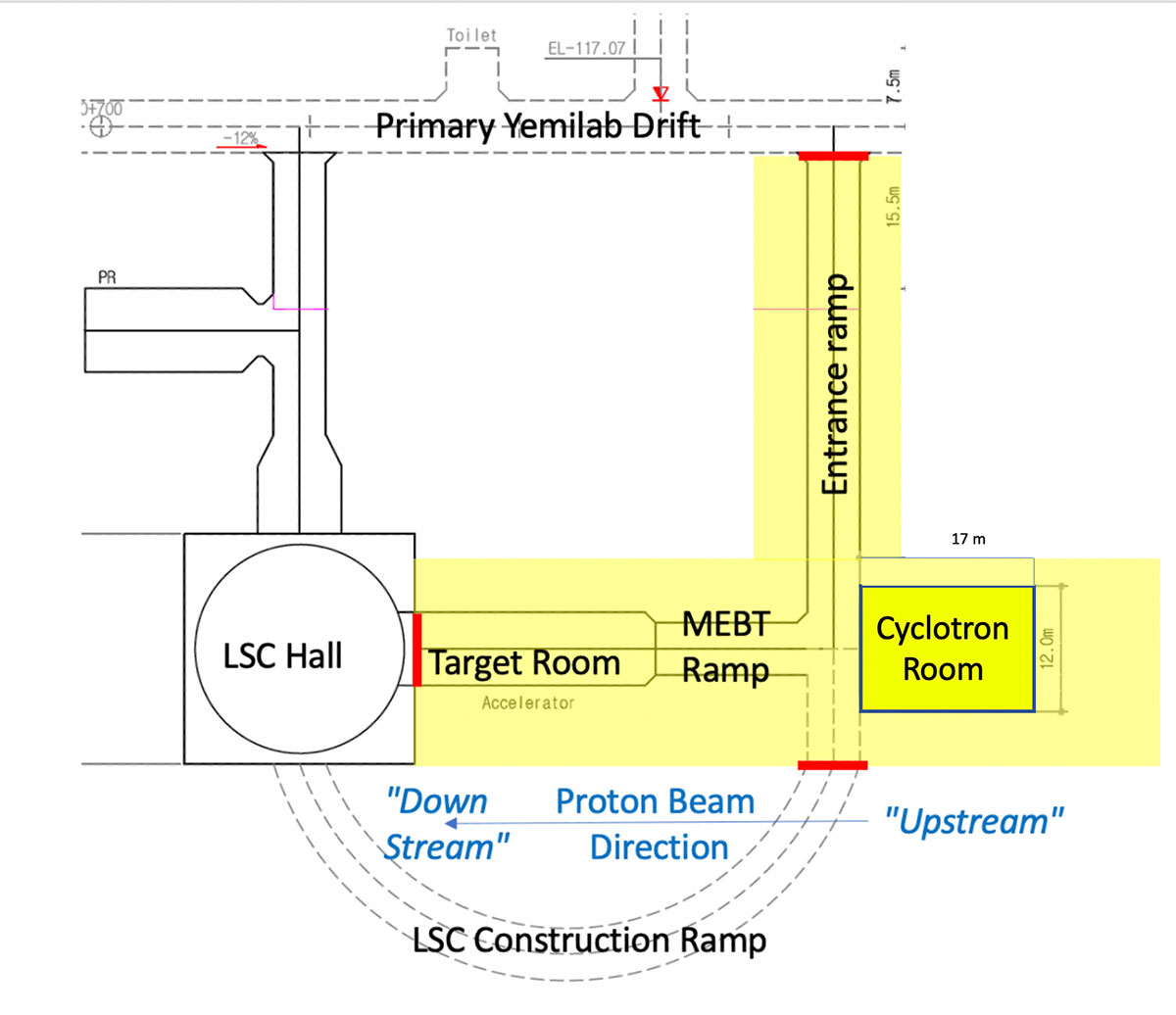}
\end{center}  
\caption{\label{IsoDARYemilayout} The layout of IsoDAR@Yemilab in the Yemilab caverns in 3D (left) \cite{NF02Whitepaper} and plan view (right). Yellow indicates the area for IsoDAR installation and equipment. The beam travels from right (cyclotron) to left (target). }
\end{figure}

The primary goal of IsoDAR is to produce a high intensity $\bar \nu_e$ flux from $^8$Li beta decay.   Because this isotope has a $<1$s lifetime, the $^8$Li must be continuously produced by this facility.    That is performed in three steps.  First, a cyclotron accelerates 5 mA of H$_2^+$ ions to 60 MeV/amu.   Second, the electron is removed, resulting in a 10 mA beam of 60 MeV protons that are transported to the neutrino source.   Third, at the source, the protons drive production of neutron productions that subsequently slow and capture on $^7$Li to produce the desired $^8$Li isotope.     The flux is isotropic, so to maximize the number of $\bar \nu_e$ entering the LSC, the source is placed approximately 7 m from the outer surface of the LSC.   The design of the facility is shown in Fig.~\ref{IsoDARYemilayout}.  The excavation of the IsoDAR cavern complex is complete.

Ref.~\cite{Alonso:2022uar} provides a description of papers associated with the IsoDAR facility.  In particular, 
Ref.~\cite{alonso:isodar_jinst} describes the components and installation of IsoDAR in detail.   Therefore, in this section we provide only a brief overview of the components of the facility.  The installation of IsoDAR is arranged to not interfere with installation of the LSC.

The physics requires 10~mA of protons that are produced using a cyclotron that accelerates 5~mA of H${_2^+}$ ions to 60~MeV/amu~\cite{winklehner2021order, winklehner:nima, winklehner:rfq}.  This compact cyclotron is an example of a new generation of cyclotrons accelerating as much as 10 times the current of existing compact cyclotrons~\cite{IBA}.   
The cyclotron will be installed in the Cyclotron Room  (see Fig.~\ref{IsoDARYemilayout}). 

\begin{figure}[tb]
\centering
\includegraphics[width=3.in]{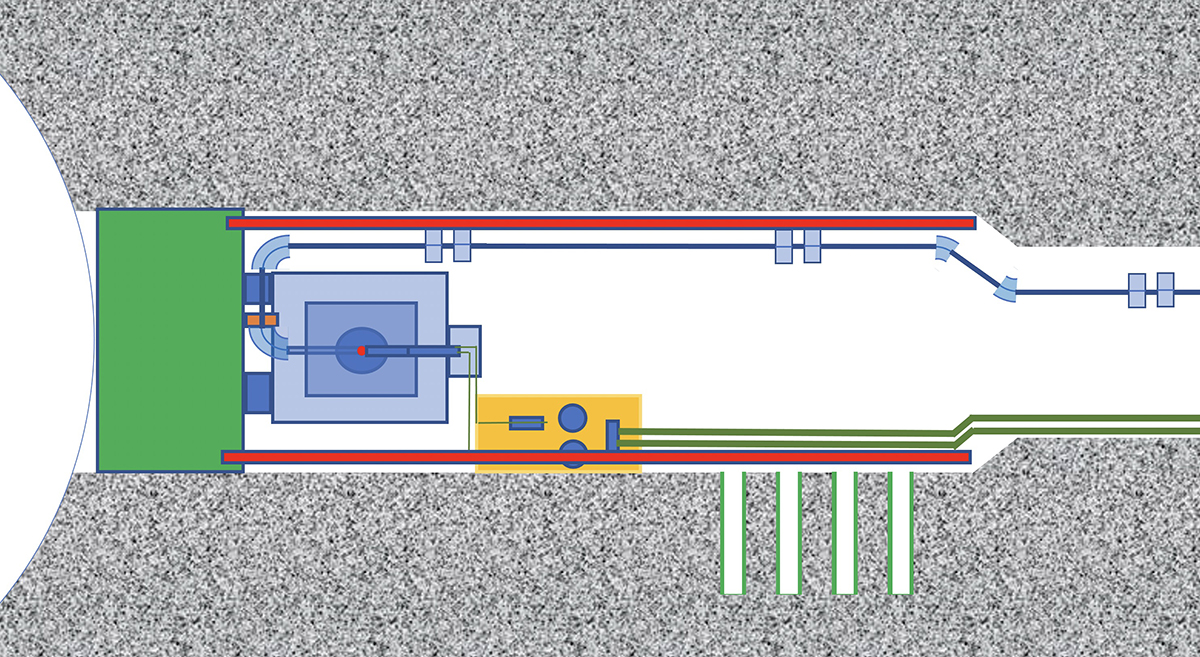}~\includegraphics[width=2.5in]{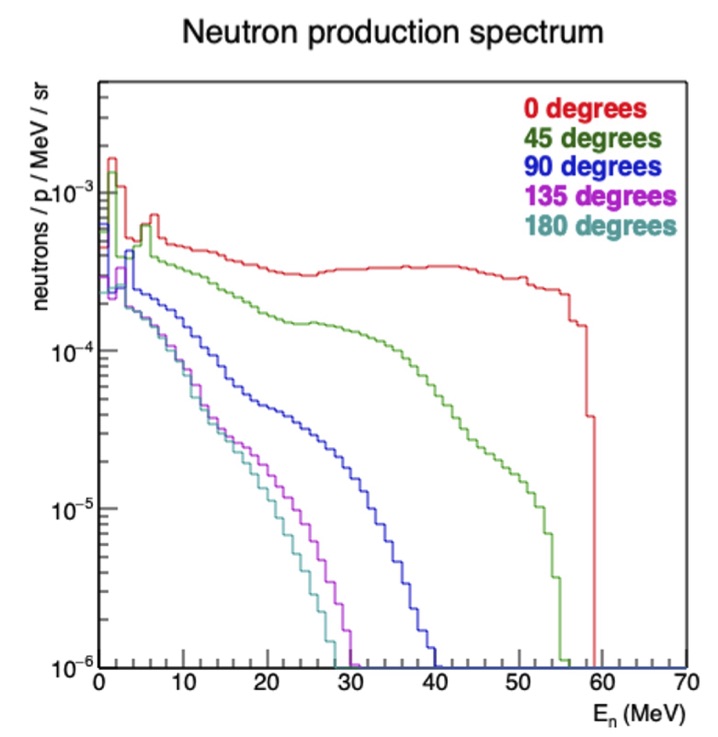}
\caption{{Left: Proposed IsoDAR Target Room. Beamline (blue line) enters from the right and is directed by two $90^{\circ}$  bends to the target within shielding blocks (blue) at left. The LSC is protected by an additional shielding wall (green). The utilities skid is indicated in orange.  Right: the angular distribution of neutrons emerging from the target.  High energy neutrons are directed away from the LSC into an absorber.
 }
\label{target_room}}
\end{figure}

The extracted H${_2^+}$ ions are passed through a thin carbon stripping foil that removes the binding electron and converts the beam to protons.   This represents the start of the  Medium Energy Beam Transport (MEBT) line that brings the beam into the Target Room (see Fig.~\ref{target_room}).    To reach this space, the beamline will cross the top of the entrance to the LSC construction ramp, which allows the closed-off ramp to be re-opened if necessary in the future.  The beampipe is also designed to be easily removed at that point if necessary for LSC access.  At the end of the MEBT, the beam is bent through two 90$^\circ$ bends (see Fig.~\ref{target_room}, Left) to strike the target. 
 
The target consists of nested beryllium hemispheres cooled by flowing heavy water, which is used in place of light water to reduce neutron absorption. The target produces neutrons that flow into a sleeve of about 1.5 meter diameter, filled with a mixture of beryllium (${\approx}75\%$) and enriched ($>$99.99\%)  $^7$Li (${\approx}25\%$). The beryllium serves as a neutron multiplier.  GEANT4 calculations estimate an overall yield of about 0.0146 $\bar \nu _e$ per incident proton.  The target and sleeve are surrounded by shielding, seen in Fig.~\ref{target_room}, Left, 
in light and dark blue, to absorb neutrons. 
Fig.~\ref{target_room} (Right) illustrates the rationale for the two 90$^\circ$ bends in the beam line.  Having the beam striking the target pointing away from the LSC reduces the fast neutrons going towards the LSC.

%% file: Facility/linac.tex
\subsection{Linac facility}
\label{linac}
In Yemilab a 100 MeV electron beam can be used to search for feebly interacting particles such as dark photons, axion-like particles, and low-mass dark matter. For the primary electron beam production, a warm RF linear accelerator is foreseen. The linac should be able to provide 100 MeV beam with
an average power of 100 kW, which translates into 1 mA average current, whereas the maximum peak current will reach $\sim$0.33A.  
The electron source is based on 1A peak emission triode gun, operating at 9-15kV voltage. In the first preliminary design the accelerating part consists of the buncher and 9 constant-gradient, standing wave accelerating structures. The buncher and the first section will be mounted in focusing solenoids. The Linac will be  
 powered by 10 x 5 MW, 15 kW klystrons, operating at 2998 MHz frequency. 
The pulse length will last 12 us and feature a repetition rate up to 300 Hz. 
Detailed beam optics studies will be performed to specify the positions of required focusing quadrupole magnets in dublet or triplet layout. Additionally, special attention will be given to beam dynamics calculations and delivered power optimization. A dedicated magnetic chicane can be used between the buncher and the first accelerating section to minimize the energy spread, in order to avoid the acceleration of out-of-phase particles. However, the maximum power on the target is foreseen, thus too strong removal of the particles from the beam is not acceptable. 
The accelerator system will include also all auxiliary systems like a water cooling system and vacuum pumps to ensure the vacuum level <10$^{-8}$ Torr.

\begin{figure}[b]
\begin{center}
{\includegraphics[width=1.0\textwidth]{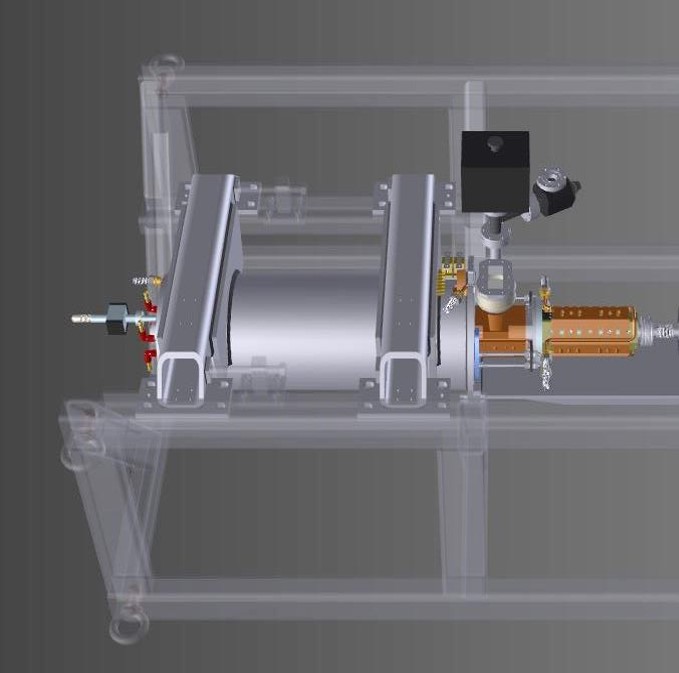}}
\end{center}
\caption{Design of the first section of the 100 MeV Linac.}
\end{figure}

%% file: Physics/physics.tex
\begin{figure}[h]
\begin{center}
\includegraphics[width=.9\textwidth]{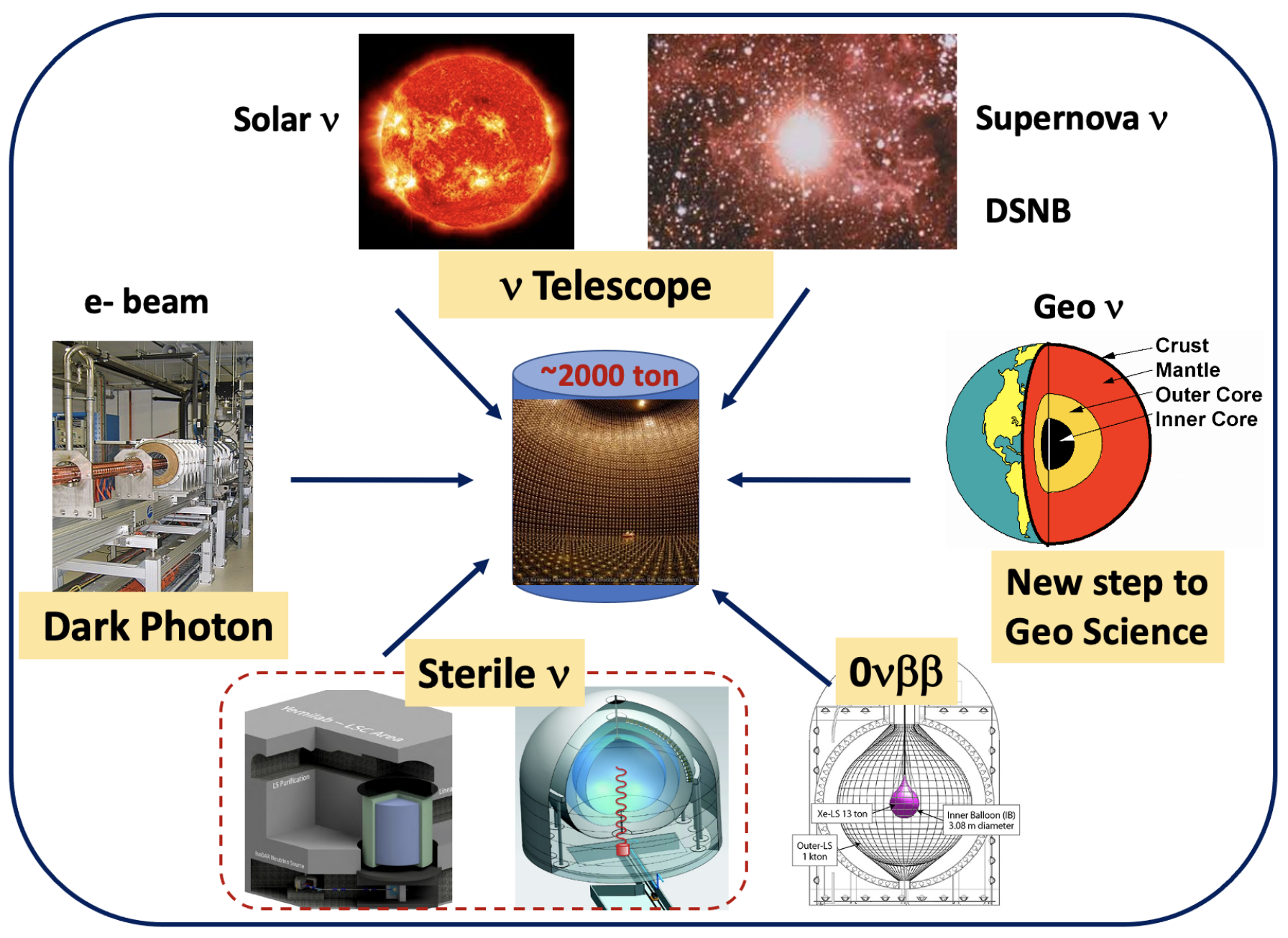}
\end{center}
\caption{The LSC physics program. With additional facilities such as a Linac or IsoDAR source, searches for dark photons and sterile neutrinos are possible, respectively. A search for $0\nu\beta\beta$ could be considered after all the other studies are completed.}
\label{f:program}
\end{figure}

With a $\sim$2~kiloton target neutrino detector, several very interesting physics opportunities can be purused, ranging from particle physics to astroparticle physics.  
Figure~\ref{f:program} shows the physics program of the $\sim$2~kiloton target neutrino detector in Yemilab. 
The main physics programs are solar neutrino physics, sterile neutrino, and dark photon searches but the detection of Supernova burst neutrinos and geo-neutrinos are also possible. 
In a later stage, the detector can be upgraded to focus on a $0\nu\beta\beta$ decay search. 

In this section, physics capabilities for each topic in Fig.~\ref{f:program} are discussed in the following order: solar neutrinos, geo and reactor neutrinos, supernova neutrinos, DSNB neutrinos, IsoDAR physics, sterile neutrino search with radioactive sources, and light dark photon search. 

%% file: Physics/Solar.tex
\subsection{Solar neutrinos}

The Sun serves as a natural factory of low-energy neutrinos, ranging from sub-MeV to MeV energies, which are produced through the nuclear fusion process that takes place in its core. Every second, several tens of billions of solar neutrinos pass through our bodies (per cm$^2$).
The nuclear fusion process in the Sun's core can be broadly categorized into two main processes: the proton-proton ($pp$) chain and the Carbon-Nitrogen-Oxygen (CNO) cycle. 

\begin{figure}[h]
\begin{center}
\includegraphics[width=.45\textwidth]{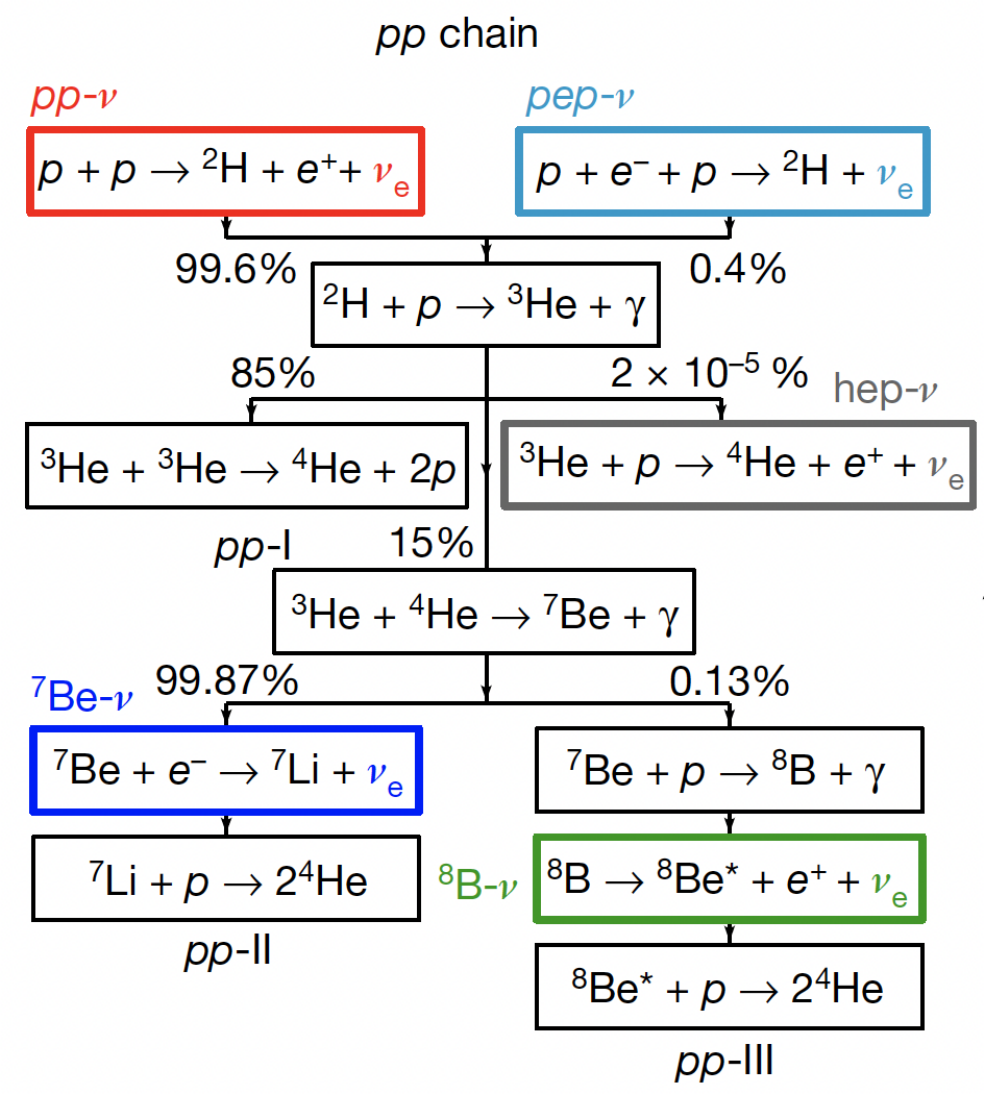} \hspace{1.cm}
\includegraphics[width=.45\textwidth]{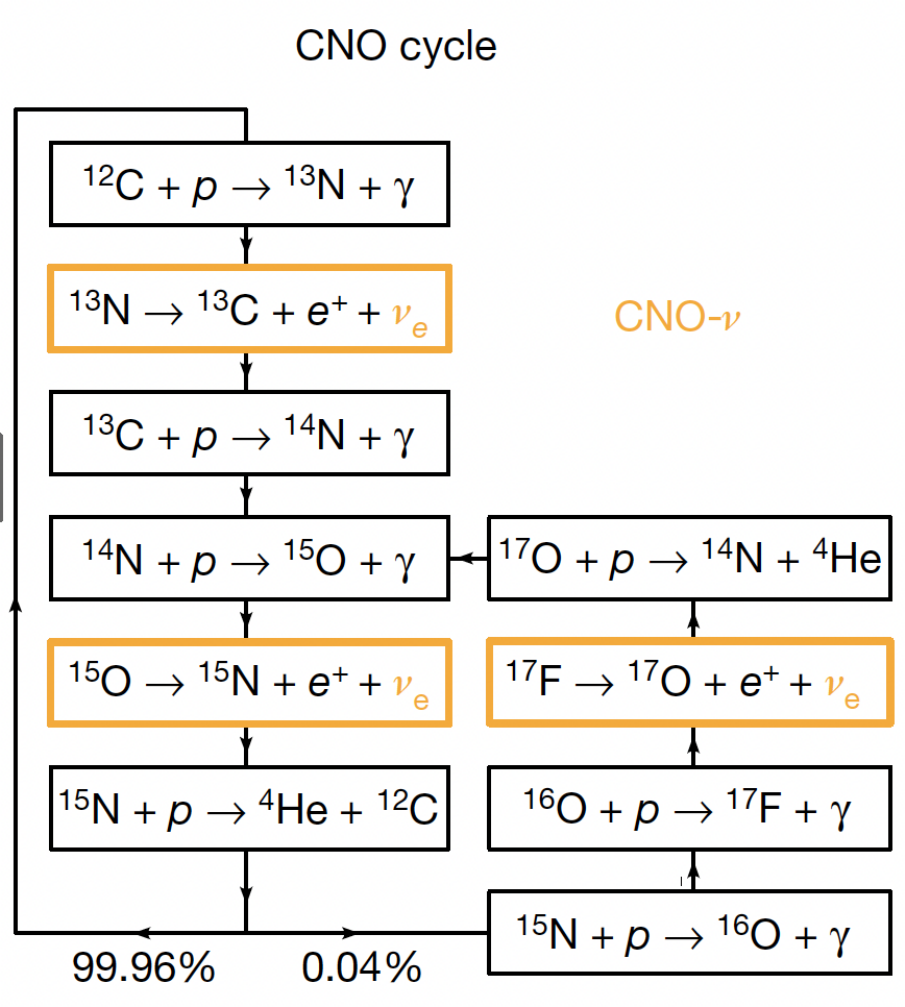}
\end{center}
\caption{$pp$ chain (left) and CNO cycle (right) of the solar fusion reaction where different types of solar neutrinos are produced. Adapted from Borexino Nature article in 2018~\cite{BOREXINO:2018ohr}.}
\label{f:pp_cno}
\end{figure}

\begin{figure}[h]
\begin{center}
\includegraphics[width=.8\textwidth]{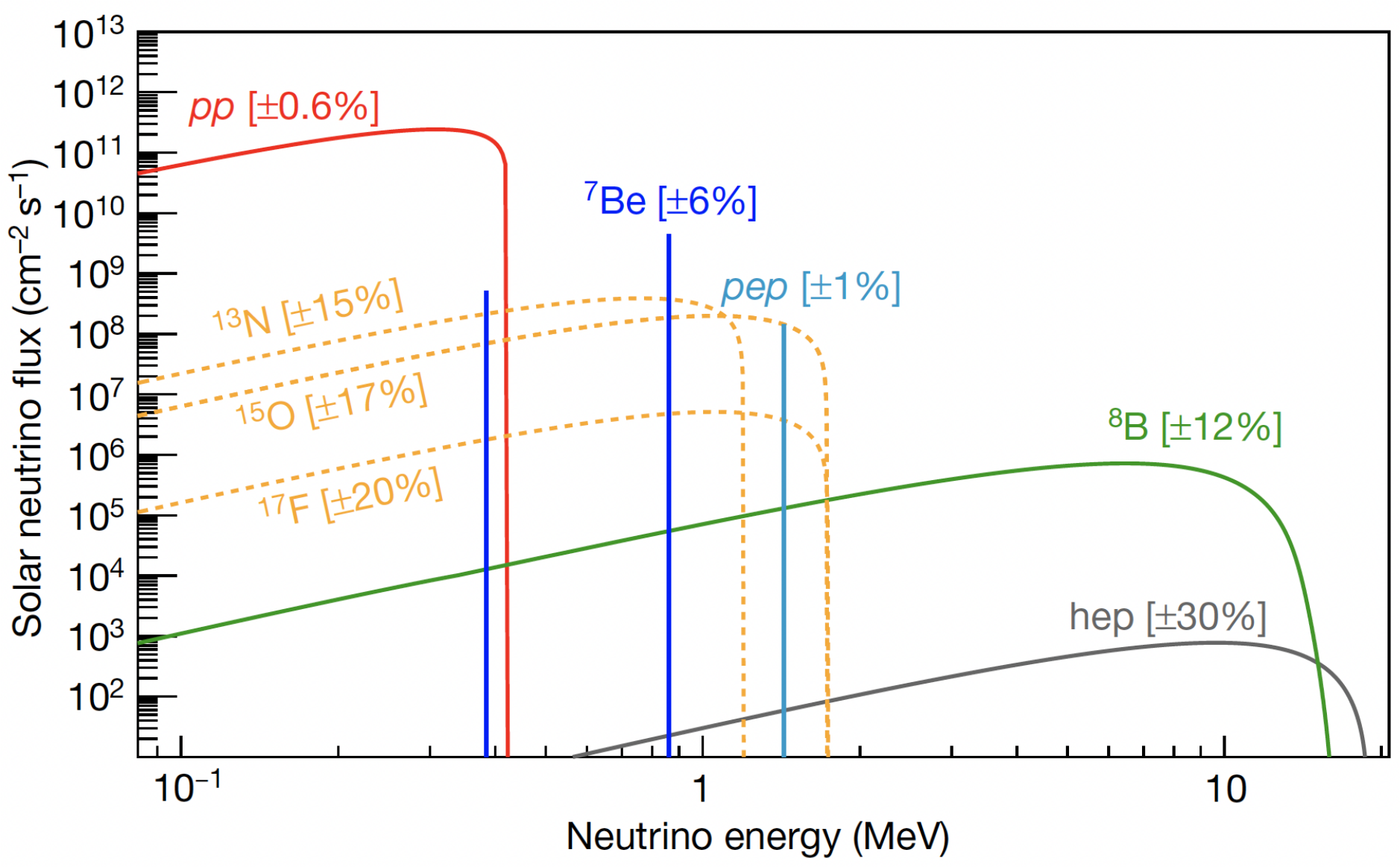}
\end{center}
\caption{Solar neutrino flux. Uncertainties
correspond to the B16-GS98 SSM~\cite{Vinyoles:2016djt}.
}
\label{f:solar_flux}
\end{figure}

The $pp$ chain is the dominant process in the Sun, producing more than $99\%$ of the solar neutrinos. In the $pp$ chain, there are five types of neutrinos produced, depending on the types of fusion reactions: $pp$, $^7$Be, $pep$, $hep$, and $^8$B neutrinos, as shown in the left panel of Fig.~\ref{f:pp_cno}. Unlike $^7$Be ($E_\nu$ = 384, 862 keV) and $pep$ ($E_\nu$ = 1.44 MeV) neutrinos, which are produced at fixed energies, $pp$, $hep$, and $^8$B neutrinos have broad energy distributions due to three-body kinematics with Q-values of 423 keV, 18.8 MeV, and 14.6 MeV, respectively.
The CNO cycle, as shown in the right panel of Fig.~\ref{f:pp_cno}, is a subdominant process in solar fusion, making 
more challenging to measure neutrinos from it.
Nevertheless, measuring CNO neutrinos is important as it provides a direct probe for determining solar metallicity.
Figure~\ref{f:solar_flux} shows solar neutrino fluxes produced from the $pp$ chain and the CNO cycle based on the standard solar model (SSM)\cite{Serenelli:2016dgz, Vinyoles:2016djt}. 

Solar neutrinos have been measured since 1960s, starting with the Homestake experiment~\cite{Cleveland:1998nv}, followed by Kamiokande-II~\cite{hirata89}, GALLEX~\cite{GALLEX:1998kcz}, SAGE~\cite{SAGE:2002fps}, Super-Kamiokande~\cite{Super-Kamiokande:2016yck}, SNO~\cite{SNO:2011ajh}, and Borexino~\cite{Borexino:2017uhp}. The deficit in solar neutrino flux observed by these experiments compared to the standard solar model (SSM) had been a long-standing problem in solar neutrino physics in the past, but was finally explained by the upgraded SNO experiment (with 1 kton of heavy water) in 2002~\cite{SNO:2002tuh}.
Borexino is the most modern and state-of-the-art solar neutrino experiment with a very low radioactive background, and it has measured all neutrinos from the $pp$ chain except for $hep$ neutrinos. In 2020, Borexino achieved the measurement of CNO neutrinos for the first time~\cite{BOREXINO:2022abl}. In particular, the recent achievement of CNO neutrino measurement by Borexino in 2020~\cite{BOREXINO:2022abl} has provided valuable insights into the understanding of solar fusion processes and opened up new avenues for studying the composition and dynamics of the solar core. Although significant progress has been made, further research in solar neutrino physics is still needed to unravel the remaining questions, such as the precise determination of neutrino oscillation parameters, the detection of rare neutrino processes like the hep neutrinos, and the investigation of new physics beyond the standard solar model.


In the following subsections, the sensitivities of the LSC detector for solar neutrino flux and solar metallicity measurements are discussed.

\subsubsection{Solar Neutrino Flux Measurements}


The study of solar neutrino flux requires consideration of three key ingredients: branching of chains, termination of chains, and the spatial distribution of neutrino sources. Branching fraction and the number of unterminated chains, which are sources of uncertainties, strongly depend on the physical conditions (temperature, density, and chemical composition) in the Sun \cite{Vinyoles:2016djt}. The neutrino fluxes can be calculated with respect to the $pp$ neutrino flux, which is the most dominant. For instance, the flux of $^7$Be, which is the second most dominant, can be estimated using the fraction equation and solar luminosity \cite{Vinyoles:2016djt}. The branching of $pep$ neutrinos depends strongly on the electron density, while the ratio of fluxes of $^8$B and $^7$Be neutrinos strongly depends on temperature. Furthermore, the branching of $^7$Be to $pp$ neutrinos depends on the ratio of $^3$He to $^4$He and does not depend on the temperature. By considering the branching and luminosity constraints, it is possible to reproduce the different fluxes. Table \ref{t:solarNu_fluxes} presents the neutrino fluxes and their theoretical uncertainties for models with high and low metallicities, and compares them with the expected measurements at LSC in Yemilab for 5 years of data taking. The central value and the systematic uncertainties are assumed as those of Borexino \cite{BOREXINO:2018ohr}. It is noteworthy that LSC at Yemilab, is expected to yield significant improvements in solar neutrino flux measurement compared to Borexino. The larger fiducial mass, higher statistics, and better energy resolution of LSC at Yemilab will enable more accurate flux measurements.

\begin{table}[t]
\centering
\scalebox{1.0}{}
\begin{tabular}{|l|l|l|l|}
\hline\hline
Solar $\nu$ type & Rate & Flux & Flux-SSM prediction  \\
           & (counts/day/100 ton) & ($cm^{-1}s^{-1}$) & ($cm^{-1}s^{-1}$) \\
\hline
$pp$       & $134\pm2^{+6}_{-10}$  & $(6.1\pm0.08^{+0.3}_{-0.5})$x $10^{10}$ & 5.98(1.0$\pm$0.006)x $10^{10}$ (HZ)  \\
  & & & 6.03(1.0$\pm$0.005)x $10^{10}$ (LZ) \\
\hline
$^7$Be      & $48.3\pm0.2^{+0.4}_{-0.7}$   & $(4.99\pm0.02^{+0.06}_{-0.08})$x $10^{9}$ & 4.93(1.0$\pm$0.006)x $10^{9}$ (HZ) \\
& & & $4.50(1.0\pm0.006)$x $10^{9}$ (LZ)\\
\hline
$pep$ (HZ) & $2.43\pm0.06^{+0.15}_{-0.22}$  & $(1.27\pm0.03^{+0.08}_{-0.12})$x $10^{8}$ & 1.44(1.0$\pm$0.01)x $10^{8}$ (HZ)  \\
& & & $1.46(1.0\pm0.0009)$x $10^{8}$ (LZ)\\
$pep$ (LZ)   & $2.65\pm0.06^{+0.15}_{-0.22}$ & $(1.39\pm0.03^{+0.08}_{-0.13})x 10^{8}$ & 1.44(1.0$\pm$0.01)x $10^{8}$ (HZ) \\
& & & $1.46(1.0\pm0.009)$x $10^{8}$ (LZ)\\
\hline
$^8$B      & $0.223^{+0.002}_{-0.003}\pm0.006$        & ($5.68\pm0.06\pm{0.03})$x $10^{6}$ & 5.46(1.0$\pm$0.12)x $10^{6}$ (HZ) \\
& & & $4.50(1.0\pm0.12)$x $10^{6}$ (LZ)\\
\hline
CNO    &        &   & $4.88(1.0\pm0.11)$x $10^{8}$ (HZ)  \\
& & & $3.51(1.0\pm0.10)$x $10^{8}$ (LZ)\\
\hline
$hep$     & any      & <2.2x $10^5$ (90\% C.L.)   & 7.98(1.0$\pm$0.30)x $10^{3}$ (HZ)  \\
& & & $8.25(1.0\pm0.12)$x $10^{3}$ (LZ)\\
\hline\hline
\end{tabular}
\caption{
Expected solar neutrino measurements for 5 years of operation at LSC in Yemilab assuming the same central values and systematic uncertainties as those of Borexino~\cite{BOREXINO:2018ohr}. 
}
\label{t:solarNu_fluxes}
\end{table}

\begin{figure}[h]
\begin{center}
\includegraphics[width=1.\textwidth]{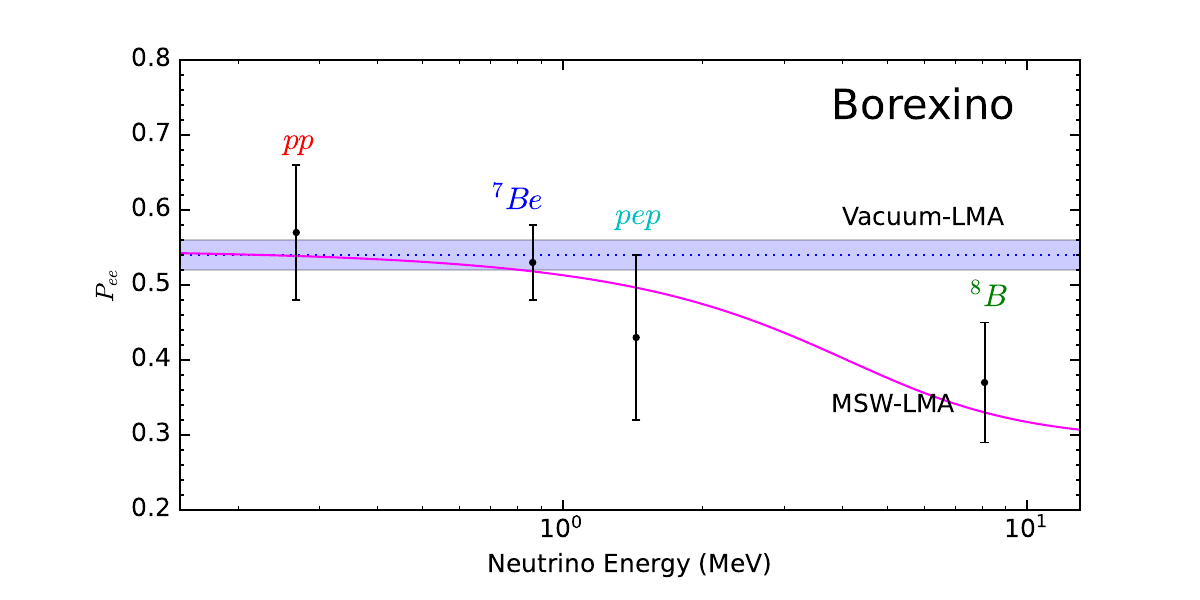}\\
\includegraphics[width=1.\textwidth]{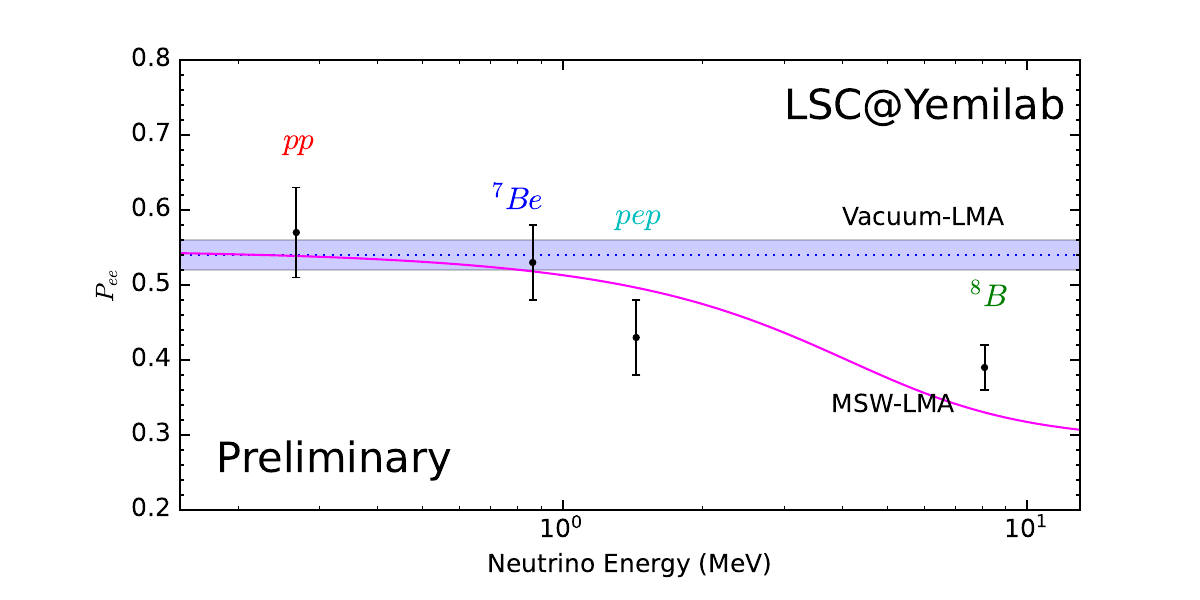}
\end{center}
\caption{Solar Neutrino survival probability $P_{ee}$ at Borexino (top) and Yemilab (bottom). 
The violet band indicates the vacuum-LMA solution. Data points indicate Borexino results for $pp$ (red), $^7$Be (blue), $pep$ (cyan), and $^8$B (green) assuming HZ-SSM. Systematic uncertainties are assumed to be the same as those of Borexino~\cite{BOREXINO:2018ohr}.}
\label{f:solar_Pee}
\end{figure}

\subsubsection{Propagation and flavor conversion of solar neutrino; MSW-LMA solution}

For electron neutrinos produced in the Sun's central regions and detected on Earth, oscillation is generally negligible since the mass eigenstates lose coherence during their journey to the surface of the Earth. The evolution of the flavor-states, denoted by $\nu_f \equiv (\nu_e, \nu_\mu, \nu_\tau)^T$, is described by the equation:

\begin{equation}
  i \frac{d \nu_f}{dx} = (H_{0} + H_{ {\rm mat}}^{ {\rm MSW}}) \nu_f \,,
\end{equation}
Here, $H_{0}$ is the vacuum Hamiltonian, and $H_{ {\rm mat}}^{ {\rm MSW}} = {\rm diag}(V_e, 0, 0)$
is the diagonal matrix of matter potentials with $V_e = \sqrt{2} G_F N_e$.  $G_F$ is the Fermi
constant and $N_e$ is the number density of electrons.  Since the wave packets of different mass eigenstate components of the solar neutrino spectrum  have different sizes, they lose coherence as they travel to the Earth at different group velocities. The probability of finding $\nu_e$ at the time of arrival ($t_E$) for the MSW-LMA solution is given by \cite{Maltoni:2015kca, Bakhti:2020tcj}:

\begin{equation}
  \label{eq:patearth}
  P_{ee} = |\langle \nu_e | \nu (t_E) \rangle |^2
  =   c_{13}^2 c_{13}^{m2} P_2^{\rm ad}
  + s_{13}^2 s_{13}^{m2} \,,
\end{equation}
where
\begin{align}
  \label{eq:adform}
  P_2^\text{ad} &= \frac{1}{2}(1 + \cos 2\theta_{12} \cos 2 \theta_{12}^m) \,,
\end{align}
and mixing angle in matter $\theta_{12}^m$ is given by
\begin{equation}
  \cos 2 \theta_{12}^m =
  \frac{\cos 2\theta_{12} - c_{13}^2 \epsilon_{12}}{\sqrt{(\cos 2\theta_{12}
      - c_{13}^2 \epsilon_{12})^2 + \sin^2 2\theta_{12}}}\,,
\end{equation}
with the parameter
\begin{equation}
  \label{eq:eps12}
  \epsilon_{12} \equiv \frac{2V_e E}{\Delta m_{21}^2} \,.
\end{equation}
The averaged value of solar neutrino survival probability, $P_{ee}$ over the flux production region as a function of neutrino energy for MSW-LMA solution is shown in Fig.~\ref{f:solar_Pee} by the purple curve. 
The top panel shows the solar neutrino survival probability at Borexino and the bottom panel shows the corresponding expected measurement at Yemilab. The error bars include experimental and theoretical uncertainties with the systematic uncertainties being assumed to be same as those of Borexino. The violet band represents the vacuum-LMA case with the oscillation parameters being fixed as given by \cite{Esteban:2018azc}. Data points indicate Borexino results for $pp$ (red), $^7$ Be (blue), $pep$ (cyan), and $^8$B (green) assuming HZ-SSM. The error bars exclusively encompass statistical errors.
As can be observed, reducing uncertainties at Yemilab is crucial to achieve precise and reliable measurement of $P_{ee}$.

At Yemilab, solar neutrinos are detected through $\nu-e$ scattering using a liquid scintillator detector. This detection method allows for the detection of neutrinos of all flavors, including $\nu_e, \nu_\mu,$ and $\nu_\tau$, via the following interaction:

\begin{equation}
\nu_{e,\mu,\tau} + e^- \rightarrow \nu_{e,\mu,\tau} + e^-
\end{equation}

\begin{figure}[h]
\begin{center}
\includegraphics[width=1.\textwidth]{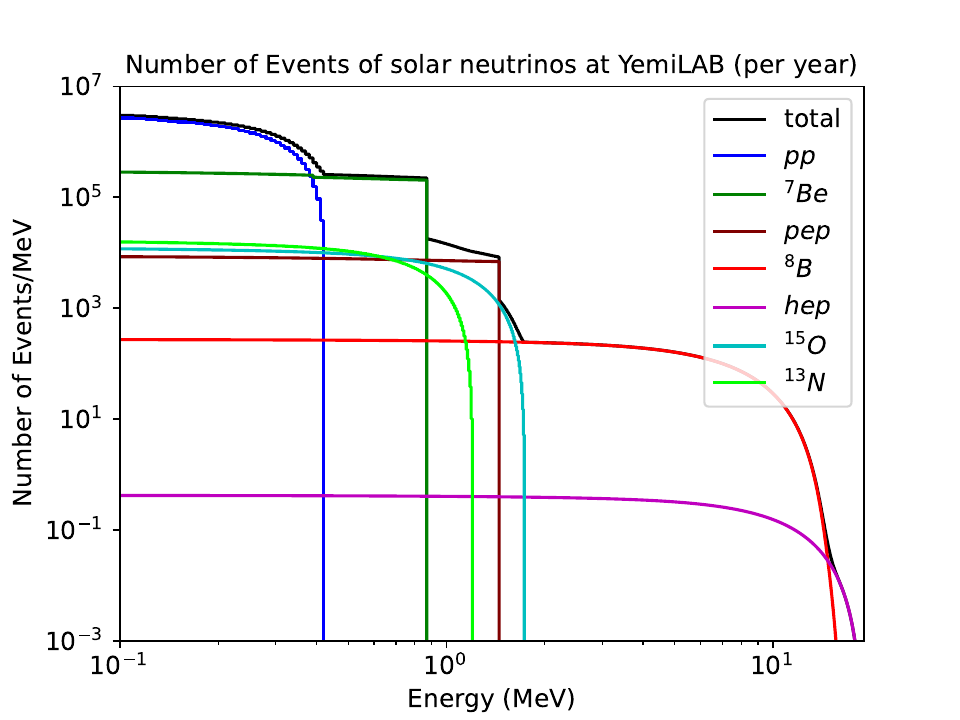}\\
\end{center}
\caption{ The expected number of annual events of scattered electrons per MeV due to solar neutrinos as a function of the kinetic energy of the electron, assuming the best-fit values of the solar neutrino data,   $\Delta m^2_{21} = 6.11 \times 10^{-5}$ and $\sin^2\theta_{12} = 0.306$ \cite{nakajima2020recent}, at Yemilab. The total number of events is represented by the black curve in the plot.}
\label{noe_solar}
\end{figure}

The expected energy spectra of the detected neutrinos as a function of the kinetic energy of the electron at LSC at Yemilab are shown in Figure \ref{noe_solar}. The oscillation parameters are set at $\sin^2{\theta_{12}} = 0.306$ and $\Delta m^2 _{21}= 6.11 \times 10^{-5}$~\cite{nakajima2020recent}, reflecting the best fit values of solar neutrino oscillation parameters based on the combined data from SK and SNO solar neutrino observatories. Our analysis assumes the absence of background events and assumes perfect energy resolution.

\begin{figure}[h]
\begin{center}
\includegraphics[width=0.49 \textwidth]{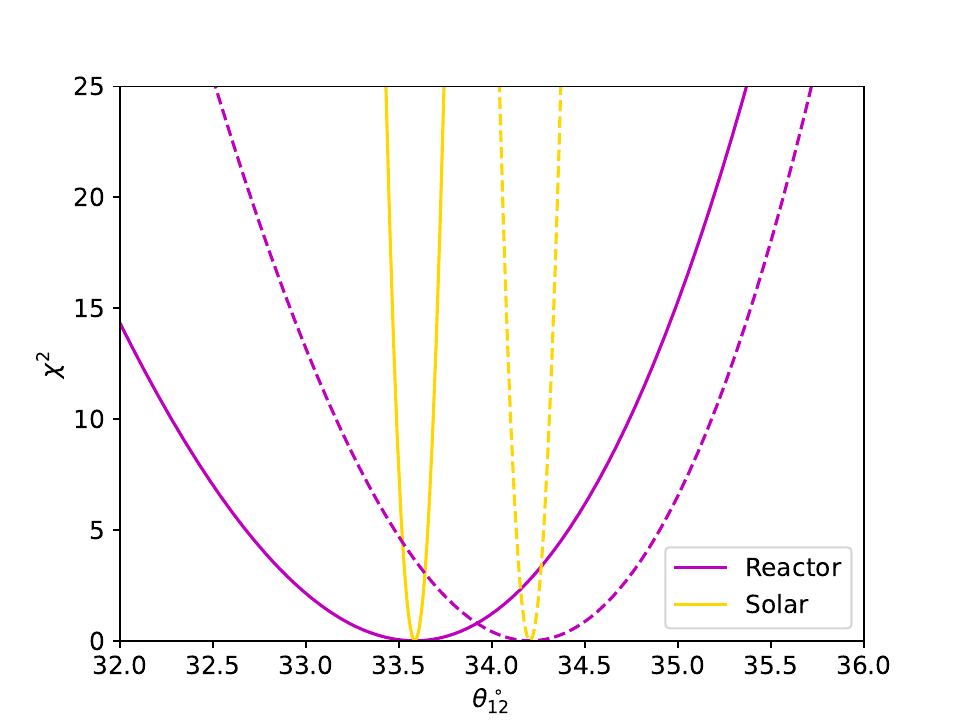}
\includegraphics[width=0.49 \textwidth]{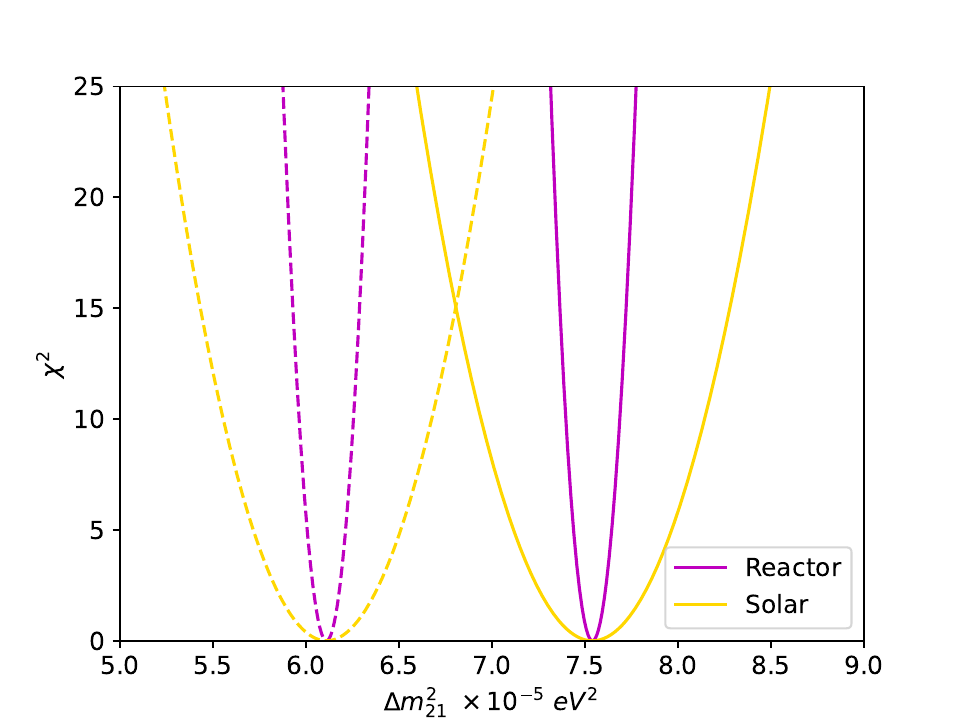}
\end{center}
\caption{\label{constraints_1d} The constraints on solar neutrino oscillation parameters $\theta_{12}$ and $\Delta m^2_{21}$ using detection of solar and  reactor neutrinos in LSC at Yemilab for five years. The dashed curves are generated under the assumption that the best fit values of SK/SNO solar data are the true values, while the solid curves are plotted assuming the best fit values of KamLAND as the true values. The magenta and gold curves correspond to the reactor and solar data, respectively.
As it is demonstrated, solar neutrino in LSC at Yemilab determines the value of $\theta_{12}$ with the highest precision. Reactor neutrinos will determine the value of $\Delta m^2_{21}$ with a higher precision.}
\end{figure}

As it is well known, the current measurements of the solar neutrino mixing angle $\theta_{12}$ and the solar neutrino mass-squared difference $\Delta m^2_{21}$ are inconsistent with the results from the KamLAND experiment at around 1 or 2 $\sigma$ C.L.~\cite{Maltoni:2015kca, nakajima2020recent}. Several potential solutions to the tension between solar neutrino experiment measurements and the KamLAND results exist. These include:  
\begin{itemize}
\item  Altering Solar Models:  The solar neutrino flux predictions depend on the employed solar models, which involve assumptions regarding the Sun's internal structure and properties. Refinements to these models, such as the incorporation of three-dimensional (3D) models for the Sun, the consideration of solar magnetic field effects, and adjustments to conduction and atomic diffusion models, could potentially affect the predicted neutrino fluxes and their flavor conversion in a way that resolves the discrepancy.
\item Introduction of New Physics Beyond the Standard Model:  This could encompass the inclusion of Non Standard Interactions (NSI) between neutrinos or adding sterile neutrino(s) which can affect the flavor conversion of solar neutrinos and improve the discrepancy.   
\item Further investigations: including more precise experimental measurements, improved theoretical calculations, and detailed analyses, are required to fully understand the nature of the tension between solar neutrino experiment measurements and the KamLAND results, and to determine the most likely solution(s).
\end{itemize}

It has been recently shown that LSC at Yemilab has remarkable potential for accurately determining the solar neutrino parameter $\theta_{12}$ \cite{Bakhti:2023vzn}. This potential arises from its capability to detect solar neutrinos, benefiting from several factors such as a low energy threshold leading to a large number of events, a minimized background, and the ability to detect pp neutrinos. In the same study, it has been demonstrated that by combining the data from reactor neutrinos originating from the Hanul power plant with LSC at Yemilab, we can achieve a simultaneous determination of the solar neutrino parameters, $\theta_{12}$ and $\Delta m^2_{21}$, with an accuracy up to the percent level. Interestingly, for smaller values of $\Delta m^2_{21} < 6 \times 10^{-5}$,  LSC at Yemilab demonstrates the most exceptional capability in accurately determining the precise value of $\Delta m^2_{21}$ \cite{Bakhti:2023vzn}. For more details about reactor neutrino detection at Yemilab please see chapter \ref{sec:georeactor}.
Figure \ref{constraints_1d} shows the potential of Yemilab in constraining the solar neutrino oscillation parameters, $\theta_{12}$ and $\Delta m^2_{21}$, through the detection of both solar and reactor neutrinos over a five year period. The dashed curves represent the assumption that the best fit values from SK/SNO solar data are the true values \cite{nakajima2020recent}, while the solid curves assume the best fit values from KamLAND as the true values. The magenta and gold curves correspond to the reactor and solar data, respectively. In our analysis, we have employed the same background modeling as the Jinping experiment for solar neutrino analysis \cite{Jinping:2016iiq}. For the reactor neutrino analysis, we have considered the background modeling similar to JUNO, with a reduced number of background events due to the better shielding and deeper overburden \cite{JUNO:2015zny, Bakhti:2014pva}. 

As can be observed in the figure~\ref{constraints_1d}, the detection of solar neutrinos in LSC at Yemilab provides the most precise determination of $\theta_{12}$. As mentioned previously, the key advantage of LSC at Yemilab is its low energy threshold, allowing for the detection of $pp$ neutrinos and $^7$Be neutrinos with high statistical significance, amounting to several hundred thousand events per year. This exceptional capability plays a crucial role in precisely determining $\theta_{12}$. On the other hand, reactor neutrinos offer high precision in determining the value of $\Delta m^2_{21}$.

In the following section, we will focus on introducing new physics such as NSI and sterile neutrinos and investigate how they can affect neutrino flavor conversion.

\subsubsection{Impact of NSI on solar neutrino survival probability}

Non-Standard Interactions (NSIs) offer a framework beyond neutrino mass to address the solar neutrino problem by modifying the chiral couplings of the neutrino and electron and altering the $P_{ee}$. NSIs can affect solar neutrinos during propagation and detection, with neutrino-flavor-diagonal NSIs being particularly sensitive to detection. This chapter focuses on how NSIs can modify solar neutrino survival probability. The neutral current NSI is expressed as an effective four-fermion operator:

~~~~~~~~~~~~~~~~~~~~~~~~~~~~~

\begin{equation}
\label{eq:def}
\mathcal{L}_\text{NSI} =
- 2\sqrt{2} G_F \epsilon_{\alpha\beta}^{fP}
(\bar\nu_{\alpha} \gamma^\mu \nu_{\beta})
(\bar{f} \gamma_\mu P f) \,,
\end{equation}
where $G_{F}$ is the Fermi constant, $f$ denotes a charged fermion, $P=(L,R)$ and
$\epsilon_{\alpha\beta}^{fP}$ are dimensionless parameters encoding the
deviation from standard interactions.
 In the presence of NSI and including the standard matter effect (MSW), the total Hamiltonian can be written as
\begin{equation}
H = H_{0} + H_\mathrm{mat}^\mathrm{MSW} + H_\mathrm{mat}^\mathrm{NSI}
\label{H_tot}
\end{equation}
where $H_0$ and $H_\mathrm{mat}^\mathrm{MSW}$ represent the vacuum and standard matter Hamiltonians, respectively. $H_\mathrm{mat}^\mathrm{NSI}$ represents the NSI Hamiltonian and is given as


\begin{equation}
\label{eq:hmatNSI}
~~H_\mathrm{mat}^\mathrm{NSI}=
\sqrt{2} G_F \sum_{f=e,u,d} N_f
\begin{pmatrix}
\epsilon_{ee}^f & \epsilon_{e\mu}^f & \epsilon_{e\tau}^f
\\
\epsilon_{e\mu}^{f*} & \epsilon_{\mu\mu}^f & \epsilon_{\mu\tau}^f
\\
\epsilon_{e\tau}^{f*} & \epsilon_{\mu\tau}^{f*} & \epsilon_{\tau\tau}^f
\end{pmatrix}.
\end{equation}
For solar neutrinos,  $2\times2$ effective Hamiltonian is a good approximation (Since $\frac{\Delta m^2_{31}}{E_\nu}\gg G_F N_e$). Thus,  vacuum and matter Hamiltonian can be written:

\begin{align}
\label{eq:hvacsol}
H_\text{0}
&= \frac{\Delta m^2_{21}}{4 E_\nu}
\begin{pmatrix}
-\cos 2\theta_{12} & \sin 2\theta_{12} \\
\hphantom{+} \sin 2\theta_{12} & \cos 2\theta_{12}
\end{pmatrix} ,
\\
\label{eq:hmatsol}
H_\text{mat}^\text{eff}
&=H_\mathrm{mat}^\mathrm{MSW} + H_\mathrm{mat}^\mathrm{NSI}
= \sqrt{2} G_F N_e(r)
\begin{pmatrix}
c_{13}^2 & 0 \\
0 & 0
\end{pmatrix}
+ \sqrt{2} G_F \sum_f N_f(r)
\begin{pmatrix}
-\epsilon_D^{f\hphantom{*}} & \epsilon_N^f \\
\hphantom{+} \epsilon_N^{f*} & \epsilon_D^f
\end{pmatrix} .
\end{align}
The coefficients $\epsilon_D^f$ and $\epsilon_N^f$ are given with respect to the original
parameters $\epsilon_{\alpha\beta}^f$ as the following \cite{Gonzalez-Garcia:2013usa,Bakhti:2020hbz}
\begin{align}
\label{eq:epsD}
\epsilon_D^f =
-\frac{c_{13}^2}{2} \big( \epsilon_{ee}^f - \epsilon_{\mu\mu}^f \big)
+ \frac{s_{23}^2 - s_{13}^2 c_{23}^2}{2}
\big( \epsilon_{\tau\tau}^f - \epsilon_{\mu\mu}^f \big)~~~~~~~~~~~~~~~~~~~~~~~~~~~~~~~~~~~
\\
+ \text{Re}\left[ c_{13} s_{13}e^{i\delta} \big( s_{23} \, \epsilon_{e\mu}^f
+ c_{23} \, \epsilon_{e\tau}^f \big)
- \big( 1 + s_{13}^2 \big) c_{23} s_{23} \epsilon_{\mu\tau}^f \right]
\,
\end{align}
\begin{align}
\label{eq:epsN}
\epsilon_N^f =
c_{13} \big( c_{23} \, \epsilon_{e\mu}^f - s_{23} \, \epsilon_{e\tau}^f \big)
+ s_{13} e^{-i\delta} \left[
s_{23}^2 \, \epsilon_{\mu\tau}^f - c_{23}^2 \, \epsilon_{\mu\tau}^{f*}
+ c_{23} s_{23} \big( \epsilon_{\tau\tau}^f - \epsilon_{\mu\mu}^f \big)
\right]\,.
\end{align}

\begin{figure}[h]
\hspace{0cm}
\includegraphics[width=1.\textwidth, height=0.55\textwidth]{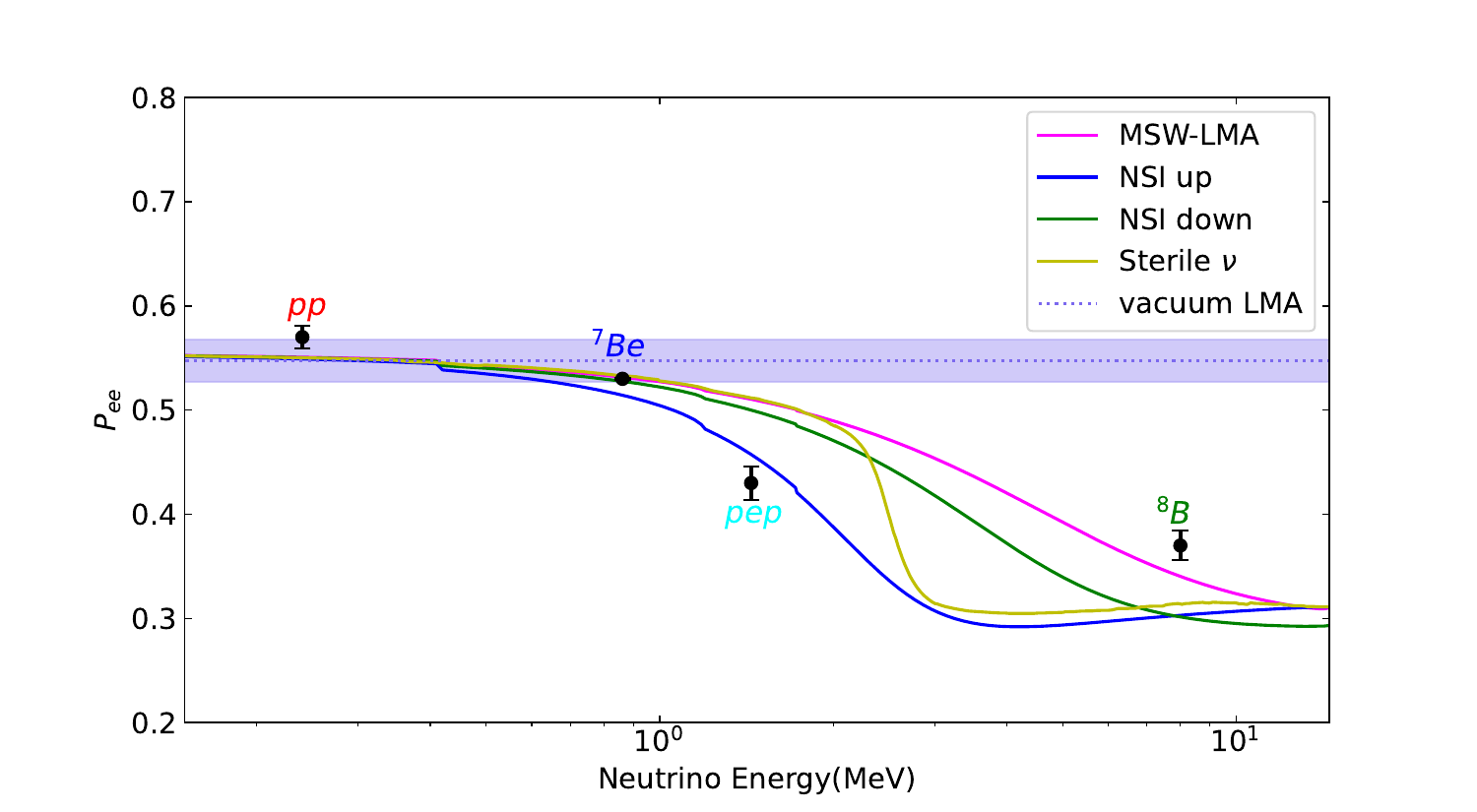} 
\caption[...]{ $P_{ee } $ vs. neutrino energy. We have assumed
$\sin^2\theta_{12}=0.31$, $\sin^2\theta_{13}=0.0244$, and $\Delta m_{21}^2=7.5\times10^{-5}~{\rm eV^2}$. The purple curve indicated the standard MSW-LMA solution. Blue and green curves indicate non-standard interaction with up type  ($\varepsilon_D^u = -0.22$, $\varepsilon_N^u =-0.30$) and down type  ($\varepsilon_D^d = -0.12$, and $\varepsilon_N^d = -0.16$) quarks, respectively. Yellow curve corresponds to the sterile neutrino case 
 assuming $\Delta m^2_{01}=2\times10^{-5}{\rm eV^2}$ and $\sin^22\theta_{01}=0.005$. 
The error bars represent solely statistical errors.}
\label{Pee}
\end{figure}

Then the effective Hamiltonian can be diagonalized as~\cite{Liao:2017awz}
\begin{align}
U'=\left(\begin{array}{cc}
\cos\tilde{\theta}_{12} & \sin\tilde{\theta}_{12} e^{-i\phi}
\\
-\sin\tilde{\theta}_{12} e^{i\phi} & \cos\tilde{\theta}_{12}
\end{array}\right)\,,
\end{align}
where
\begin{align}
\tan2\tilde{\theta}_{12}=\frac{|\sin2\theta_{12}+2\hat{A}_s\epsilon_N|}{\cos2\theta_{12}-\hat{A}_s(c_{13}^2-2\epsilon_D)}\,,
\end{align}
and
\begin{align}
\phi=-\text{Arg}\left(\sin2\theta_{12}+2\hat{A}_s\epsilon_N\right)\,.
\end{align}
and finally, the solar neutrino oscillation probability  is given by \cite{Bakhti:2020hbz, Liao:2017awz}
\begin{equation}
P_{ee} (E) = \frac{1}{2} {c_{13}^4}\left[1 + \cos 2 \theta_{12} \cos 2 \tilde{\theta}_{12}(E) \right] + s_{13}^4
\end{equation}
where $\hat{A}_s = 2 \sqrt{2} N^s _e E_{\nu} / \Delta m^2 _{21}$. Figure \ref{Pee} illustrates the impact of non-standard interactions on the solar neutrino survival probability. The purple curve represents standard MSW matter effect, while the blue and green curves depict non-standard interactions with up-type and down-type quarks, respectively. The oscillation parameters are fixed at $\sin^2\theta_{12}=0.31$, $\sin^2\theta_{13}=0.0244$, and $\Delta m_{21}^2=7.5\times10^{-5}~{\rm eV^2}$. The error bars indicate statistical uncertainties. As can be observed, including NSI can significantly modify the 
solar neutrino survival probability. Furthermore, the consideration of NSI can have impact on the detection of solar neutrinos in addition to their propagation.\cite{Bakhti:2020hbz}.

\subsubsection{Impact of Sterile neutrino on solar neutrino survival probability}

The existence of sterile neutrinos can have a significant impact on the survival probability of solar neutrinos.
In the presence of a sterile neutrino, the four flavor eigenstates are denoted as $\nu_f = (\nu_s, \nu_e, \nu_\mu, \nu_\tau)$ and the mass eigenstates as $\nu_i$, where $i = 0, 1, 2, 3$. The sterile neutrino $\nu_s$ is primarily present in the mass eigenstate $\nu_0$ with mass $m_0$.

The evolution equation for solar neutrinos in the presence of a sterile neutrino can be written as:

      \begin{equation}
            \label{eq.schr_4v}
            i \frac{d}{dx} |\nu_{\alpha}\rangle = H_{f} |\nu_{\alpha}\rangle \quad , \quad \alpha = s, e, \mu, \tau\,,
        \end{equation}
    with the Hamiltonian
        \begin{align}
        \label{eq.Hf_4v}
            H_{f}   &= U \mathrm{diag}(\frac{\Delta m_{01}^{2}}{2E_{\nu}}, 0, \frac{\Delta m_{21}^{2}}{2E_{\nu}}, \frac{\Delta m_{31}^{2}}{2E_{\nu}}) U^{\dagger} + V\,, 
          \end{align}  
  where $E_\nu$ is the neutrino energy, $\Delta m_{ij}^2$ are the mass-squared differences, 
  and the potential is given by
          \begin{align}
            V       &= \mathrm{diag}(0, V_{CC}+V_{NC}, V_{NC}, V_{NC}) \\
                    &= \sqrt{2}G_{F}~\mathrm{diag}(0, N_{e}-N_{n}/2, -N_{n}/2, -N_{n}/2)\,.
        \end{align}
In the above formula,  $N_{e}$ and $N_{n}$  are the number density of electron and neutron, respectively.  $V_{CC}$ and $V_{NC}$ are the charged-current  and neutral-current  potentials. The mixing matrix $U$ is given as ~\cite{deHolanda:2010am,deHolanda:2003tx}
    \begin{eqnarray}
        \label{eq.U}
        U\equiv\left( \begin{matrix}1 & 0\cr 0 & U^{3\nu}\end{matrix}
        \right)\cdot U_S\,,
    \end{eqnarray}
    where $U^{3\nu}$ is the standard three-neutrino mixing matrix, (PMNS matrix) and is given by $U^{3\nu} = R_{23}(\theta_{23}) \cdot R_{13}(\theta_{13},\delta_{cp}) \cdot R_{12}(\theta_{12})$ and $U_{S}$ is the sterile mixing matrix.


    
    
Finally, after some calculations, the survival probability   ($\nu_{e} \rightarrow \nu_{e}$)  on the Earth can be written as
    \begin{equation}
        \label{eq:pee_nonad}
            P_{ee} = \left|\sum_{i=0}^{3}U_{ei}e^{-i\frac{\Delta m_{i1}^{2}}{2E_{\nu}}L_{0}}A_{ei}\right|^{2}\,,
    \end{equation}
    where $L_{0} \simeq 1.5 \times 10^{11} \ \mathrm{m}$ is the distance between the Earth and the Sun and  $A_{ei}$ is the amplitude of $\nu_{e} \rightarrow \nu_{i}$ transition inside the Sun.  Considering the case in which the coherence effect can be ignored, the survival probability becomes \cite{deHolanda:2010am,deHolanda:2003tx}
         \begin{equation}
     \label{eq.pee_gen}
     P_{ee} = \sum_{i=0}^{3} \left|U_{e i}\right|^{2}\left|A_{ei}\right|^{2}\,.
     \end{equation}
     $A_{ei}$ can be calculated using  elements of the effective mixing matrix in the center of the Sun ($A_{e i}=U_{e i}^{M_{0}}$). Assuming  adiabaticity condition is satisfied,  the survival probability can be written as
    \begin{equation}
    \label{eq.pee_ad_4v_l}
    P_{ee} = \sum_{i=0}^{3} \left|U_{e i}\right|^{2}\left|U_{ei}^{M_{0}}\right|^{2}\,
    \end{equation}
where superscript  $M_{0}$ represent the effective parameter at center of the Sun. 

The yellow curve in Fig.~\ref{Pee} represents the results of the numerical calculation for the electron survival probability $P_{ee}$ as a function of neutrino energy. One can observe that the presence of a sterile neutrino can significantly affect the electron survival probability. In the case of a sterile neutrino, there is a resonance around 1-2 MeV due to the dependence of the resonance on $\Delta m_{01}^2$, in addition to the standard resonance condition that occurs above 5 MeV. However, this new resonance is absent in the cases of NSI or MSW-LMA. 
~~~~~~~~~~~~~~~~~~~~~~~~~~~~~~~~~~~~~~~~~~~~~~~~~~~~~~~~~~~~~~~~~~~~~~~~~~

\subsubsection{Solar Metallicity}

The solar metallicity problem remains an important challenge in astrophysics, and addressing it will require future solar experiments with higher precision and accuracy.
The metallicity of the Sun, represented by $Z$, is the abundance of elements heavier than helium present in the Sun. Determining solar metallicity is crucial for understanding the evolutionary history of not only the Sun, but also other stars. However, the Solar metallicity problem arises due to the inconsistency between helioseismic observations and predictions of solar models obtained from photospheric abundances. Solar models can only fit helioseismic data if the metallicity is set to be higher than the measured values from the photosphere. 
One promising avenue for progress is the use of spectroscopic techniques to study the solar atmosphere and measure the abundance of key elements more precisely. Another approach is to study the properties of solar neutrinos, which can provide valuable information about the processes occurring in the solar interior. 
Solar metallicity problem has a direct and indirect impact on neutrino flux. For instance, in the case of N and O neutrinos, fluxes depend on the metallicity linearly. Thus, low metallicity (measured from the photosphere) can decrease the flux by up to $40\%$. Indirectly, the effect of metallicity on other types of neutrino fluxes can occur through changes in temperature and density. To determine solar metallicity, reducing uncertainties in the measurement of CNO neutrino flux is necessary \cite{Serenelli:2016nms, Vinyoles:2016djt}. 

Overall,  a precise measurement of the flux of solar neutrinos from the $pp$ chain and CNO cycle would effectively resolve the discrepancy between high-metallicity (HZ) and low-metallicity (LZ) Standard Solar Models (SSMs).
The Yemilab experiment is expected to play a crucial role in measuring the CNO neutrino fluxes with higher accuracy. In addition, Yemilab benefits from a lower energy threshold compared to Borexino, which allows for a better detection of $pp$ neutrinos. Higher precision measurements of the $^7$Be flux and CNO flux measurement can help to determine the solar metallicity accurately. Future experiments such as the Super-Kamiokande with Gadolinium  and Jiangmen Underground Neutrino Observatory (JUNO) detectors are expected to measure the low-energy solar neutrino fluxes with unprecedented precision, enabling more accurate constraints on the solar metallicity. A combined analysis of these experiments will significantly help to solve the solar metallicity problem and improve our understanding of the Sun's evolutionary history. 
 
Moreover, it is essential to continue developing and refining theoretical models of the Sun. Such models can help to elucidate the mechanisms responsible for the observed discrepancies between helioseismic data and predictions based on photospheric abundances. Ultimately, a better understanding of solar metallicity will not only shed light on the evolution of the Sun but also have important implications for our understanding of the composition and evolution of other stars in the universe

\begin{figure}[h]
\begin{center}
\includegraphics[width=1.\textwidth]{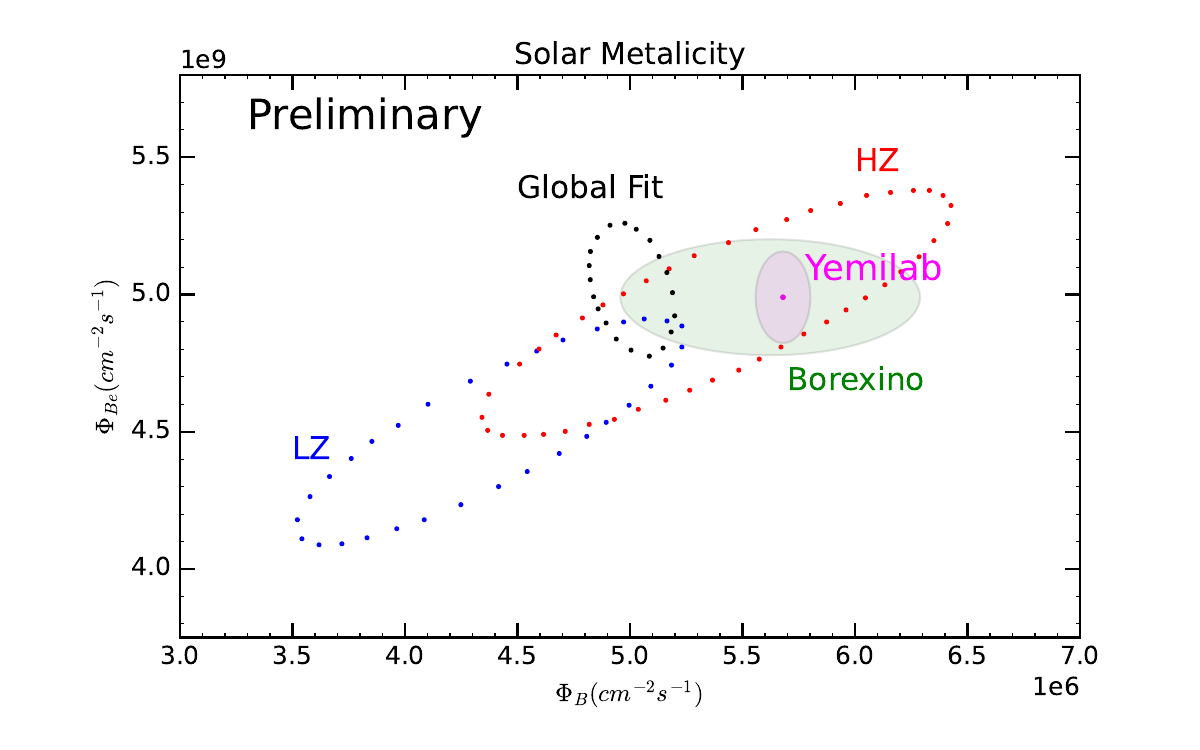}
\end{center}
\caption{Sensitivity on solar metallicity at Yemilab LSC (magenta) together with Borexino measurement (green)~\cite{BOREXINO:2018ohr} and theoretically allowed regions for high (HZ) and low (LZ) metallicity.}
\label{f:solar_YM}
\end{figure}

The results of the global fit, which includes data from Borexino and other solar experiments along with KamLAND, are shown in dotted black contour in Figure \ref{f:solar_YM}. The oscillation parameters $\theta_{12}$, $\Delta m_{12}^2$, and the $ ^7\text{Be}$ and $ ^8\text{B}$ neutrino fluxes are set to be free in the fit \cite{BOREXINO:2018ohr}. The dotted blue and dotted red regions represent the theoretical predictions of the Standard Solar Model (SSM) for low-metallicity (LZ) and high-metallicity (HZ), respectively. It can be observed that the Borexino results, when combined with other solar-neutrino experiments, weaken the previously observed hint towards HZ. The Yemilab experiment is expected to further improve the sensitivity and provide a more accurate measurement of solar metallicity.






%% file: Physics/Geo_Reactor.tex
\subsection{Geo and reactor neutrinos}\label{sec:georeactor}





Electron antineutrinos ($\bar{\nu}_e$) are produced naturally by the slow decay of $^{40}$K, $^{232}$Th, and $^{235,238}$U within the rocky layers of the Earth and artificially by the fast decay of nuclear isotopes within man-made nuclear reactors. These antineutrinos, which have typical energy from $1$ to $10$ MeV, are called geo and reactor neutrinos, respectively. Detecting the interactions of geo and reactor neutrinos within the LSC target contributes to applied antineutrino physics.  

\subsubsection{Signal estimates}
We use the online neutrino signal estimator at reactors.geoneutrinos.org for the spectra and numbers of $\bar{\nu}_e$ inverse beta decay (IBD; $\bar{\nu}_e+p \rightarrow e^+ + n$) and $\bar{\nu}_e$-e elastic scattering (ES; $\bar{\nu}_e + e^- \rightarrow \bar{\nu}_e + e^-$) interactions in the LSC. The numbers of interactions assume $1.85 \times 10^{32}$ free proton targets and $7.40 \times 10^{32}$ atomic electron targets, corresponding to a target composition equivalent to CH$_2$ ($4$ atomic electrons per free proton). Efficiencies are not included.

The estimate of the geo-neutrino signals are $60.6 \pm 13.6$ IBD and $819 \pm 174$ ES per year. It is $72$\% from the crust and $28$\% from the mantle for the IBD signal, and $74$\% from the crust and $26$\% from the mantle for the ES signal. Upon approval to construct the LSC detector, it is advisable to commission a detailed geological study of the distribution of uranium and thorium in the local crust, which dominates the geo-neutrino signal. Measuring the geo-neutrino signal at LSC would help constrain models of the bulk silicate earth, which largely define our planet's heat flow and thermal evolution.

The estimate of the reactor signal depends on the operation of nuclear power plants. There are eight cores in the closest complexes, which are at a distance of $\sim 65$ km. Table~\ref{t:Core-info} lists information on the locations and IBD signals associated to these cores. Together they provide $1488 \pm 46$ IBD per year and $690 \pm 25$ ES per year when running at a load factor (LF) of $100$\%. Using the average annual LFs for 2021, obtained from the IAEA, all other operational cores provide a signal of $458 \pm 14$ IBD per year and $123 \pm 4$ ES per year.

Several opportunities for demonstrating reactor monitoring capabilities are present for LSC. With sufficient detection efficiency and energy resolution, the imprint of the neutrino oscillations on the IBD spectrum would allow an estimate of the distance to the closest reactor complexes. This analysis is analogous to measuring the neutrino mass-squared differences. Moreover, with sufficient position resolution of the energy depositions of the positron and neutron from IBD reactions, the aggregate displacements would allow an estimate of the direction to the closest reactor complexes. Taken together, the distance and direction estimates, which are determined solely from the antineutrino measurements, would reveal the location of these reactor complexes out of the background of all other reactors in the world.

\begin{figure}[h]
\begin{center}
\includegraphics[width=1.\textwidth]{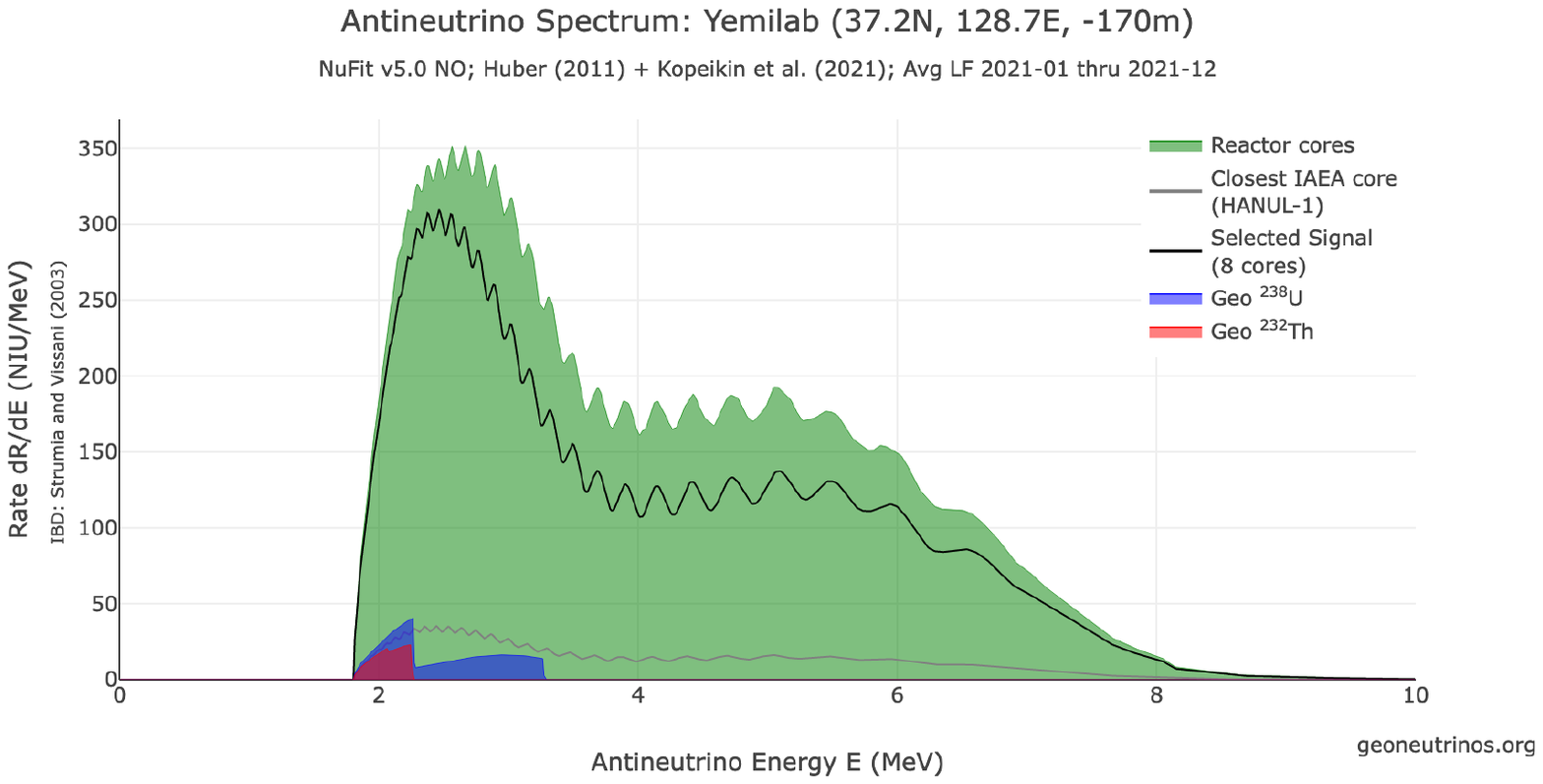}
\end{center}
\caption{Geo- and reactor antineutrino IBD spectra at the Yemilab site~\cite{Dye:2015bsw}.}
\label{f:Yemi_georeac_pIBD}
\end{figure}

\begin{figure}[h]
\begin{center}
\includegraphics[width=1.\textwidth]{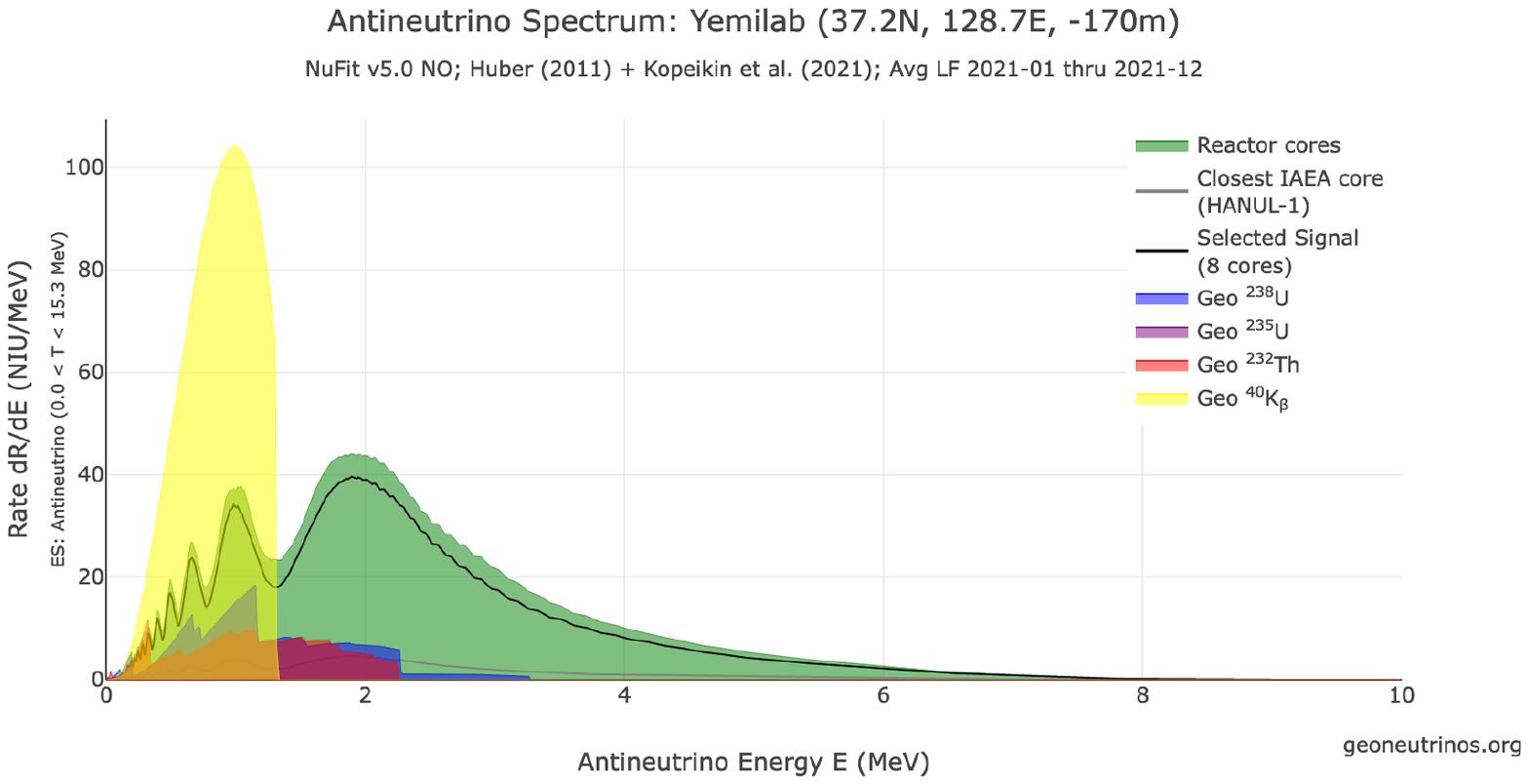}
\end{center}
\caption{Geo- and reactor antineutrino ES spectra at the Yemilab site~\cite{Dye:2015bsw}.}
\label{f:Yemi_georeac_eES}
\end{figure}

The signal rates given above assume $100$ \% detection efficiency. See Fig.~\ref{f:Yemi_georeac_pIBD} and Fig.~\ref{f:Yemi_georeac_eES} for the geo- and reactor antineutrino spectra of IBD and ES, respectively.

\begin{table}[!tbp]
\centering
\begin{tabular}{|c|c|c|c|c|c|c|}
\hline
Reactor Core & $P_\textrm{th}$ (MW) & Lat (N) & Lon (E) & $h$ (m) & $d$ (km) & $R$ (/y)\\
\hline
\hline
Hanul-1 & $2775$ & $37.09578$ & $129.37844$ & $82$ & $64.71$ & $169.2\pm5.2$ \\
Hanul-2 & $2775$ & $37.09520$ & $129.37913$ & $82$ & $64.78$ & $168.9\pm5.2$ \\
Hanul-3 & $2825$ & $37.09342$ & $129.38136$ & $82$ & $65.01$ & $171.0\pm5.3$ \\
Hanul-4 & $2825$ & $37.09225$ & $129.38274$ & $82$ & $65.15$ & $170.5\pm5.2$ \\
Hanul-5 & $2825$ & $37.09041$ & $129.38475$ & $82$ & $65.36$ & $169.7\pm5.2$ \\
Hanul-6 & $2825$ & $37.08919$ & $129.38616$ & $82$ & $65.51$ & $169.3\pm5.2$ \\
Shin-Hanul-1 & $3983$ & $37.084597$ & $129.388714$ & $82$ & $65.82$ & $238.3\pm7.2$ \\
Shin-Hanul-2 & $3983$ & $37.083576$ & $129.389782$ & $82$ & $65.93$ & $236.4\pm7.2$ \\
\hline
\hline
\end{tabular}
\captionof{table}{\label{t:Core-info} Local reactor core information from reactors.geoneutrinos.org. Rate $R$ assumes IBD reactions on $1.85 \times 10^{32}$ targets with $100$\% detection efficiency.}
\end{table}

%% file: Physics/SN.tex
\subsection{Supernova neutrinos}

The lives of massive stars with an initial mass of at least $\sim\SI{8}{M_{sol}}$ end in supernovae---giant explosions that produce neutron stars and black holes and enrich their cosmic environment with heavier chemical elements that are necessary for life as we know it.

Approximately 99\,\% of the total explosion energy of $\mathcal{O}(\SI{e53}{erg})$ are emitted in the form of neutrinos with energies of \SIrange{1}{100}{MeV}. Most neutrinos escape directly from their production region near the centre of the explosion, making them a unique messenger that transports information about the supernova explosion mechanism to Earth.

\begin{table}[htbp]
\begin{center}
\begin{tabular}{cccc}
Interaction Channel & \SI{11.2}{M_{sol}} & \SI{27.0}{M_{sol}} & \SI{40.0}{M_{sol}} \\
\hline
IBD & 366/368 & 690/671 & 625/380 \\
$\nu_e + ^{12}$C CC & 8/6 & 19/16 & 37/32 \\
$\bar{\nu}_e + ^{12}$C CC & 7/8 & 18/19 & 29/27 \\
$\nu + ^{12}$C NC & 24/24 & 54/54 & 73/73 \\
$\nu$+e scattering & 24/24 & 40/40 & 21/22 \\
\hline
\textbf{Total} & 429/430 & 821/800 & 785/534 \\
\end{tabular}
\end{center}

\caption{Number of events expected in the Target volume for the main interaction channels. We use simulations of two progenitors with \SI{11.2}{M_{sol}} and \SI{27}{M_{sol}} that form neutron stars (s11.2c and s27.0c progenitors from \cite{Mirizzi:2015eza}) as well as one black-hole forming \SI{40}{M_{sol}} progenitor \cite{OConnor2015}. We assume adiabatic MSW flavor transitions with normal (left number) or inverted (right number) mass ordering.}
\label{tab:sn-events}
\end{table}

We use SNEWPY~\cite{Baxter2021,SNEWS:2021ezc} and SNOwGLoBES~\cite{SNOwGLoBES} to estimate the number of events expected in LSC for a galactic supernova burst at a fiducial distance of \SI{10}{kpc} in its main interaction channel.
Table~\ref{tab:sn-events} shows the results for different supernova models and mass orderings.

Depending on the progenitor and neutrino mass ordering, the LSC will observe about 430--820 events. It will be most sensitive to $\bar{\nu}_e$, with inverse beta decay (IBD, $\bar{\nu}_e + p \rightarrow n + e^+$) making up about 80\,\% of observed events.
Subdominant channels, which may make up a few per cent of observed events each, include charged-current (CC) interactions of $\nu_e$ or $\bar{\nu}_e$ on $^{12}$C nuclei, neutral-current (NC) interactions of any neutrino flavor on $^{12}$C nuclei, and neutrino-electron scattering.

Additionally, the narrow time structure of a supernova burst and the resulting reduction in backgrounds may enable LSC to also observe neutrino interactions in the surrounding buffer volume.
Including this buffer volume---filled with mineral oil whose composition is assumed to be C$_n$H$_{2n}$, similar to the liquid scintillator in the target volume---increases the number of expected events by about 45\,\% for each of the investigated models, reaching a total of 620--1200 events in the combined volume.

LSC would be able to identify a supernova anywhere within the Milky Way.
For a supernova in the Large Magellanic Cloud at a distance of \SI{50}{kpc}~\cite{Pietrzynski2013}---the location of SN1987A---LSC would observe 17--33 (25--48) events in the target (combined) volume.
This ability to identify the neutrino signal from a supernova outside the Milky Way is shared by only a handful of current or planned neutrino detectors.
Accordingly, LSC could make important contributions to the next-generation Supernova Early Warning System (SNEWS 2.0)~\cite{SNEWS:2020tbu}, which aims to maximize the science yield of multi-messenger observations of supernovae.


\subsection{Pre-supernova neutrinos}
For several days before a supernova, the progenitor star emits a steadily increasing flux of neutrinos from silicon burning.
These so called “pre-supernova neutrinos” can act as an early warning for the supernova neutrino burst itself.

The energy of pre-supernova neutrinos is below the thresholds for CC and NC interactions on carbon nuclei, eliminating these subdominant channels.
In the main interaction channel, inverse beta decay (IBD), LSC is able to detect both the prompt signal from the positron and the delayed signal from neutron capture on hydrogen nuclei.
The temporal and spatial coincidence of both signals is a powerful tool to reduce backgrounds and enable the LSC to be sensitive to even a small number of pre-supernova events.
Therefore, we only consider IBD as a detection channel for pre-supernova neutrinos.

With a target mass of \SI{2.26}{kt}, the LSC would be about three times larger than KamLAND-Zen, whose fiducial volume for pre-supernova searches is about \SI{0.65}{kt}~\cite{KamLAND:2015dbn}.\footnote{The KamLAND detector has a nominal target mass of \SI{1}{kt}. However, in 2011 an inner balloon containing Xe-loaded liquid scintillator for $0\nu\beta\beta$ searches was deployed, reducing the fiducial volume.}
Meanwhile, in LSC the background rate from reactor neutrinos is expected to be about 1 per day per kt LS, as described in section~\ref{sec:georeactor}---approximately equal to the KamLAND background rate while Japanese nuclear reactors are running (referred to as the “high-reactor flux period” in~\cite{KamLAND:2015dbn}). 

As a result, LSC’s sensitivity to pre-supernova neutrinos will surpass that of KamLAND and reach approximately \SI{1}{kpc} for optimistic progenitor models.
This would make it one of the world’s most sensitive detectors in this area, alongside the near-future JUNO detector.
Precise quantitative estimates of the sensitivity are highly dependent on detector background levels and low-energy threshold. They would therefore be premature at this stage and will be presented once LSC progresses to a more advanced stage.


%% file: Physics/DSNB.tex
\subsection{Diffuse Supernova Neutrinos}

The Diffuse Supernova Neutrino Background (DSNB) is the faint background flux of neutrinos created by all core-collapse Supernovae (SNe) throughout the visible Universe \cite{Ando:2004hc,Beacom:2010kk,Vissani:2011kx,Lunardini:2012ne,Nakazato:2015rya,Horiuchi:2017qja,Priya:2017bmm,Moller:2018kpn,Riya:2020wpw,Kresse:2020nto}. Its detection promises to reveal information on both the red-shift dependent SN rate and a cosmic average of the very variable neutrino spectra expected from different types of core-collapse SNe. With an expected flux of $\leq10^2$\,cm$^{-2}$s$^{-1}$, only very large neutrino detectors on the scales of kilotons can hope to collect a handful of events over the periods of years. At the same time, excellent background discrimination capabilities are required to suppress the otherwise dominant background created by the Neutral-Current (NC) interactions of atmospheric neutrinos. While on the lower end of the required liquid-scintillator (LS) mass, a decade of measurement done in the Yemilab neutrino detector might result in the positive measurement of a signal. Especially when using water-based liquid scintillator (WbLS) \cite{Theia:2019non,Sawatzki:2020mpb}, it could provide vital information to improve and cross-link the background predictions of larger water-Cherenkov (SuperKamiokande-Gd) and organic scintillator detectors (JUNO) \cite{Super-Kamiokande:2021jaq,JUNO:2022lpc}.

\begin{table}[b!]
	\centering	
	\begin{tabular}{l|cccc}
        \hline
         & \multicolumn{4}{c}{\bf{Rate}[18.5 $\text{kt} \times \text{yr}$]}\\
         \hline
		{\bf{Signal }} & $E_{\rm vis}$ window & muon veto  & PSD  & TFC cut  \\
		\hline	
		12 MeV & 2.0  &  1.9  &  1.6  & 1.5 \\
		15 MeV & 2.6  &  2.4  &  2.1  & 2.0 \\
		18 MeV & 3.2  &  3.0  &  2.6  & 2.4 \\
		21 MeV & 3.6  &  3.4  &  3.0  & 2.8 \\
		\hline	
		{\bf{Backgrounds }} &       &       &       &  \\
		\hline	
		Fast neutron    & 1.6  & 1.5   &    0.03   &  0.03 \\
		Atm-$\nu$ CC    & 0.3  & 0.2   &   0.2   &  0.2 \\
		Atm-$\nu$ NC without ${}^{11}\mathrm{C}$
		& 32    & 30 &  0.11   & 0.11 \\
		Atm-$\nu$ NC with ${}^{11}\mathrm{C}$
		& 23 &  22 &  0.5   & 0.11 \\
		\hline
		{\bf{Total backgrounds }} & 57 & 54 &   0.7  & 0.5 \\
        \hline
	\end{tabular}%
	\caption{\label{ts2} Event rates of the DSNB signal and corresponding backgrounds in the FV with the prompt energy in [12, 30] MeV. For the DSNB signal, we assume a black hole fraction of 0.27, the SN rate at $z=0$ of $ 1.0 \times 10^{-4} \, \text{yr}^{-1}\text{Mpc}^{-3}$, and mean energies of the SN neutrino spectrum of 12, 15, 18 and 21 MeV. Adopted from Ref.~\cite{JUNO:2022lpc}.}
\end{table}%

\paragraph*{DSNB signal rates.} We estimate the expected DSNB signal and background rates based on the recent study performed for JUNO in Ref.~\cite{JUNO:2022lpc}. There are two critical parameters: Detector performance and fiducial mass. For the former, we assume similar levels of light collection, i.e.~more than 1,000\,pe/MeV as well as pulse shaping capabilities comparable to the LS foreseen for JUNO (i.e.~LAB with 2.5\,g/l PPO). For the latter, we define a fiducial volume only slightly smaller than the active volume, i.e.~14\,m in height and 7\,m in radius, corresponding to 2150\,m$^3$ or 1.85\,kt of LS. Both choices might be somewhat optimistic but are indicative of the best performance that can be expected for the Yemilab detector.

\begin{figure}[t]
\begin{center}
\includegraphics[width=1.0\textwidth]{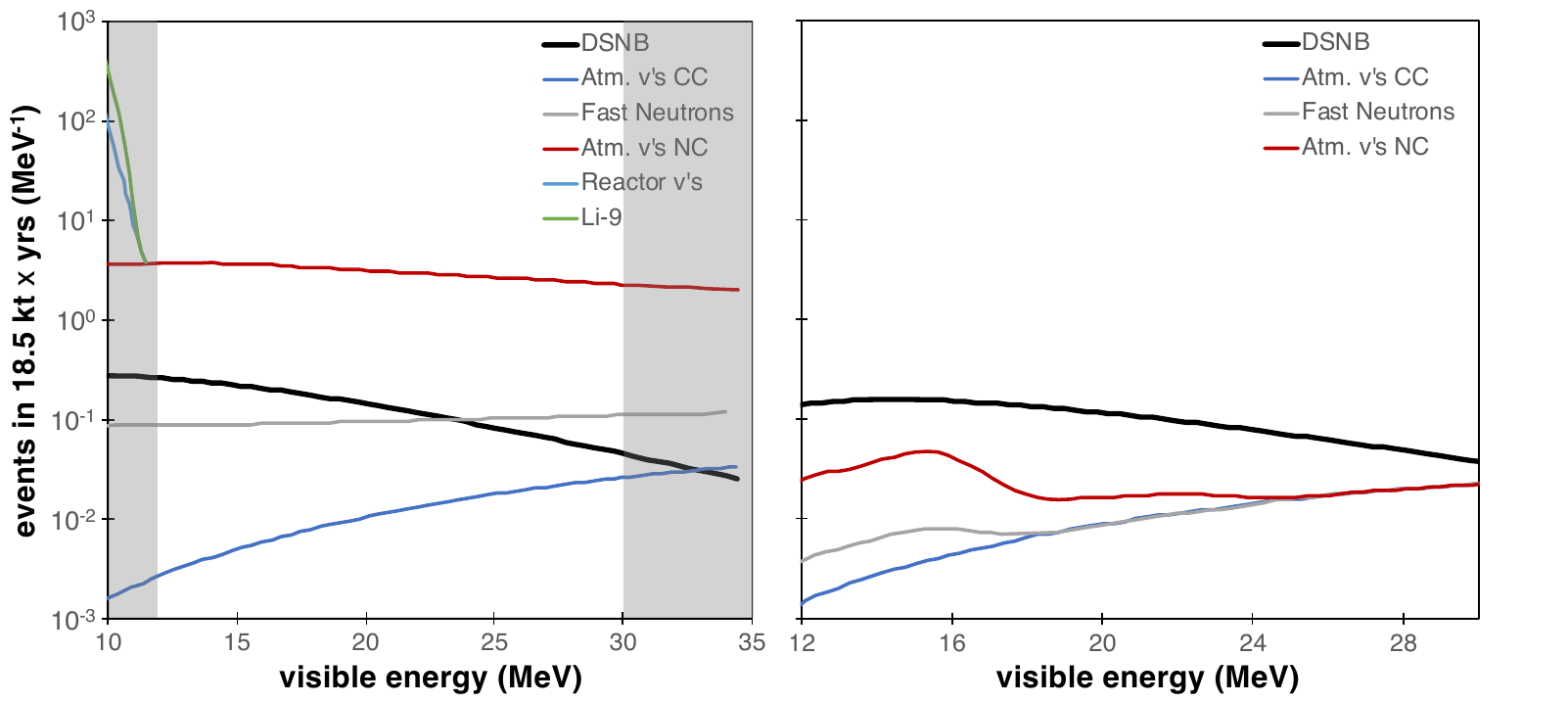}
\end{center}
\caption{DSNB signal and background spectra. The left panel shows the DSNB signal and background spectra, indicating the observation window (12-30\,MeV). The right panel shows the observation window after background reduction. Rates are scaled from Ref.~\cite{JUNO:2022lpc}.}
\label{f:dsnb}
\end{figure}

Based on these assumptions, the corresponding signal and background rates expected for the Yemilab Detector can be scaled from Ref.~\cite{JUNO:2022lpc} and are given in Table \ref{ts2}. The primary channel of signal detection is of the $\bar\nu_e$ component of the DSNB via Inverse Beta Decays (IBDs). The IBD signal rate depends on the underlying $z$-dependent Supernova rate and the average mean energy of the SN neutrino spectrum. While the first factor introduces a simple linear scaling relation, the scaling of rates with mean energy is nearly quadratic, as exemplified by the range of values given in the upper half of Table \ref{ts2}. 
As depicted in Figure \ref{f:dsnb}, the DSNB observation is limited to an energy window from about 12 to 30\,MeV. The lower end is defined by the ample background rate from the IBD of reactor antineutrinos, the upper end by the Charged-Current (CC) interactions of low-energy atmospheric $\bar\nu_e$'s. The corresponding DSNB rate estimate is 2.0-3.6 events for 10 years of exposure but might vary within about a factor 2. Note that this estimate already includes a fraction of SNe (0.27) that end up as Black Holes, featuring both higher neutrino fluxes and mean energies \cite{JUNO:2022lpc}. 

\paragraph*{Background rates and background suppression.} Beyond the irreducible antineutrino backgrounds, detection within the observation window has to overcome several other background species that mimic the time correlation and energy levels of true IBD events. The two crucial components are fast neutrons induced by cosmic muons crossing the surrounding rock volume, and Neutral-Current (NC) interactions of atmospheric neutrinos on the carbon nuclei of the LS. While the corresponding atmospheric neutrinos feature a much higher energy, the scintillation of nuclear decay fragments from $^{12}$C often falls into the DSNB energy window, creating a background 1-2 orders of magnitudes larger than the DSNB signal. This background was first observed by the KamLAND experiment \cite{KamLAND:2011bnd}. As shown in a relatively recent Borexino analysis (albeit at low statistics), these events are quite effectively removed by cuts on the event pulse shape \cite{Borexino:2019wln}. This has been studied in detail for JUNO, suggesting that a highly efficient background suppression is possible for a moderate loss in signal efficiency ($\sim$20\%) \cite{JUNO:2022lpc}.

The corresponding situation is exemplified in both Table \ref{ts2} and Figure \ref{f:dsnb}. After application of several background reduction cuts $-$ i.e.~a veto of fast cosmogenic decays, pulse shape discrimination (PSD) and the veto of atmospheric NC events that can be based on the $\beta^+$-decay of the long-lived nuclear fragment $^{11}$C $-$ the corresponding DSNB signal efficiency is still $\sim$75\%, while the background residual inside the observation window is reduced to $\sim$1\%. Correspondingly, we obtain about 2 DSNB signal events over 10 years over $\sim$0.5 residual background events.

\paragraph*{Discussion on Sensitivity.} Based on the expected signal and background levels, the statistical median sensitivity for a 10 year measurement in an LS detector at Yemilab is expected to be on the level of 97.5\% C.L. This is short of a $3\sigma$ observation but could be improved if the DSNB signal rate is higher than the expectation or the background levels are even lower. For both, using a WbLS instead of an organic LS target might be highly interesting. In addition to a larger target mass (2.15\,kt) because of the greater density, the possibility to utilize additional discrimination potential given by the (missing) Cherenkov light output of nuclear fragments compared to IBD positrons might lead to a substantial further reduction of background levels and such an increase in sensitivity \cite{Sawatzki:2020mpb}.

However, the main benefit of a measurement at Yemilab with a WbLS target would be to link the background predictions for atmospheric $\nu$ NC interactions for water and organic scintillator targets, i.e.~the current Super-Kamiokande with gadolinium phase and the upcoming measurement in JUNO. For both, a reduction of systematic uncertainties on the NC background might prove to be crucial to fully tap their potential for DSNB discovery. A scintillating water target (i.e.~with interactions on oxygen instead of carbon and providing Cherenkov in addition to scintillation light) would provide highly complementary information, arguably leading to a substantial reduction in systematic uncertainties.

%% file: Physics/IsoDAR.tex

\subsection{IsoDAR physics}
\label{IsoPhysText}

IsoDAR@Yemilab will be the first short-baseline pairing of an accelerator-based neutrino source with a multi-kiloton liquid scintillator detector.   The novel new design opens the opportunity for powerful and first-of-their-kind searches for signatures of beyond Standard Model (BSM) physics.   The physics program includes studies of exotic neutrino oscillations and decays; non-standard interactions; new scalar, pseudoscalar, and vector particles that decay to photons, electrons, or neutrinos; and mirror neutrons.
Because this is an entirely new approach, IsoDAR@Yemilab is rich with discovery potential.

The novel IsoDAR neutrino source is driven by a 60~MeV/amu cyclotron producing 5~mA of H$_2^+$. 
After beam extraction, the ions are stripped of their electron to form a proton beam~\cite{Winklehner:2021qzp} which is then transported to a target of $^9$Be, producing neutrons.  The neutrons enter 
a surrounding  isotopically-pure $^7$Li sleeve, where neutron
capture results in $^8$Li \cite{Bungau:2018spu}.   This isotope $\beta$-decays with a half-life of 839~milliseconds to create a high-intensity decay-at-rest (DAR) $\bar \nu_e$ flux, with peak at $\sim 6$~MeV, as seen in Fig.~\ref{IsoDARfluxes} (left).   Monoenergetic photons are also produced in the source, as seen in Fig.~\ref{IsoDARfluxes} (right).  These particle fluxes form the foundation of the IsoDAR@Yemilab physics program.

This compact antineutrino source is designed to be installed underground~\cite{Alonso:2022mup} in an already-excavated cavern as shown in Fig.~\ref{IsoDARYemilayout}.
The source will be aligned with the vertical center of the LSC, at 17~m from the detector mid-point.  Due to this proximity, during a live-time (total run time) of 4~(5)~years , the cyclotron will deliver $1.97\cdot 10^{24}$ protons on target, producing  $1.15\cdot 10^{23}$ $\bar \nu_e$.   This will yield  $1.67\cdot 10^6$ reconstructed $\bar{\nu}_e$ inverse beta decay (IBD; $\bar{\nu}_e+p \rightarrow e^+ + n$) events assuming~92\% efficiency, and 6980 $\bar{\nu}_e$-e elastic scattering (ES; $\bar{\nu}_e + e^- \rightarrow \bar{\nu}_e + e^-$) events assuming 32\% efficiency above 3~MeV visible energy.  These antineutrino events can be used for a unique program of searches for deviations from the well-predicted $\bar \nu_e$ flux and cross sections.  The LSC can also potentially observe exotic particles produced in the IsoDAR source that decay, via the signature of an exponential change in rate as a function of distance from the target.    The production region of the source has $\sim 40$~cm 1$\sigma$ radial extent, dominating the uncertainty in the reconstruction of distance traveled, $L$, but is still sufficiently compact for the physics goals when compared to the total $L$ of 9.5 to 26.6~m in consideration of the source extent and detector size.

\begin{figure}[t]       
\begin{center}
\includegraphics[width=6.3in]{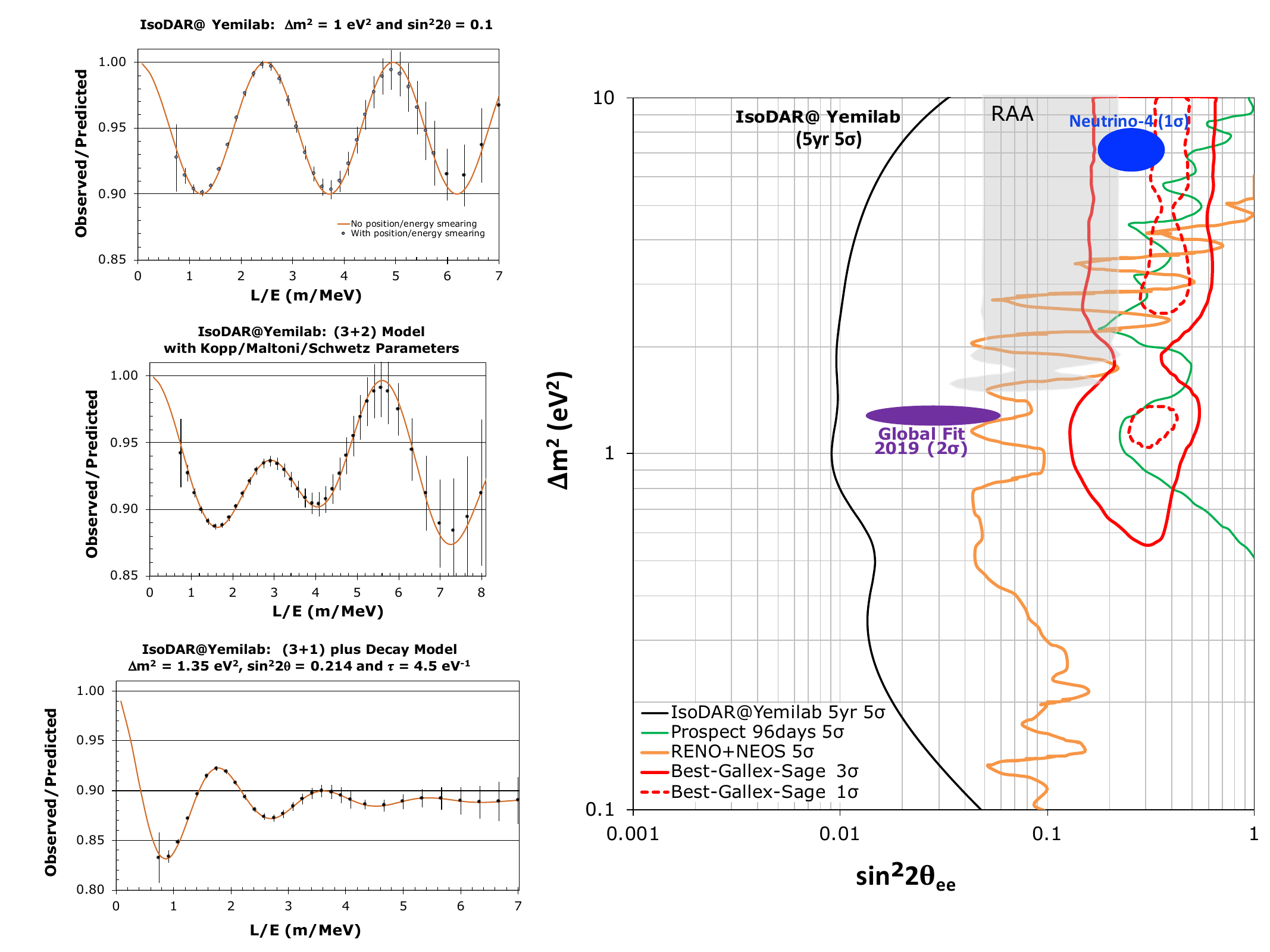}
\end{center}  
\vspace{-.5cm}
\caption{\label{IsoDARdisappearance} Left:  Examples of three representative oscillation wave scenarios involving $\bar \nu_e$ disappearance that be reconstructed. IsoDAR@Yemilab is unique in the wide $L/E$ range that can be covered.   Right: The state of the electron-flavor anomalies in February 2022, with the level of limit or allowed region indicated in the legend, along with the 5$\sigma$ sensitivity of IsoDAR@Yemilab for 4~(5)~years of live- (run-) time. These plots are from Ref.~\cite{Alonso:2021kyu}}. 
\end{figure}

IsoDAR can resolve the question of whether the short-baseline anomalies are due to oscillations involving a new, light sterile neutrino~\cite{Diaz:2019fwt}, and although the source has proven to have much wider applications, this still remains the flagship measurement.    This is a search for $\bar \nu_e$ disappearance using the IBD sample.  The very-high-statistics, well-understood, single-isotope flux, excellent resolution of the detector in antineutrino energy $E$, and the large range of $L$ that can be covered, will allow reconstruction of antineutrino oscillation waves with precision beyond any existing or planned experiment. Fig.~\ref{IsoDARdisappearance}, left, provides three examples for popular models explaining the short baseline anomalies:  3+1 oscillations (top),  3+2 oscillations (middle), and 3+1+decay (bottom).   For more examples of oscillation waves allowed from fits to today's experiments, see Ref.~\cite{Alonso:2021kyu}.   These examples demonstrate the unique clarity that IsoDAR@Yemilab will bring to understanding the possible oscillation waves.

 IsoDAR@Yemilab will allow us to identify the underlying neutrino phenomenology in an agnostic manner, as opposed to our present method of postulating a model and then comparing results within this prediction.  The problem with our present method of guess-and-compare is illustrated in Fig.~\ref{IsoDARdisappearance}, right.  This plot shows electron-flavor disappearance limits and anomalies within a 3+1 interpretation; it is readily apparent that the results do not agree. Both muon-flavor- and appearance-based experimental results show similar internal incongruities within 3+1, and the global fits show substantial tension between measurements~\cite{Diaz:2019fwt}.   With these results, one cannot conclude if some or all experiments suffer from unknown backgrounds or systematic effects; if the 3+1 physics, based on plane-waves, is over simplified~\cite{Arguelles:2022bvt}; or if some other underlying model is the source of the anomalies.   The confusion that arises from using a poor model for comparison is hampering our ability to take the next steps in finding new physics.
 
 The IsoDAR@Yemilab 5$\sigma$ 3+1 sensitivity, also shown on this plot, has excellent coverage across all anomalous signals.    The sensitivity is statistics limited--the systematic uncertainty for an energy-dependent (a.k.a. ``shape'') analysis is negligible due to the well understood $^8$Li flux and IBD cross section and very low IBD backgrounds \cite{Alonso:2021kyu}. In general, if one or more of the short-baseline anomalies is due to new physics, IsoDAR@Yemilab will make a discovery. 

IsoDAR@Yemilab also has high discovery potential using the ES sample, which will be about a factor of 7~times larger than any neutrino or antineutrino elastic scattering sample so far collected.   Antineutrino-electron scattering is a purely leptonic interaction, where the Standard Model (SM) parameters are very well predicted.  Thus, this is an ideal process to search for new particles appearing in loops that adjust the interaction rate, leading to ``non-standard interactions'' (NSIs).

\begin{figure}[t]       
\begin{center}
\includegraphics[width=5.in]{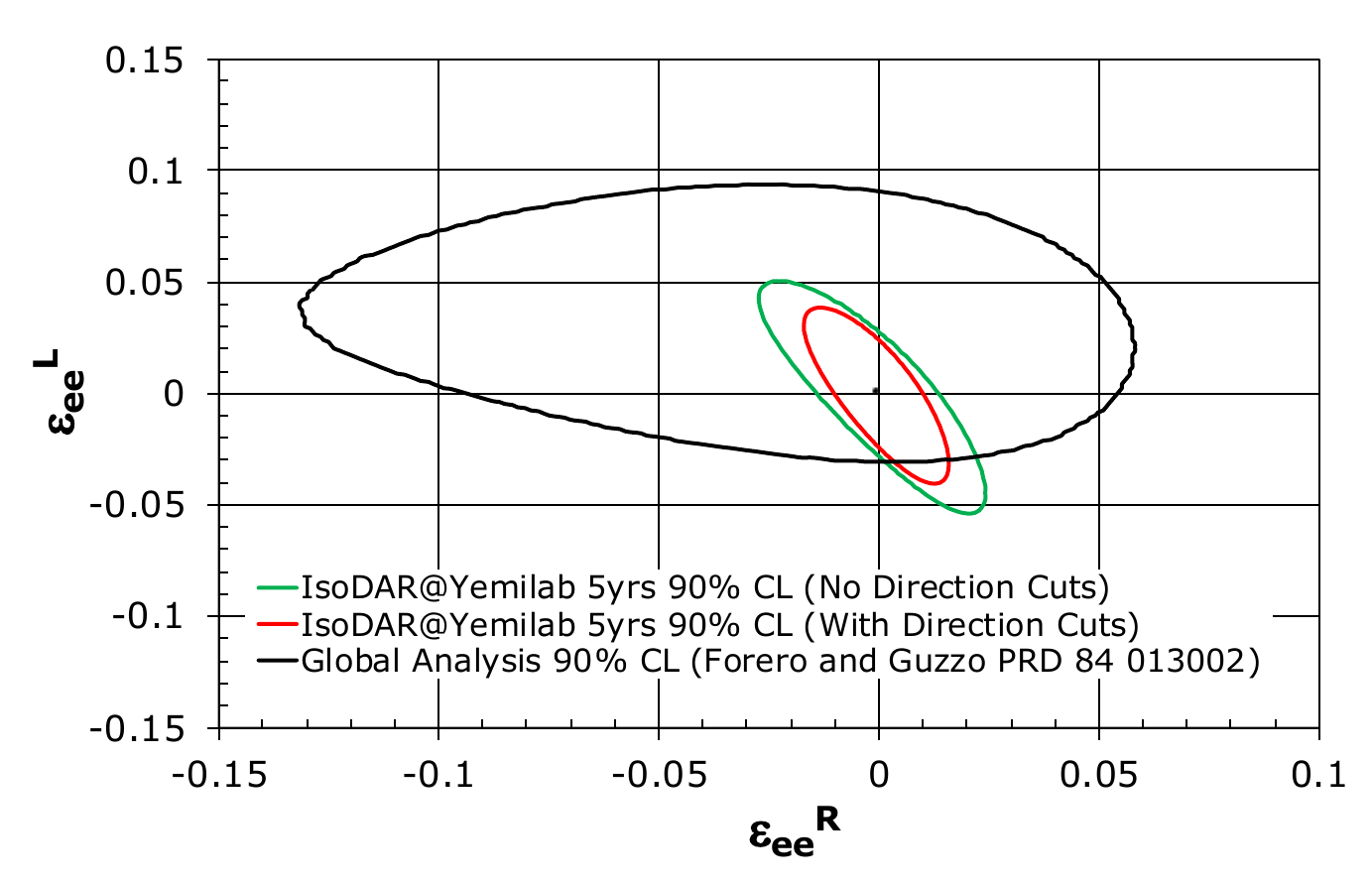}
\end{center}  
\vspace{-.5cm}
\caption{\label{IsoDARee}  
Sensitivity to adjustments in the left and right handed couplings due to NSIs, compared to existing world limits.   See text for further discussion. The SM prediction corresponds to (0,0).
Figure, from Ref.~\cite{Alonso:2021kyu}, is for 4 (5) year live- (total-) time for IsoDAR@Yemilab.}
\end{figure}

In the case of NSIs affecting electron-flavor ES (``$ee$''), new physics can adjust the SM couplings such that $g \rightarrow (1+\varepsilon_{ee}) g$, and the adjustment may be different for the left- and right-handed case.  Thus, NSI limits are presented in the left- and right-handed coupling plane, with (0,0) as the SM prediction, as seen in Fig.~\ref{IsoDARee}.  The IsoDAR expectation~\cite{Alonso:2021kyu} is shown in green for $>3$ MeV electrons.   
As can be seen, this NSI search will be a major improvement over the world limit (shown in black).  The difference in angle of the major axis of the allowed region is because the existing limit is dominated by neutrino-electron scatters while the IsoDAR sensitivity will come entirely from antineutrino-electron scattering.

Notably, the design of the LSC may allow for directional signal positron/electron reconstruction using the prompt hits observed in the PMTs.  This additional information can improve the NSI search, as seen in Fig.~\ref{IsoDARee}, red.   The directional analysis takes advantage of the fact that $\bar \nu_e$-electron scattering produces an electron that is highly forward going, hence lying along the line of the antineutrino trajectory from the IsoDAR source.   On the other hand, most backgrounds to the ES events are directed isotropically, and will not reconstruct to point back to the source.  See Ref.~\cite{Alonso:2021kyu} for further details on how Cherenkov light can improve this analysis.

NSIs are currently a focus of attention in the neutrino community.    The observation of coherent neutrino scattering \cite{COHERENT:2020iec} has opened up a channel for neutrino-quark NSI searches that has led to many experiments world-wide.   The IsoDAR antineutrino-electron search, which is unique to Yemilab, complements the coherent scattering NSI search program.

\begin{figure}[t]       
\begin{center}
\includegraphics[width=6.3in]{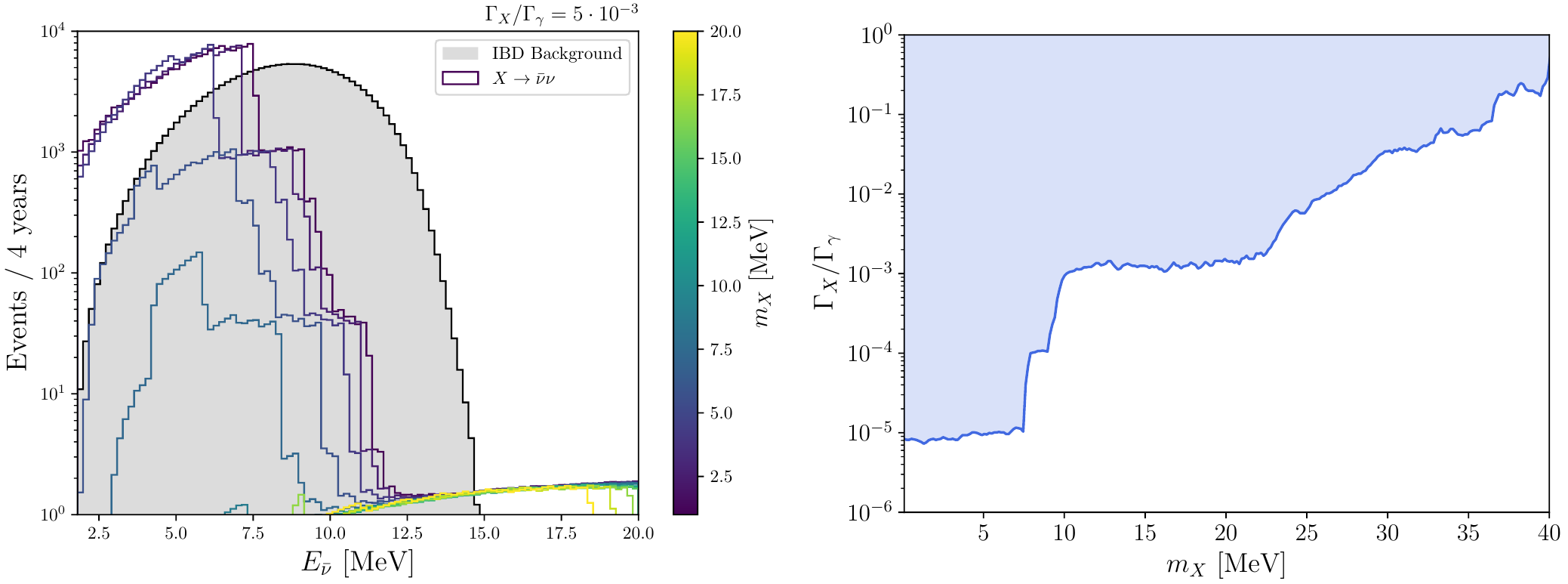}
\end{center}  
\vspace{-.5cm}
\caption{\label{IsoDARbump}  
Left:  Example IBD event peaks from $X \rightarrow \nu \bar \nu$, where the colors correspond to masses indicated by the side-bar.   The rate assumes 0.5\% transmutation of the photon to the $X$.  The IBD spectrum from $^8$Li decays is shown in gray.   Right: 90\% CL limit for the fraction of photon conversions to $X$ from observing the peak above the IBD background.  The blue region is excluded.}
\end{figure}

Along with new physics that affects neutrino properties, IsoDAR@Yemilab can also search for new particles that couple to photons produced in the target and sleeve, that subsequently decay.   IsoDAR@Yemilab will be the first experiment that can test for decays that produce $\nu_e \bar \nu_e$, through observation of an excess of IBD interactions. Also,  decays to photons or electrons produce an electromagnetic signature that can appear as an excess in the ES sample.   Because the IsoDAR photons have strong monoenergetic peaks, as seen in Fig.~\ref{IsoDARfluxes}, right, the excesses form bumps that can allow extraction of the mass of the particle~\cite{Alonso:2021kyu}.

As an example, consider the case of a new particle, $X$, produced by mixing with photons, that promptly decays to $\nu_e \bar \nu_e$.   Fig.~\ref{IsoDARbump}, left, shows the expected spectral shape of IBD events from this source, summed over all kinematically-accessible photons, assuming 0.5\% mixing.  The predicted ``bumps'' are shown for several example $X$ masses (with the value indicated by the color scale on the right).  The bump would appear as an excess on top of the IBD spectrum from the $^8$Li flux, which is shown in gray.    The IsoDAR sensitivity to photon-to-$X$ mixing, as a function of the mass, is shown in  Fig.~\ref{IsoDARbump}, right.   No other experimental exclusions are presented because this search will be first-of-its-kind.

A recent paper explores a search for high-mass axion-like-particles (ALPs) produced through photon mixing~\cite{waitesAxionlikeParticleProduction2023}.  
Relatively long-lived ALPs may decay or interact in the LSC detector, producing an additional contribution in the ES sample.   This type of search is similar to other beam dump studies, however, it leverages the monoenergetic photon peaks unique to IsoDAR.   Preliminary results indicate that IsoDAR@Yemilab will be able to close the gap commonly called ``the cosmological triangle''~\cite{Brdar:2020dpr}, which also includes open space for the QCD axion.

While these examples have focused on mixing with photons, hidden sector models also allow mixing between the SM neutron and a new baryon, $n^\prime$.  IsoDAR@Yemilab has sensitivity for a search for $n \rightarrow n^\prime \rightarrow n$ that significantly exceeds the sensitivity of existing beam dump, reactor experiment, and ultra-cold neutron experiments, as described in Ref.~\cite{Hostert:2022ntu}.   

In conclusion, the combination of the IsoDAR source and LSC detector brings a powerful program of BSM physics searches to Yemilab.    Because IsoDAR@Yemilab is so unique, many of these searches will be first-of-a-kind.

%% file: Physics/Source.tex
\subsection{Sterile neutrino search with radioactive sources}
Neutrino generators in the form of powerful radioactive sources coupled to large neutrino detectors have been already used in several experiments and carefully studied. Both as electron neutrinos sources (e.g. $^{51}$Cr or $^{37}Ar$) or anti-neutrino sources (e.g. $^{144}$Ce), they may provide excellent sensitivity to short distance neutrino oscillations and to the search of new physics, particularly by careful measurement of the Weinberg anlge at very low energy or by searching for a non-zero neutrino magnetic moment, which is not expected to be measurable in the case of Majorana neutrinos.

Radioactive sources produce pure neutrino or anti-neutrino beams, with no contamination from other flavors. Besides, electron capture neutrino generators yield monochromatic neutrinos, a feature that strongly enhance the sensitivity to oscillation effects or to any other physics mechanisms that affect the electron neutrino survival probability as a function of the distance between the source and the detection.

A large liquid scintillator detector with low radioactive background and good position reconstruction resolution can be an excellent way to exploit both the disappearance of flux because of oscillations and the appearance of oscillation waves as a function of the distance, as demonstrated in \cite{Borexino:2013xxa}. Besides, absolute calibration of the neutrino or anti-neutrino flux is possible to better than 1\% level, provided that a very careful calorimetric measurement is performed and that a precise correspondence between heat and neutrino flux is established.  
While the first condition was proved to be possible in \cite{Altenmuller:2018edv}, the second condition requires a lot of care and it is not the same for neutrino and anti-neutrino sources.

For neutrino sources, obtained via electron capture decays of suitable isotopes, the main problem comes from possible radioactive contaminants. The decay nuclei is normally obtained through irradiation in a suitable nuclear reactor of a stable isotope, which become the interesting isotope by neutron capture. This is the case both of $^{51}$Cr or $^{37}Ar$. However, any realistic sample will not be totally pure, and some other contaminants will be unavoidably be activated. Very low contamination or, alternatively, well known contamination, are needed in order to translate the measured heat into a precise neutrino flux.

In case of anti-neutrino fluxes, obtained from nuclei that decay via beta transitions, the problem is even more difficult; the high energy part of the neutrino spectrum - that above the typical Reines and Cowan reaction threshold of 1.8 MeV - corresponds to the low electron energy spectrum, the most difficult to measure precisely. While spectral knowledge at the level of several \% can be relatively easily obtained both experimentally or through theory, high precision knowledge is hard and requires a specific program in order to achieve better than 1\% knowledge. A high precision experiment requires such a program, which includes the construction of a careful spectrometric or bolometric electron calorimeter coupled to a sourced designed to minimise spurious effects at low electron energy. Besides this non-negligible difficulty, the contamination problems mentioned for the neutrino source should be faced also in case of the anti-neutrinos, for the very same reasons. 

\begin{figure}[h]
\begin{center}
\includegraphics[width=0.48\textwidth]{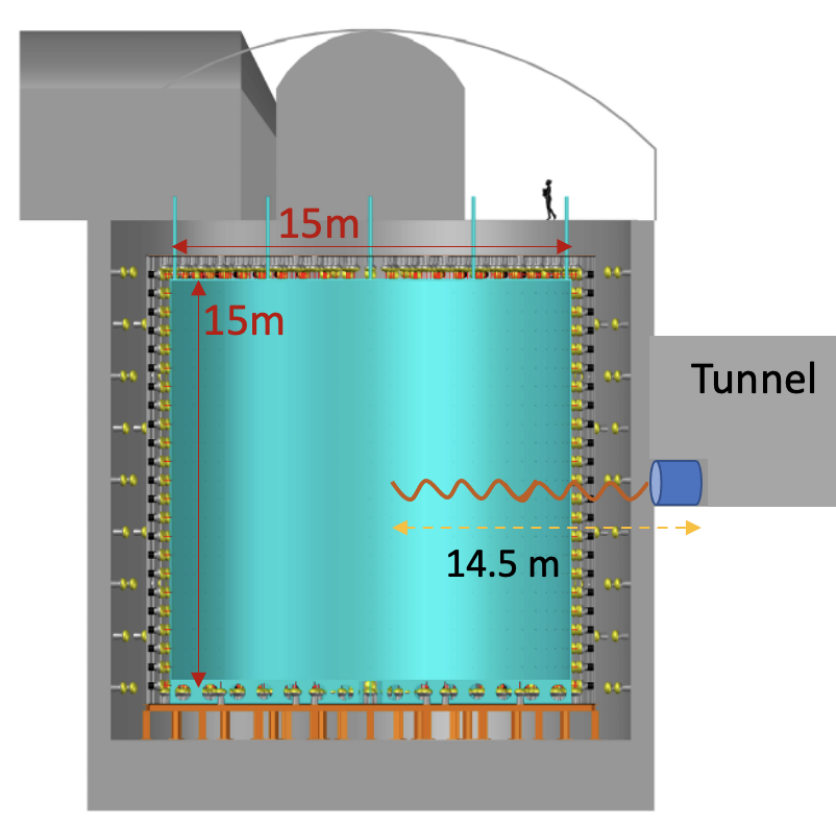}
\includegraphics[width=.48\textwidth]{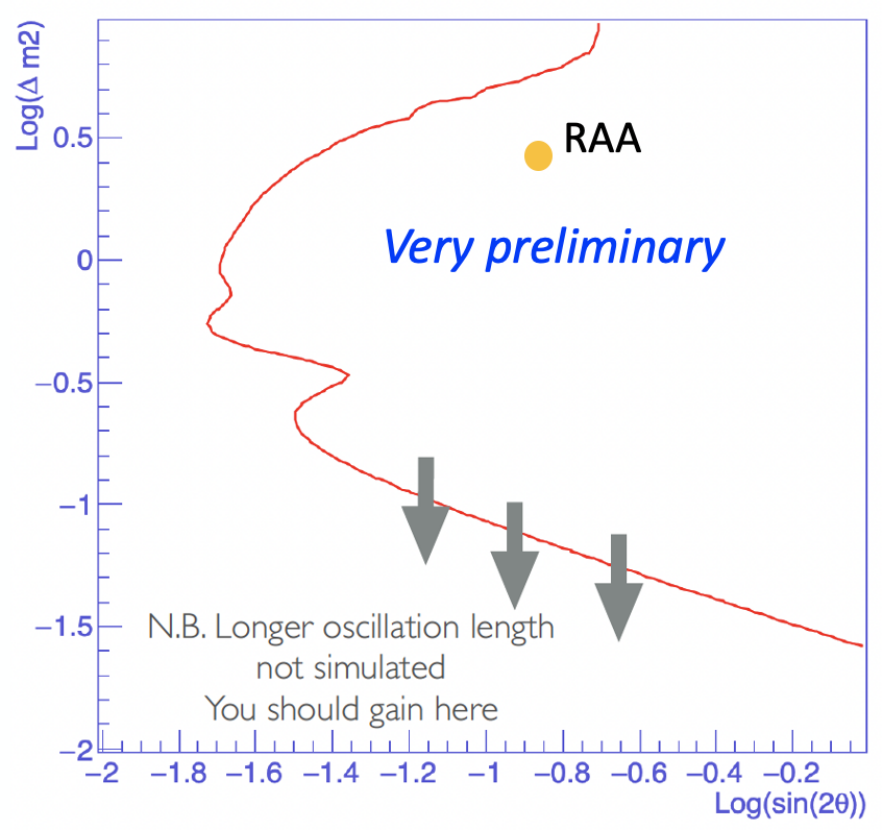}
\end{center}
\caption{Layout of radio-active source and the LSC detector (left) and sensitivity on sterile neutrinos using radio-active Ce-144 source (right).}
\label{f:source_yemi}
\end{figure}

Assuming that the source can be calibrated to better than 1\%, an experiment at Yemilab with a source located like in Fig. \ref{f:source_yemi}-left may yield a sensitivity better than the one shown in Fig. \ref{f:source_yemi}-right.

\hspace{0.5cm}

%% file: Physics/DP.tex

\subsection{Electron dump physics}

Many extensions of the Standard Model predict the existence of feebly or very weakly interacting particles such as dark photon and axion-like particles, especially in the context of portal scenarios in which these particles mediate the interactions between the Standard Model particles and dark-sector particles including dark matter (see, e.g., Ref.~\cite{Batell:2022xau}). 
The expected production rates of these particles are small, and therefore, fixed target experiments or beam-dump-type experiments using a highly intensified beam are well motivated in the search for them and to study related phenomenology. 
The Yemilab linac facility described in Section~\ref{linac} features a high-intensity electron beam, allowing for various physics opportunities regarding the search for feebly interacting particles.

\subsubsection{Light dark photon search}\label{sec:darkphoton}

Dark photons are hypothetical particles that have the potential to explain a number of experimental anomalies such as muon $g-2$~\cite{Muong-2:2021ojo} and ATMOKI anomaly~\cite{Krasznahorkay:2015iga, Krasznahorkay:2021joi, Krasznahorkay:2022pxs}. 
Since the pioneering attempt of dark photon search at SLAC in the late 1980's, a variety of experiments using beam dumps, fixed targets, or colliders have been looking for dark photons over time~\cite{Fabbrichesi:2020wbt}. 
For the last decade, even some dedicated experiments have been proposed and conducted.  
So far there has been no experiment searching for dark photon in the deep underground that provides an environment to reduce cosmic muon-related background as suggested in Ref.~\cite{izaguirre2015mev}. 
The large neutrino detector in Yemilab would be the first deep underground detector for the dark photon search using an electron beam from a linac.

\begin{figure*}[h]
\begin{center}
\includegraphics[width=0.85\textwidth]{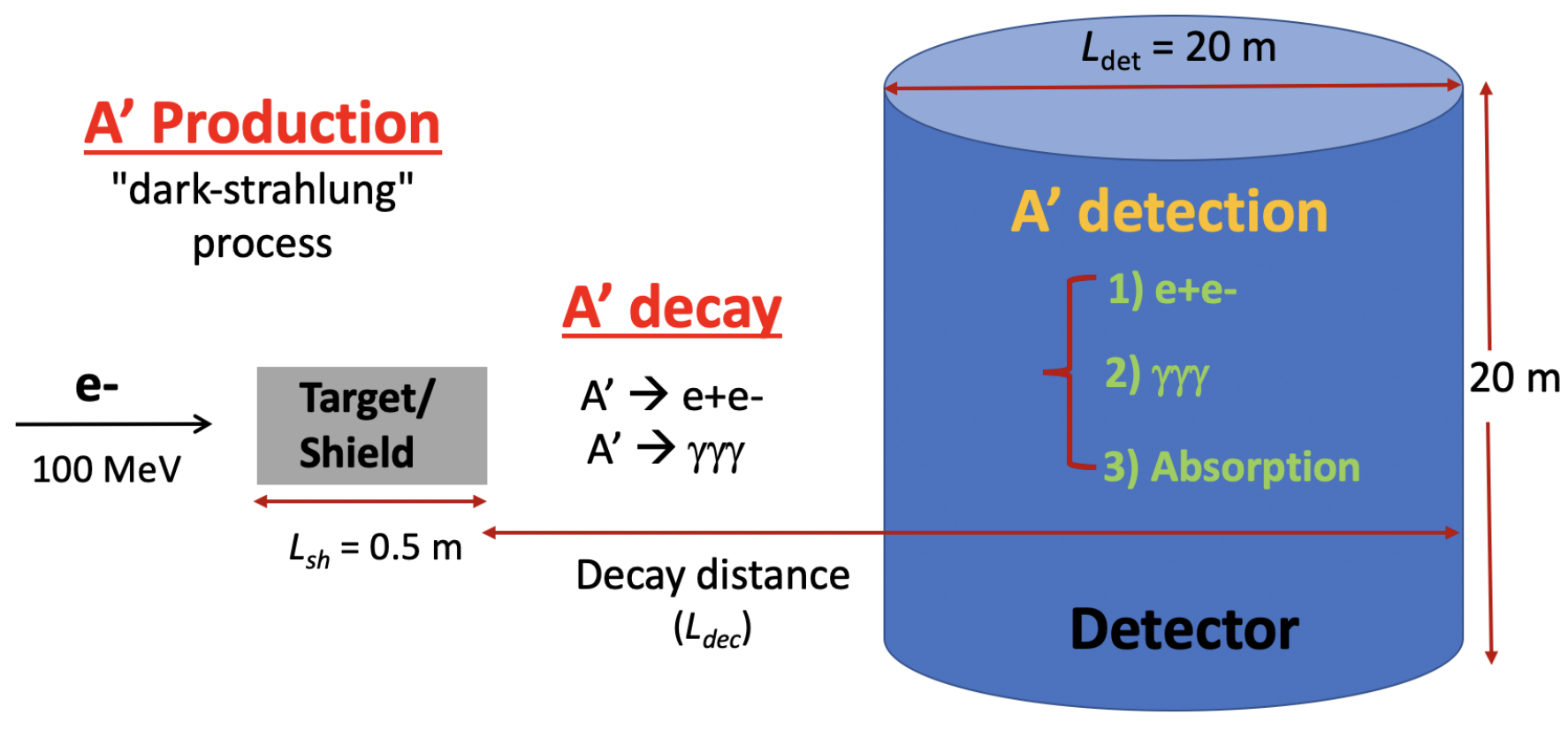}
\end{center}
\caption{
A schematic diagram showing a possible experimental configuration for a dark photon search at Yemilab. Taken from Ref.~\cite{seo2021dark}. 
The $L_{sh}$ ($L_{det}$) represents horizontal length of target \& shield (detector) in the direction of $e^-$ beam.
Decay distance, $L_{dec}$, is set as 20~m in our study but it is flexible to change.
}
\label{f:setup}
\end{figure*}

The dark photon of interest (denoted by $A'$) is mixed with the ordinary photon, which is parameterized by kinetic mixing parameter $\epsilon$~\cite{Holdom:1985ag, Huh:2007zw, Chun:2010ve}. 
The interaction Lagrangian $\mathcal{L}_{\rm int}$ contains the following operator,
\begin{equation}
    \mathcal{L}_{\rm int} \supset -\epsilon e Q_f \Bar{f}\gamma^\mu f A'_\mu\,,
\end{equation}
where $f$ and $Q_f$ denote the charged fermion and its electromagnetic charge in the SM, resepcticely. 
Given this interaction Lagrangian, the dark photon search at Yemilab consists of two steps: production of dark photons through $A'$ bremsstrahlung process, i.e., $e^- + Z \to e^- + Z + A'$, by an electron beam striking a thick tungsten target and subsequent detection in a 3 kiloton-scale neutrino detector through their visible decays or “absorption”. 
For sub-MeV dark photons, photon-to-dark-photon oscillation ($\gamma \rightarrow A'$) production/detection are effective and included in our analysis. 
Figure~\ref{f:setup} shows an illustration of an experimental setup for dark photon search at Yemilab. 

%

Our study~\cite{seo2021dark} shows that a combination of a 3 kiloton-scale neutrino detector and an electron beam at Yemilab could constrain dark photon kinetic mixing parameters with the world’s best direct search sensitivity for sub-MeV-to-MeV range
dark photons produced via $A'$ bremsstrahlung or oscillations. 
Dark photon signals are searched by counting the excess of events in beam-ON data subtracted by beam-OFF data. 
The expected number of dark photon signal events are obtained from Eq. (3.7) in Ref.~\cite{seo2021dark}, where the energy threshold of the detector (E$_{cut}$) is set at 5~MeV in order to safely remove all radiogenic backgrounds. 
The total number of background events is estimated as $\sim$1000/year which are mostly from a combination of $^8$B solar neutrino, cosmic muon, and neutron backgrounds~\cite{seo2021dark}. 

\begin{figure*}[h]
\centering
\includegraphics[width=1.0\textwidth]{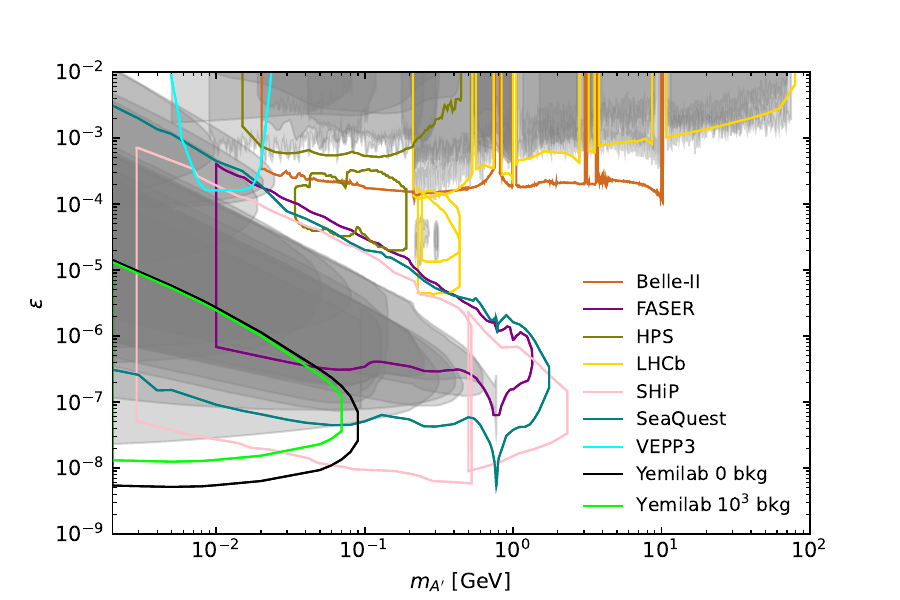}
\captionof{figure}{\label{f:dp_comp}
Comparison of existing limits (shaded gray) and some future projections (colored lines) for DP ($m_{A^{\prime}} > 2m_e$) searches, drawn with darkcast framework \cite{Ilten:2018crw} at \href{https://gitlab.com/philten/darkcast}{https://gitlab.com/philten/darkcast}.
Yemilab 95\% C.L. sensitivities with zero ($10^3$) background events are shown in a black (green) line.
Other future projections are drawn for Belle-II \cite{Kou:2018nap} (chocolate), FASER \cite{Ariga:2018uku} (purple), HPS \cite{Baltzell:2016eee} (olive),
LHCb \cite{Ilten:2016tkc} (gold), SHiP \cite{Alekhin:2015byh} (pink), SeaQuest \cite{Gardner:2015wea} (teal), and VEPP3 \cite{Wojtsekhowski:2012zq} (cyan\
). 
}
\end{figure*}

\begin{figure*}[h]
\centering 
\includegraphics[width=.9\textwidth]{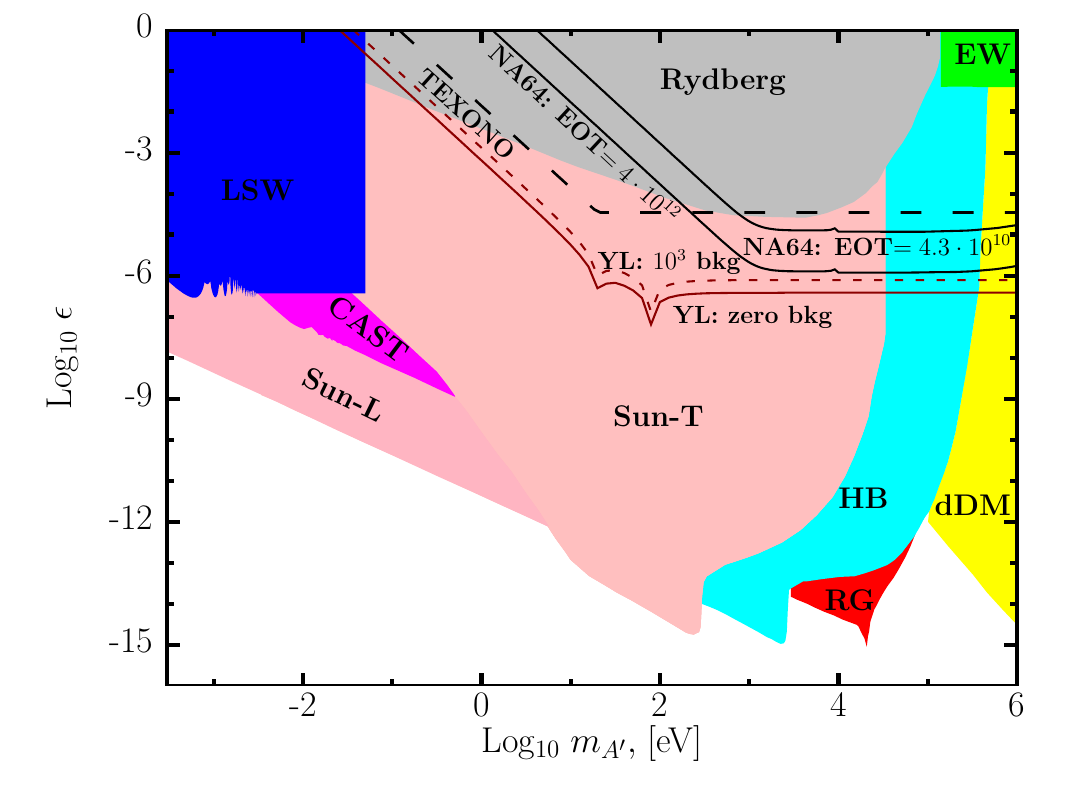}
\captionof{figure}{\label{f:dp_sens_osc}
  The dark photon sensitivity from the $\gamma \leftrightarrow A^{\prime}$ oscillation for $m_{A^{\prime}} < 2m_e$ at the Yemilab (YL) neutrino detector for one year of data taking with a 100~MeV-100~kW $e^-$ beam (zero background: solid red line, $10^3$ background: dotted red line) on a tungsten target (50~cm), 
  compared to those of the recent direct search experiments, TEXONO (dashed black line)~\cite{Danilov:2018bks} and NA64 (two solid black lines)~\cite{Demidov:2018odn}.
  Details on the limits from the helioscopic/astrophysical observations and the other experiments are found in~\cite{Redondo:2013lna,Hewett:2012ns,An:2014twa}. Taken from Ref.~\cite{seo2021dark}. 
}
\end{figure*}

Based on the expected number of signal and background events, sensitivities for dark photon search at Yemilab are obtained, and they are shown in Fig.~\ref{f:dp_comp} and Fig.~\ref{f:dp_sens_osc} together with those of other existing and future experiments. 
Sensitivities with zero background is also shown for comparisons to $\sim$1000/year background events. 
As shown in Fig.~\ref{f:dp_comp} where dark photon mass greater than $2m_e$, we estimate the best sensitivity of $\epsilon^2 >O(10^{-17})(O(10^{-16}))$ at 95\% C.L. is achieved for dark photon masses between $2m_e$ and $\sim60 (\sim30)$~MeV assuming $\sim1000$ (zero) background events.
As shown in Fig.~\ref{f:dp_sens_osc} where the dark photon mass is less than $2m_e$, the best direct search sensitivity of $\epsilon^2>1.5\times 10^{-13}(6.1\times10^{-13})$ at 95\% C.L. is obtained. 

The search for heavier dark photons would be possible by increasing the electron beam energy, but due to limited space in the deep underground Yemilab, this would be very challenging.

\subsubsection{Axion-like particle search}

Axions are one of the well-motivated and extensively investigated extensions of the SM as they can not only address the strong CP problem but serve as a dark-matter candidate. 
Theoretical studies have branched away from investigating pure QCD axions to more generic pseudo-scalars (often called axion-like particles or ALPs) which have properties similar to those of the QCD axions. 
Likewise, experimental programs for searching for axions have incorporated the ALPs into their scope. 

Depending on the underlying model details, ALPs can interact with a variety of SM particles. 
Among them, the parameter space of the ALP interacting with the SM photon has been most extensively explored, while the ALPs interacting with the SM electron are receiving growing attention. 
The related interaction Lagrangian contains the following operators:
\begin{equation}
    \mathcal{L}_{\rm int} \supset \frac{1}{4}g_{a\gamma} a F_{\mu\nu} \Tilde{F}^{\mu\nu} + i g_{ae} a \bar{e}\gamma_5 e\,,
\end{equation}
where $a$ and $e$ denote the ALP and SM electron, respectively, and $F_{\mu\nu}$ and $\tilde{F}_{\mu\nu}$ are the usual electromagnetic field strength tensor and its dual. 
$g_{a\gamma}$ and $g_{ae}$ parameterize the interaction strengths of the ALP with the photon and the electron, respectively. 
In general, ALPs have non-zero couplings to both the photon and the electron (and other species, e.g., quarks, gluons, and nucleons), but many of the searches have taken simplified-model approaches, i.e., the couplings are turned on one at a time. 
We follow the same strategy.

The ALPs interacting with photons can be produced via the Primakoff process, $\gamma + A \to a + A$, with $A$ being a tungsten atom inside the target. 
Once an electron beam impinges on the tungsten target, a number of secondary photons can be produced in the process of electromagnetic showering. 
For a more precise estimate of the photon flux, we perform a dedicated electron dump simulation using the \texttt{GEANT} code package~\cite{GEANT4:2002zbu}. 
The ALPs produced by the Primakoff process reach the detector and preferentially decay to a pair of photons or scatter off a nucleus inside the detector (i.e., the inverse Primakoff process, $a+A \to \gamma + A$)~\cite{Dent:2019ueq,Brdar:2020dpr}, depending on the parameter region of interest.
The LSC detector observes scintillation light induced by either decay or scattered photons. 

\begin{figure}[t]
    \centering
    \includegraphics[width=0.7\textwidth]{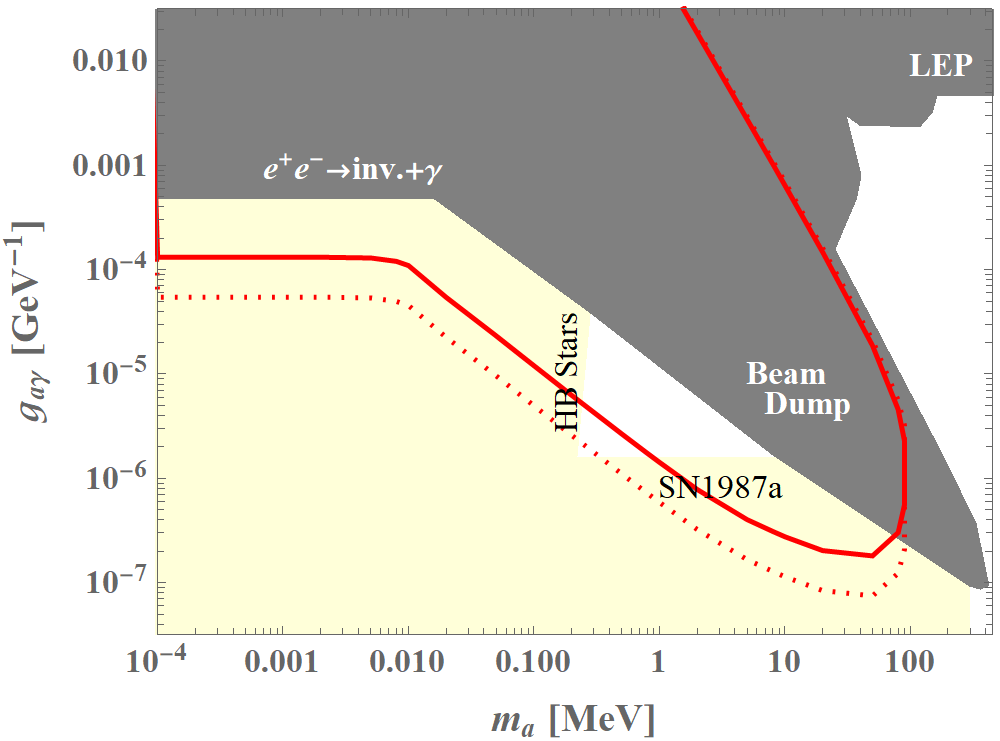} 
    \caption{The expected sensitivity reaches of ALPs interacting with the SM photon at the Yemilab neutrino detector for $g_{a\gamma}$ as a function of ALP mass $m_a$. 
    A one-year exposure is considered under the assumptions of zero backgrounds (dashed red line) and $\sim1000$ backgrounds (solid red line). A 100~MeV-100~kW electron beam is assumed to impinge on a tungsten target (50 cm). 
    The existing limits are shown based on the limits compiled in Refs. e.g.,~\cite{Bauer:2018uxu,Lanfranchi:2020crw}.}
    \label{fig:alpsearch}
\end{figure}

Figure~\ref{fig:alpsearch} shows the 90\% C.L. sensitivity reach of the LSC detector for $g_{a\gamma}$ as a function of the ALP mass $m_a$. 
Considering again a 100~MeV-100kW electron beam, we estimate the limits in the basis of expected statistical error only, assuming 0 (red dotted line) and $\sim1000$ (red solid line) background events for a year of exposure. 
We also show the existing excluded regions based on the limits compiled in Refs. e.g., \cite{Bauer:2018uxu,Lanfranchi:2020crw}. 
The gray-shaded regions show the limits from the laboratory-produced ALP searches, while the regions constrained by astrophysical ALP searches (e.g., HB stars, supernovae, etc) are yellow-shaded.
Our initial estimates suggest that LSC can allow for the exploration of the so-called ``cosmological triangle'' region surrounded by the beam-dump, HB stars, and SN1987a limits. 
In addition, LSC can probe the regions constrained only by astrophysical considerations, which depend on the underlying assumptions~\cite{Jaeckel:2006xm,Khoury:2003aq,Masso:2005ym,Masso:2006gc,Dupays:2006dp,Mohapatra:2006pv,Brax:2007ak,Dev:2021ofc,Fortin:2021cog}, and set the model-independent limits.

The secondary photons are capable of producing the ALPs via a Compton-like process, $\gamma +e^- \to a + e^-$, in the presence of non-vanishing $g_{ae}$. Again, the produced ALPs reach the detector and leave the experimental signature in two ways: they preferentially decay to an electron-positron pair if the ALP mass is greater than twice the electron mass or undergo an inverse Compton-like scattering process, $a + e^- \to \gamma + e^-$ if the decay is kinematically forbidden. 

\begin{figure}[t]
    \centering
    \includegraphics[width=0.7\textwidth]{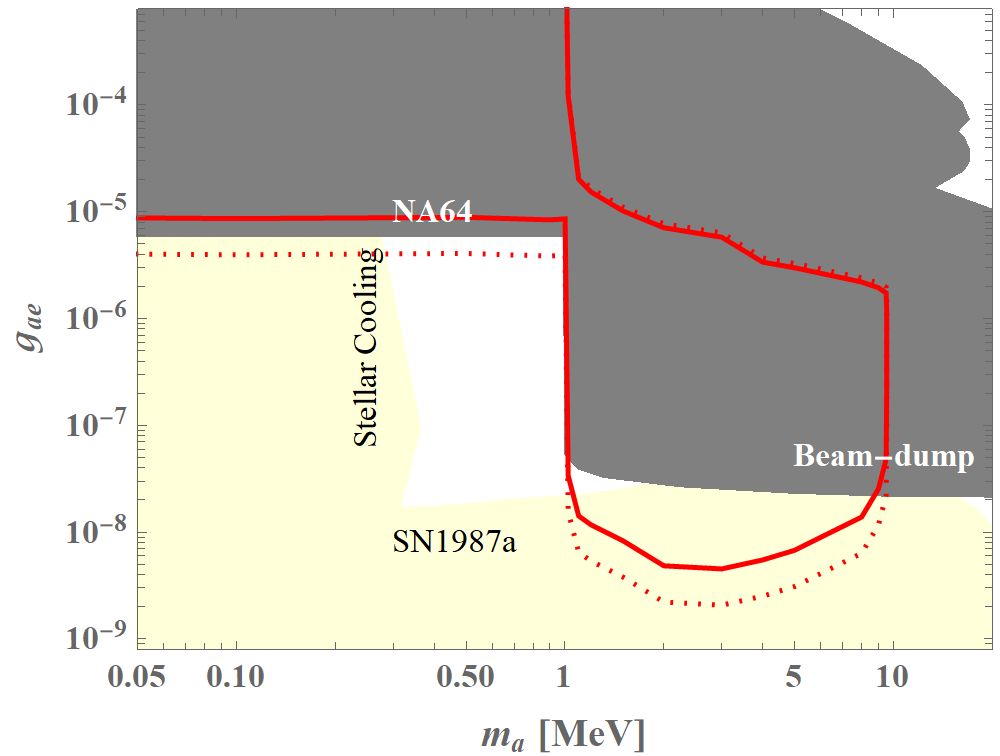}
    \caption{The expected sensitivity reaches of ALPs interacting with the electron at the Yemilab neutrino detector for $g_{ae}$ as a function of ALP mass $m_a$. 
    A one-year exposure is considered under the assumptions of zero backgrounds (dashed red line) and $\sim1000$ backgrounds (solid red line). 
    A 100~MeV-100~kW electron beam is assumed to impinge on a tungsten target (50 cm). 
    The existing limits are shown based on the limits compiled in Ref. e.g.,~\cite{CCM:2021lhc}. }
    \label{fig:alpgaee}
\end{figure}

Figure~\ref{fig:alpgaee} displays the sensitivity reaches of the LSC detector for $g_{ae}$ as a function of $m_a$. 
As before, the red solid (dashed) line shows the sensitivity curve under the assumptions of $\sim1000$ (zero) background events and a one-year exposure.
The currently excluded regions also appear based on the limits compiled in Ref. e.g.,~\cite{CCM:2021lhc}. 
Again the gray-shaded regions show the limits from the laboratory-produced ALP searches, while the regions constrained by astrophysical ALP considerations (e.g., stellar cooling, supernovae, etc) are yellow-shaded. 
Our initial estimates suggest that LSC can explore the regions covered only by astrophysical considerations, which again depend on the underlying assumptions, and allow us to set the laboratory-based model-independent limits.

\subsubsection{Light dark matter search}

One of the well-motivated dark-matter candidates is (MeV-range) vector-portal dark matter. 
The underlying idea is that dark-sector particles including dark matter can interact with the SM particles via a vector mediator, for example, dark photon in Section~\ref{sec:darkphoton}.
The interaction Lagrangian is extended to include the coupling of the dark photon to dark-matter species $\chi$. 
While the main search strategy is straightforwardly relevant to other types of dark matter, we here assume that $\chi$ is fermionic for purposes of illustration, having the following interaction structure, 
\begin{equation}
    \mathcal{L}_{\rm int} \supset -\epsilon e \bar{f}\gamma^\mu f A'_\mu - g_D \bar{\chi} \gamma^\mu \chi A'_\mu\,,
\end{equation}
where $g_D$ parameterizes the dark-sector coupling strength. 

Under this model setup, dark photons are produced in the tungsten target by the electron beam bremsstrahlung mentioned in Section~\ref{sec:darkphoton}. 
To explore the full capability of the Yemilab linac facility, we include $A'$ bremsstrahlung of secondary electrons and positrons in electromagnetic showers, associated production of $A'$, i.e., $e^++e^- \to A' + \gamma$, and the resonance production via the secondary positrons, i.e., $e^+ +e^- \to A'$. 
Once produced, $A'$ promptly decays to a pair of dark-matter particles, i.e., $A' \to \bar{\chi} \chi$, and the produced dark matter entering the detector can scatter off the electron, leaving an electron recoil, i.e., $\chi + e^- \to \chi + e^-$.

\begin{figure}[t]
    \centering
    \includegraphics[width=0.7\textwidth]{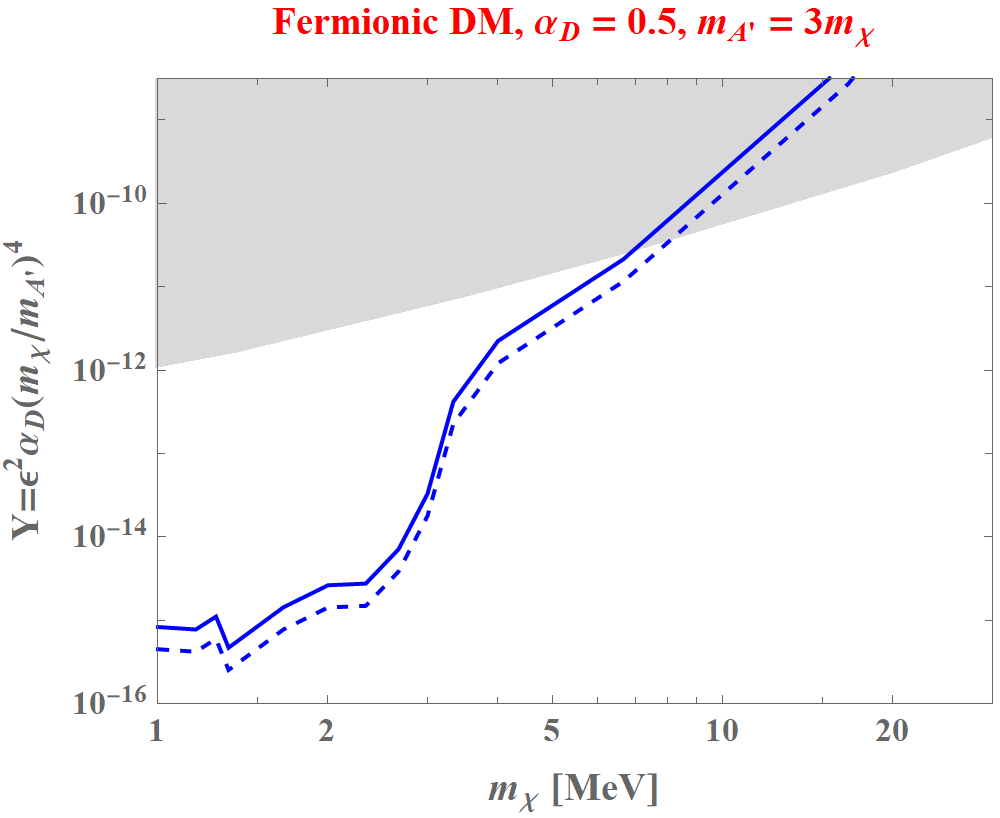}
    \caption{The expected light dark-matter sensitivity reaches at the Yemilab neutrino detector for $Y=\epsilon^2\alpha_D(m_\chi/m_{A'})^4$ as a function of dark-matter mass $m_\chi$. 
    A one-year exposure is considered under the assumptions of zero backgrounds (dashed blue line) and $\sim1000$ backgrounds (solid blue line). 
    A 100~MeV-100~kW electron beam is assumed to impinge on a tungsten target (50~cm). 
    The leading limits are from the NA64 measurements~\cite{Andreev:2021fzd} and they are shown by the gray-shaded region. }
    \label{fig:LSC-ldm}
\end{figure}

Our result is shown in Fig.~\ref{fig:LSC-ldm}. 
The sensitivity of the LSC is estimated for $Y=\epsilon^2\alpha_D(m_\chi/m_{A'})^4$ as a function of dark-matter mass with the dark-sector coupling $\alpha_D\equiv g_D^2/(4\pi)$ to be 0.5 and $m_{A'}=3m_\chi$.  
As in previous subsections, we consider a 100~MeV-100~kW electron beam striking a tungsten target and one-year exposure of the neutrino detector to the beam. 
Again, two background assumptions are taken into account: zero backgrounds (blue dashed line) and $\sim1000$ backgrounds (blue solid line). 
The most stringent limits are from NA64~\cite{Andreev:2021fzd} and they are shown by the gray-shaded region. 
Our initial estimates suggest that LSC can probe a broad range of unexplored regions of parameter space, especially below $m_\chi\sim 7$~MeV. 

%% file: Physics/CosmogenicBSM.tex
\subsection{Cosmogenic BSM}

Signals from outer space and their detection have been playing a crucial role in particle physics, especially in searches for physics beyond the Standard Model (BSM); for example, the DM-induced neutrinos are important for indirect DM searches. 
Beyond the neutrino signals from the annihilation or decay of DM, a wide range of new models and dark-sector scenarios predicting cosmogenic and non-conventional DM signals have been suggested in the last decade, largely motivated by the null observation of WIMP signals.
Among them, scenarios of cosmogenic boosted dark matter (BDM) and related phenomenology have been most extensively investigated, including semi-annihilation, models of annihilation/decay two-component DM, and cosmic-ray/neutrino-induced BDM  (see, e.g., Ref.~\cite{Berger:2022cab}).
The underlying idea is that a small fraction of DM (or DM components) can be relativistic in the present universe, whereas the usual (cold) halo DM is somehow less sensitive to the existing experiments.
The resulting signals often have less intense and more energetic fluxes compared to the majority of halo DM.
In addition, energetic dark-sector particles can emerge in the atmosphere, e.g., energetic cosmic rays impinging on the atmosphere can copiously produce various relativistic BSM particles.
These scenarios also predict energetic cosmogenic signals.
Therefore, kiloton-scale neutrino detectors such as the LSC are excellent places to explore those new opportunities~\cite{Kim:2020ipj}.

\subsubsection{Boosted dark matter}

The novel phenomena of BDM usually arise in models beyond the minimal WIMP scenario, where a small relativistic component of DM can be produced and detected through its interactions with the SM particles.
A BDM flux can emerge from a number of dark-sector scenarios, such as dark-sector structures~\cite{DEramo:2010keq, Belanger:2011ww, Agashe:2014yua, Berger:2014sqa, Kong:2014mia, Kim:2016zjx, Chigusa:2020bgq, Toma:2021vlw, Bhattacharya:2014yha, Kopp:2015bfa, Bhattacharya:2016tma, Heurtier:2019rkz}, DM-induced nucleon decay~\cite{Davoudiasl:2010am, Huang:2013xfa}, charged cosmic-ray acceleration~\cite{Bringmann:2018cvk, Ema:2018bih, Cappiello:2019qsw, Dent:2019krz, Wang:2019jtk, Ge:2020yuf, Cao:2020bwd, Jho:2020sku, Cho:2020mnc, Dent:2020syp, Bell:2021xff, Xia:2021vbz}, cosmic-ray neutrino acceleration~\cite{Jho:2021rmn, Das:2021lcr, Chao:2021orr}, astrophysical processes~\cite{Kouvaris:2015nsa, Hu:2016xas, An:2017ojc, Emken:2017hnp, Calabrese:2021src, Wang:2021jic} or inelastic collision of cosmic rays with the atmosphere~\cite{Alvey:2019zaa, Su:2020zny}.
The detection of BDM could be a smoking gun signal for the discovery of DM when it is challenging to detect the dominant component of cold DM.
However, it requires new experimental approaches beyond conventional (cold relic) DM direct detection experiments which focus on low recoil energy (with the exception of low-mass DM~\cite{Cherry:2015oca, Giudice:2017zke, Kim:2020ipj}).
Due to the relatively small flux of BDM but the energetic states that are generically produced as the outcome of BDM interactions in detectors, large volume neutrino detectors are attractive and sensitive facilities for BDM searches.
Various phenomenological studies have demonstrated promising BDM sensitivities at dark matter and neutrino experiments,
while several dark matter and neutrino experiments have reported dedicated searches for BDM, expanding the bounds in the related models~\cite{Super-Kamiokande:2017dch, COSINE-100:2018ged, PandaX-II:2021kai, CDEX:2022fig, Super-Kamiokande:2022ncz}.

We investigate the detection potential of BDM signals in LSC taking the annihilating two-component DM model~\cite{Belanger:2011ww} as the benchmark scenario.
The dark sector includes two DM species whose stability is protected by unbroken separate symmetries such as ${\rm U}(1)'\otimes{\rm U}(1)''$ or $Z_2\otimes Z_2'$.
The heavier DM particle (say $\chi_0$) does not have direct coupling to the SM particles, while the lighter one (say $\chi_1$) directly interact with the SM particles.
In addition, the sizable direct interaction between $\chi_0$ and $\chi_1$ is allowed.

In this scenario, a relativistic $\chi_1$ is produced from the annihilation of $\chi_0$, $\chi_0\bar{\chi}_0 \rightarrow \chi_1\bar{\chi}_1$, and obtains a sizable boost factor which is given by the mass ratio between $\chi_0$ and $\chi_1$, i.e., $\gamma_1=m_0/m_1$.
Assuming that $\chi_0$ is the dominant DM species and follows the Navarro-Frenk-White (NFW) profile~\cite{Navarro:1995iw,Navarro:1996gj} in the galaxy, we obtain the expected flux of boosted $\chi_1$ near the Earth from all sky~\cite{Agashe:2014yua, Kim:2018veo}:
\begin{eqnarray}
\frac{d\Phi_1}{dE_1} &=& 
\frac{1}{4\pi}\int d\Omega \int_{\rm l.o.s.} ds \langle \sigma v \rangle _{\chi_0\bar{\chi}_0 \rightarrow \chi_1\bar{\chi}_1} \frac{dN_1}{dE_1} \left(\frac{\rho_0(r(s,\theta))}{m_0}\right)^2  \nonumber \\
&=& 1.6\times 10^{-4}~{\rm cm}^{-2}{\rm s}^{-1}\times \left(\frac{\langle \sigma v \rangle _{\chi_0\bar{\chi}_0 \rightarrow \chi_1\bar{\chi}_1}}{5\times 10^{-26}~{\rm cm}^3\,{\rm s}^{-1}} \right) \left( \frac{{\rm GeV}}{m_0}\right)^2 \frac{dN_1}{dE_1}\,, \label{eq:fluxAnnihilation}
\end{eqnarray}
where $\langle \sigma v \rangle _{\chi_0\bar{\chi}_0 \rightarrow \chi_1\bar{\chi}_1}$ is the velocity-averaged annihilation cross section in the galaxy today, $\rho_0$ is the $\chi_0$ density in our galaxy as a function of the distance $r$ to the galactic center (GC), $s$ is the line-of-sight distance to the Earth, $\theta$ is its angular direction relative to the GC-Earth axis, and $\Omega$ is the solid angle.
The annihilation of $\chi_0$ yields a pair of mono-energetic $\chi_1$ particles whose differential spectrum is simply given by
\begin{equation}
\frac{dN_1}{dE_1}=2\delta(E_1 - m_0)\,.
\end{equation}
We assume here that $\chi_0$ and its antiparticle $\bar{\chi}_0$ are distinguishable and thus their fractions are same.
Therefore, an additional pre-factor 2 is needed for the indistinguishable case. 
As variations of the annihilating galactic BDM, solar captured BDM scenarios~\cite{Berger:2014sqa, Kong:2014mia} have been also studied where the heavier (dominant) DM species $\chi_0$ can be efficiently captured in the Sun and annihilate into the relativistic lighter component $\chi_1$.
For the solar captured BDM, a sizable self-interaction of the heavier DM greatly enhances the capture rate in the Sun and results in the explorable BDM fluxes at current and future neutrino experiments~\cite{Kong:2014mia}.

We are now in the position to discuss phenomenology of the aforementioned annihilating two-component BDM model at LSC.
The expected number of signal events $N_{\rm sig}$ is given by
\begin{align}
N_{\rm sig} = \sigma_\epsilon \;\mathcal F \;  t_{\rm exp} \; N_T\, , \label{eq:Nsig}
\end{align}
where $T$ stands for the target, $\sigma_\epsilon$ is the scattering cross section for the process $\chi_1 T \to \chi_1 T$, ${\mathcal F}$ symbolizes the flux of boosted $\chi_1$, $t_{\rm exp}$ is the exposure time, and $N_T$ denotes the number of target particles in the detector fiducial volume $V_{\rm fid}$ of interest.
Note that the fiducial cross section $\sigma_\epsilon$ includes realistic effects such as the acceptance from cuts, detector response, and threshold energy $E_{\rm th}$.

We can translate the expected number of events into the search limits in terms of model parameters if the interaction between the $\chi_1$ and SM particles is specified.
In this analysis, we choose a dark photon scenario for illustration in which the relevant Lagrangian terms are summarized as
\begin{align}
\mathcal L \supset -\frac{\epsilon}{2}F_{\mu\nu}X^{\mu\nu} + g_D \bar{\chi}_1\gamma^{\mu}\chi_1 X_\mu \,, \label{eq:lagrangian}
\end{align}
where the first term implies the kinetic mixing between U(1)$_{\rm EM}$ and U(1)$_{\rm X}$ parameterized by the small number $\epsilon$, $F_{\mu \nu}$ and $X_{\mu \nu}$ are the field strength tensors for the SM photon and the dark photon, respectively.
The second term with the interaction strength parameterized by $g_D$ describes the coupling of the dark sector to the SM sector, mediated by the dark photon $X_\mu$.

\begin{figure*}[h]
\begin{center}
\includegraphics[width=0.5\textwidth]{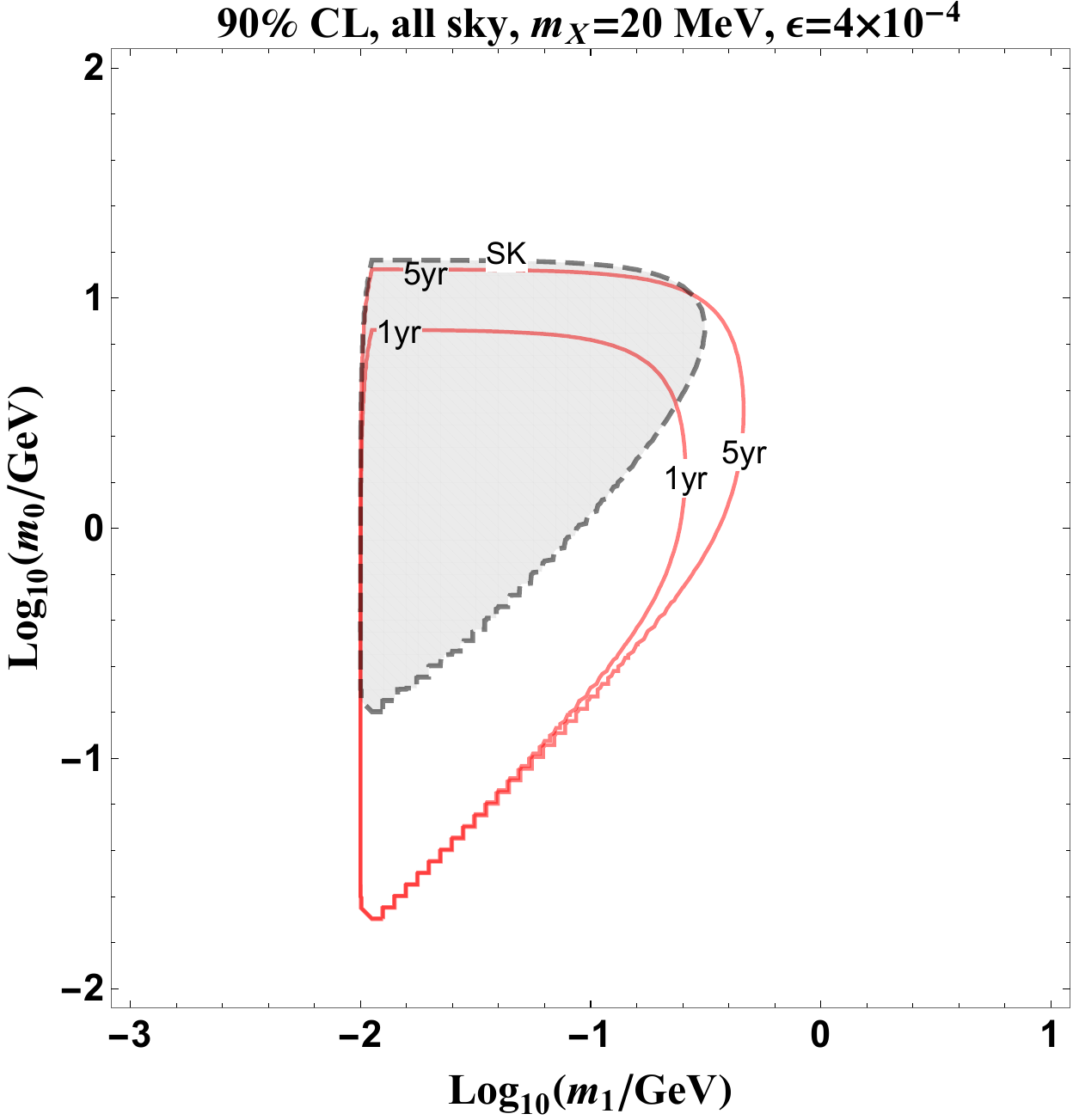}
\end{center}
\caption{
The expected $90\%$ CL sensitivities from 1-year and 5-year running of LSC with $V_{\rm fid}=2.26~{\rm kt}$ and $E_{\rm th}=3~{\rm MeV}$ for the BDM flux from all sky, assuming $\sim1000/ {\rm year}$ backgrounds.
The gray-shaded area represents
the $90\%$ exclusion limit with all sky data from SK (13.6 year)~\cite{Dziomba:2012paz, Super-Kamiokande:2015qek}, assuming $10\%$ systematic
uncertainty in the estimation of background number of events.
}
\label{f:GC-BDM_Sensitivity}
\end{figure*}

In Fig.~\ref{f:GC-BDM_Sensitivity}, we demonstrate the expected $90\%$ CL exclusion bounds from 1-year and 5-year running of 2.26 kt-$V_{\rm fid}$ for all-sky data in the standard parameterization of $m_0$ versus $m_1$, fixing $m_X=20$ MeV, $g_D=1$ and $\epsilon = 4\times 10^{-4}$.
Here we assume $\sim1000$ background events per year.
The gray-shaded area represents the $90\%$ CL exclusion bound inferred from the atmospheric neutrino measurement for the whole sky in Ref.~\cite{Dziomba:2012paz, Super-Kamiokande:2015qek} collected for 13.6 years by the SK Collaboration.
We use the fully contained single-ring $e$-like events including both sub-GeV (0-decay electron events only) and multi-GeV as a conservative estimation of a total of 10.7 years~\cite{Dziomba:2012paz} and normalize the event rate to 13.6 years~\cite{Super-Kamiokande:2015qek}.
For the analysis, we include $10\%$ systematic uncertainty in the estimation of the number of background events.
Consequently, LSC can access a large area of parameter space of the benchmark model unexplored by
the SK experiment.

\begin{figure}[t]
    \centering
    \includegraphics[width=0.49\textwidth]{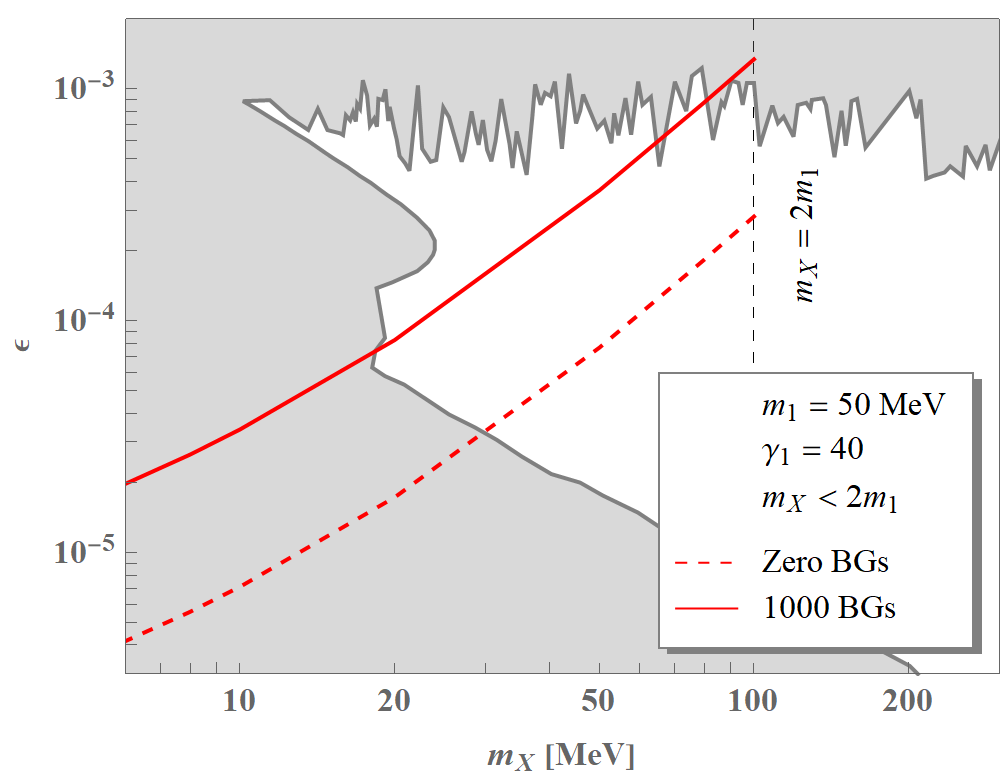}
    \includegraphics[width=0.49\textwidth]{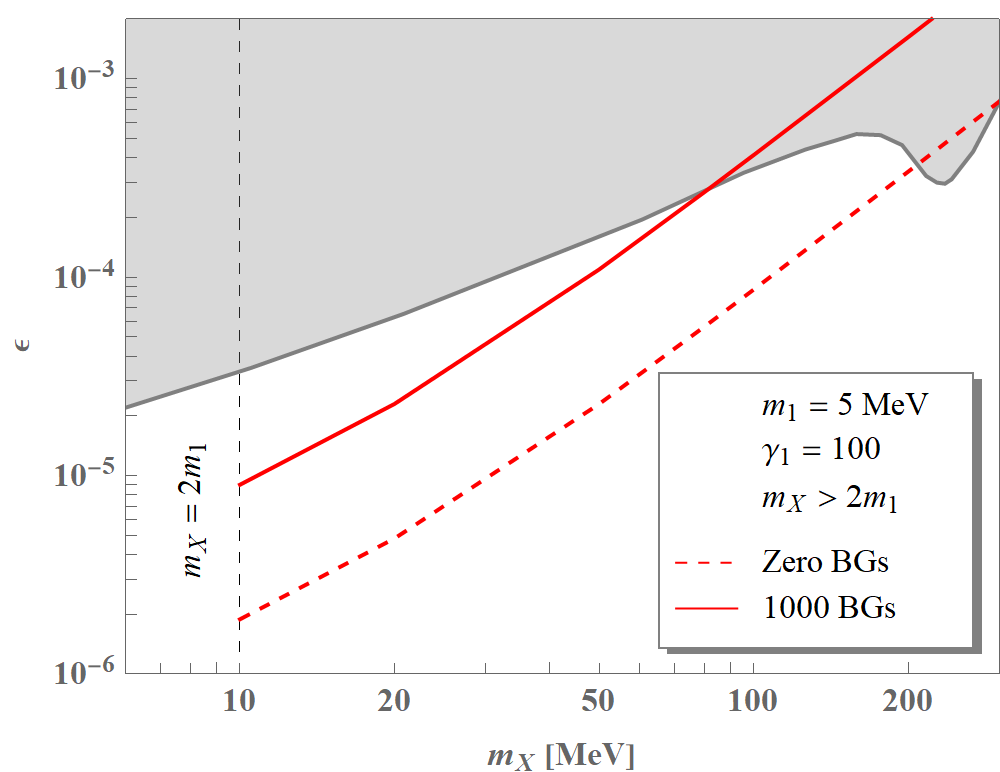}
    \caption{Experimental sensitivities of LSC in the dark photon model parameters $m_X-\epsilon$ for the cases of $m_X<2m_1$ (left) and $m_X>2m_1$ (right). 
    A one-year exposure is considered with zero backgrounds (red dashed lines) and $\sim1000$ backgrounds (red solid lines). 
    Mass hierarchies between $\chi_0$ and $\chi_1$ are shown in the legends and $g_D=1$ is kept for both cases. 
    The gray-shaded regions show the currently excluded parameter space according to the reports in Ref.~\cite{NA64:2018lsq} (left) and Ref.~\cite{Andreev:2021fzd} (right).}
    \label{fig:BDMDP}
\end{figure}

We also investigate the BDM search prospects of LSC in different parameter spaces. 
Since the Lagrangian in Eq.~\eqref{eq:lagrangian} suggests that the BDM can interact with SM particles via a dark photon, we estimate the sensitivity reaches for the kinetic mixing parameter $\epsilon$ as a function of dark photon mass $m_X$. 
Our results are shown in Fig.~\ref{fig:BDMDP}. 
Since a dark photon predominantly decays to invisible $\chi_1$ (visible SM particles) for $m_X> 2m_1$ ($m_X<2m_1$), we report the results separately. 
First, the left panel of Fig.~\ref{fig:BDMDP} shows the 90\% CL sensitivity with $m_0=2$~GeV and $m_1=50$~MeV, assuming the scenario of $m_X<2m_1$, i.e., visibly decaying dark photon.
We again consider a one-year exposure and two background assumptions, zero backgrounds (dashed red line) and $\sim1000$ backgrounds (solid red line). 
We take the existing limits compiled in Ref.~\cite{NA64:2018lsq}, showing them by a gray-shaded region. 
On the other hand, the right panel of Fig.~\ref{fig:BDMDP} shows the 90\% CL sensitivity with $m_0=500$~MeV amd $m_1=5$~MeV, assuming the scenario of $m_X>2m_1$, i.e., invisibly decaying dark photon. 
The leading limits are from the NA64 measurements~\cite{Andreev:2021fzd}. 
Considering these initial estimates with some benchmark points, we find that LSC can test a wide range of BDM scenarios in terms of dark-photon parameter space.

\subsubsection{Atmospheric collider}

Cosmic rays impinging on the atmosphere copiously produce various particles, analogous to terrestrial colliders. 
The ``atmospheric collider'' has historically played a crucial role in neutrino physics, leading to the discovery of neutrino oscillations~\cite{Super-Kamiokande:1998kpq}.
The atmospheric collider is also a natural laboratory for exploration of BSM physics, including light DM~\cite{Alvey:2019zaa, Su:2020zny}, millicharged particles~\cite{Plestid:2020kdm, ArguellesDelgado:2021lek}, monopoles~\cite{Iguro:2021xsu}, and heavy neutral leptons~\cite{Coloma:2019htx}. 
The resulting flux of boosted particles from the atmospheric collider can be readily searched for in DM and neutrino experiments. 
The beam of atmospheric collider is always ``ON'', it is robust and independent of local particle abundance, it extends over a broad spectrum from MeV to over PeV, and it is available for all terrestrial experiments, especially large volume detectors including LSC.
These features highlight atmospheric collider as an excellent laboratory for further exploration of new physics at LSC. 

%% file: summary.tex
\section{Summary \& Conclusion}

Due to the reduced muon flux, an underground lab is ideal for neutrino and rare event search experiments in (astro-)particle physics. Demand for underground labs has been increasing, and as a result, existing labs are expanding, or new underground labs are being constructed/planned.

Yemilab is the 1st deep underground lab dedicated to science in Korea, and its construction was finished in 2022. In Yemilab, a large cylindrical cavern (D: 20~m, H: 20~m) was also built, as well as two main sites for dark matter and 0$\nu\beta\beta$ experiments being installed, where a kiloton scale LS-based neutrino detector (LSC) could be installed in the future. However, the LSC detector, in principle, can be placed in any underground lab where there is a demand for physics with a kiloton-scale LS detector. 

candidate design of the detector consists of 3 cylindrical layers, from innermost to outermost, target (D: 15~m, H: 15~m) filled with LS ($\sim$2.26 kton), buffer (D: 17~m, H: 17~m) filled with mineral oil where a few thousand 20-inch PMTs to be attached to the wall of the buffer, and veto (D: 20~m, H: 20~m) filled with purified water. 

Using a $\sim$2.26 kiloton ultra-purified LS target, $\sim$8 times bigger than Borexino, precise measurements of solar neutrino fluxes are possible. This could either consolidate the standard solar model or may reveal new physics. Additionally, it is possible to precisely measure solar metallicity critical to determine the fate of the Sun, whether low or high metalicities, that has not yet been settled for several decades.

LSC can detect geoneutrinos emitted from natural radioactive decays of $^{238}$U, $^{232}$Th, and $^{40}$K, arising from the interior of the Earth. About 60 geoneutrinos per year are expected to be detected through the inverse beta decay (IBD) process, which has a drawback of a 1.8~MeV energy threshold blocking most of the geoneutrino flux and collecting a huge background from reactor neutrinos (from 460 to 1500 events/year, depending on the number of operational reactors) from the Hanul nuclear reactor complex, $\sim$60~km away. Via an elastic scattering process, geoneutrinos from $^{40}$K decay can be effectively detected, outnumbering the reactor neutrino background by a factor of $\sim$2.5. 

The LSC can participate in the Supernova Early Warning System by observing Supernova burst neutrinos. A few hundred Supernova burst neutrinos at 10~kpc are expected to be detected, where the number can vary from $\sim$400 to $\sim$800 depending on a progenitor mass. The LSC is not designed to detect DSNB (Diffuse SuperNova Background), but it is expected to observe 2 signal events over 0.5 residual background events for 10 years. The statistical median sensitivity for a 10-year measurement is expected to be 97.5\% C.L. 

With an electron linac (100~MeV, 100~kW) facility, a light dark photon search is possible, where dark photons are produced via the ``darkstrahlung'' process and detected by visible decays. A competitive limit on visible decays of dark photons whose mass from 1~MeV to 30~MeV can be achieved assuming 1000 background for one year of operation time. For sub-MeV dark photons, the competitive limit on ``direct'' dark photon search via oscillations between dark and nominal photons can be achieved down to the $\sim$eV range. 

Using an IsoDAR facility where a proton beam (60~MeV, 600~kW) from a cyclotron strikes a $^8$Be target enclosed by a $^7$Li sleeve, the contradictory measurements on sterile vs. no sterile neutrinos could be definitively resolved thanks to a wide range of L/E that could distinguish (3+1)$\nu$, (3+2)$\nu$, and more exotic scenarios, using $\sim$2 million IBD events collected over 5 calendar years.
The IsoDAR experiment can also collect about 7\,000 elastic scattering events over 
5 calendar years. These elastic scattering events allow us to precisely measure 
the weak mixing angle at the momentum transfer of a few MeV and the neutrino non-standard 
interaction parameter space ($\epsilon_{ee}^{L}$ vs. $\epsilon_{ee}^{R}$) could be 
well constrained. In addition, IsoDAR combined with the LSC is uniquely sensitive to a number of new particle searches, including axion-like-particles, mirror neutrons, and low-mass mediators. 

The LSC can also search for sterile neutrinos using a $^{144}$Ce (100 kiloCurie) radioactive source, as SOX had attempted but never realized. The exclusion sensitivity using the radioactive source in LSC is not as good as that of IsoDAR in LSC, but a comparable result is expected while further study is still needed. 

An Axion-like particle (ALP) search is also possible using the electron linac and IsoDAR facilities. An ALP may be produced from the Primakoff process, and the produced ALP can be detected by either ALP decay (to two photons) or a scattered photon from the inverse Primakoff process.  Competitive sensitivity on an ALP-photon coupling can be obtained in a direct search for ALP masses below 80~MeV.  The ALa can also be produced from a Compton-like process, and the produced ALPs can be detected either by e$^+$e$^-$ decay or an inverse Compton-like process. The competitive sensitivity on ALP-electron coupling is obtained in a direct search for ALP masses between 1 and 8~MeV. Using the electron linac facility, a light DM search is also possible. Light DM is produced from dark photon decay, where dark photons are produced by either Bremsstrahlung or resonance from e$^+$e$^-$ annihilation. The produced light DM can detected by measuring the recoil energy of an electron from its elastic scattering with the DM. Competitive sensitivity for light DM between a mass of 1 and 7~MeV is expected.

The LSC has a sensitivity to a boosted DM (BDM) search from the sky. 
The sensitivity better than existing limits is expected for mass ranges of m$_0$ in O(10~MeV - 10~GeV) and m$_1$ in O(10~MeV - 100~MeV), where m$_0$ and m$_1$ are masses of heavier and lighter DM species, respectively. 
The lighter species can be produced from self-annihilation of the heavier even in the present universe, and then boosted due to the mass difference.

After completing the studies described above, the LSC detector could be upgraded to search for $0\nu\beta\beta$ decay, but this is out of the scope of this article. 
The LSC requires some R\&D work for LS purification at the level of Borexino for a few years until the funding is approved. Detector construction could take about 4 years. If the LSC is ever built in any underground lab, including Yemilab, it could bring several world-leading results, especially when combined with linac and/or cyclotron facilities.